\documentclass[11pt,preprint]{aastex}
\usepackage{times}
\usepackage{amssymb}
\usepackage{amsmath}
\usepackage{xspace}
\usepackage{natbib}

\newcommand{\sax}{\emph{BeppoSAX}\xspace}
\newcommand{\xmm}{\emph{XMM-Newton}\xspace}
\newcommand{\xmmrgs}{\emph{XMM-Newton/RGS}\xspace}
\newcommand{\suzaku}{\emph{Suzaku}\xspace}
\newcommand{\asca}{\emph{ASCA}\xspace}
\newcommand{\ginga}{\emph{Ginga}\xspace}

\newcommand{\chandra}{\emph{Chandra}\xspace}
\newcommand{\chandrahetg}{\emph{Chandra}/HETG\xspace}

\newcommand{\exosat}{\emph{EXOSAT}\xspace}
\newcommand{\fuse}{\emph{FUSE}\xspace}

\newcommand{\vunit}{\mbox{\,km\,s$^{-1}$}}
\newcommand{\vout}{v_{\rm out}}

\newcommand{\Msun}{\hbox{$\rm\thinspace M_{\odot}$}}
\newcommand{\ls}
{\mathrel{\hbox{\rlap{\hbox{\lower4pt\hbox{$\sim$}}}\hbox{$<$}}}}
\newcommand{\gs}
{\mathrel{\hbox{\rlap{\hbox{\lower4pt\hbox{$\sim$}}}\hbox{$>$}}}}

\newcommand{\logxi}{\log (\xi/\rm{erg\,cm\,s}^{-1})}
\newcommand{\lognh}{\log (N_{\rm H}/\rm{cm}^{-2})}

\received{}
\begin{document}
\title{A High Resolution View of the Warm Absorber in the Quasar MR\,2251-178}
\shorttitle{The Warm Absorber in MR\,2251-178}
\shortauthors{Reeves et al.}
\author{J.N. Reeves\altaffilmark{1,2}, D. Porquet\altaffilmark{3}, V. Braito\altaffilmark{4}, 
J. Gofford\altaffilmark{1}, E. Nardini\altaffilmark{1}, 
T. J. Turner\altaffilmark{2}, D. M. Crenshaw\altaffilmark{5}, S. B. Kraemer\altaffilmark{6}}
\altaffiltext{1}{Astrophysics Group, School of Physical and Geographical Sciences, Keele 
University, Keele, Staffordshire, ST5 5BG, UK; j.n.reeves@keele.ac.uk}
\altaffiltext{2}{Center for Space Science and Technology, 
University of Maryland 
Baltimore County, 1000 Hilltop Circle, Baltimore, MD 21250, USA}
\altaffiltext{3}{Observatoire Astronomique de Strasbourg, Universit{\'e}
  de Strasbourg, CNRS, UMR 7550, 11 rue de l'Universit{\'e}, F-67000
  Strasbourg, France}
\altaffiltext{4}{INAF - Osservatorio Astronomico di Brera, Via Bianchi 46 I-23807 Merate (LC), Italy}
\altaffiltext{5}{Department of Physics and Astronomy, Georgia State University, Astronomy Offices, One Park Place South SE, Suite 700, Atlanta, GA 30303, USA}
\altaffiltext{6}{Institute for Astrophysics and Computational Sciences, Department of Physics, The Catholic University of America, Washington, DC 20064, USA}

\begin{abstract}

High resolution X-ray spectroscopy of the warm absorber in the nearby quasar, MR\,2251-178 ($z=0.06398$) 
is presented. The observations were carried out in 2011 
using the Chandra High Energy Transmission Grating 
and the XMM-Newton Reflection Grating Spectrometer, with net exposure times of 
approximately 400\,ks each. A multitude of absorption lines from C to Fe are detected, 
revealing at least 3 warm absorbing components ranging in 
ionization parameter from 
$\logxi = 1-3$ and with outflow velocities $\ls500$\,km\,s$^{-1}$.  
The lowest ionization absorber appears 
to vary between the Chandra and XMM-Newton observations, which implies a radial distance 
of between $9-17$\,pc from the black hole. The soft X-ray 
warm absorbers likely contribute a negligible $<0.01$\% of the bolometric output in terms of 
their kinetic power.
Several broad soft X-ray emission lines are strongly detected, most notably from He-like Oxygen, 
with FWHM velocity widths of up to 10000\,km\,s$^{-1}$, consistent with an origin from Broad Line 
Region (BLR) clouds. 
In addition to the warm absorber, gas partially covering the line of 
sight to the quasar appears to be present, of typical column density $N_{\rm H}=10^{23}$\,cm$^{-2}$. 
We suggest that the partial covering absorber may arise from the same BLR clouds responsible 
for the broad soft X-ray emission lines. 
Finally the presence of a highly ionised outflow 
in the iron K band from both 2002 and 2011 Chandra HETG observations appears to be confirmed, 
which has an outflow velocity of $-15600\pm2400$\,km\,s$^{-1}$. However  
a partial covering origin of the iron K band absorption cannot be excluded, resulting 
from low ionization material with little or no outflow velocity.

\end{abstract}


\section{Introduction}

Photo-ionized or ``warm'' absorbers are commonly observed in at least 50\% of the 
UV/X-ray spectra of Seyfert 1s and type-1 QSO and are an important constituent of AGN 
\citep[e.g.,][]{reynolds1997,crenshaw2003,porquet2004,blustin2005}. 
The Seyfert warm absorbers that are frequently observed at 
high spectral resolution with {\sl XMM-Newton} and {\sl Chandra} are now known to give rise to 
numerous narrow absorption lines, usually blue-shifted, implying outflowing winds 
of a few hundred km\,s$^{-1}$ up to a few thousand  km\,s$^{-1}$. These arise  
from various elements over a wide range of ionization
parameters, especially from carbon, nitrogen, oxygen, neon, silicon,
  sulfur, and iron \citep[e.g.,][]{kaastra2000,kaspi2002,blustin2002,mckernan2003}.

X-ray spectral signatures of the warm absorber 
range from the lowly ionized  Unresolved Transition Array (UTA) of M-shell iron ($<$ Fe\,\textsc{xvii})
at $\sim16$\AA~ \citep{sako2001,behar2001} to absorption from highly ionized 
(H-like and He-like) iron which may originate from an accretion disk wind 
\citep[e.g.,][]{reeves2004a,risaliti2005,braito2007,turner2008,tombesi2010a,tombesi2010b,gofford2013}.
These spectroscopic measurements can
reveal crucial information on the outflow kinematics, physical conditions and locations 
relative to the central continuum source -- ranging from the inner nucleus (0.01 pc) to the 
galactic disk or halo (10 kpc) -- which can ultimately unveil the
inner structure of quasars \citep{elvis2000}. 

 The warm absorption signatures observed in the soft X-ray band cover a wide
range of column densities and ionization parameters from $\lognh\sim20-23$ and
  $\logxi\sim -1-3$\footnote{The ionization parameter is defined as $\xi=L_{\rm
    ion}/nR^{2}$ (\citealt{tarter1969}), where $L_{\rm ion}$ is the
  $1-1000$ Rydberg ionizing luminosity, $n$ is the electron density
  and $R$ is the distance of the ionising source from the absorbing
  clouds.} 
These warm absorbers are thought to be typically located on fairly
  large distances from the central black hole, from
    their low ionization parameter and velocity values, 
their (relative) lack of variability, plus in some
    cases from their inferred low densities (e.g., NGC\,3783:
    \citealt{Behar2003,Krongold2005}; Mrk\,279:
    \citealt{Scott2004,Ebrero2010}; NGC\,4051:
    \citealt{Steenbrugge2009}; Mrk\,290: \citealt{Zhang2011}; and
    Mkn\,509: \citealt{Kaastra2012}). These soft X-ray warm
    absorbers can be associated with, for example, 
  a wind originating from the putative parsec scale torus
  (\citealt{blustin2005}) or the latter stages of an accretion
  disc wind which has propagated out to larger radii
  (\citealt{proga&kallman2004,Tombesi2013}). By virtue of their low outflow
  velocities the soft X-ray warm absorbers are thought to only have a
  weak feedback effect in their host galaxy. Indeed, the mechanical
  power imparted by individual warm absorption components is very low,
  typically $\lesssim0.01\%$ of an AGN's bolometric luminosity
  ($L_{\rm bol}$) (e.g., \citealt{blustin2005}), which is
  significantly lower than the $\sim0.5\%$ of $L_{\rm bol}$ thought
  necessary for feedback to affect the host galaxy
  (\citealt{hopkins&elvis2010}). However,
  \cite{crenshaw&kraemer2012} have recently shown that this
  $\sim0.5\%$ threshold can be exceeded provided the mechanical power
  is integrated over all UV and X-ray absorption components, at least
  in the case of a few moderate-luminosity local AGN.

Recent systematic archival \xmm and \suzaku studies have shown that Fe\,{\sc xxv-xxvi} absorption
 lines are present in the X-ray spectra of $\gtrsim40\%$ of
 radio-quiet AGN in the local universe with $z<0.1$
 \citep{tombesi2010a, tombesi2011a, tombesi2012a, patrick2012,
   gofford2013} and also in a sample of 30 local Broad Line Radio Galaxies
(Tombesi et al.\ 2013, in preparation) 
which thus suggests that they may represent an important addition to
the commonly held AGN unification model (e.g., \citealt{antonucci1993,
  urry&padovani1995}).  
In comparison to the soft-band absorbers these hard X-ray absorbers
generally have much more extreme parameters, with $\lognh\approx23-24$
and $\logxi\approx3-6$, and their outflow velocities relative to the
host galaxy can reach mildly relativistic values. The large inferred
velocities -- combined with the short time-scale variability sometimes
exhibited by the absorption features -- point to an origin more likely
associated with a wind which is launched from the surface of the
accretion disc itself (e.g., \citealt{pounds2003b, reeves2009,
  gofford2011, tombesi2012a}). In this scenario the inferred mass
outflow rates for disc-winds are often comparable to those of the
matter which accretes onto the central black hole and the consequent
mechanical power can also be a sizeable fraction (i.e., $\geq$ few percent)
of an AGN's bolometric luminosity (e.g., \citealt{chartas2002,
  pounds2003b, gibson2005, reeves2009, gofford2011, tombesi2012a}).

MR\,2251-178 ($z=0.06398$; \citealt{bergeron1983,canizares1978}) is one of the X-ray brightest 
AGN in the local universe ($L_{2-10\,\rm{keV}}\sim2-9\times10^{44}$\,erg\,s$^{-1}$). It was the first 
quasar identified through X-ray observations
\citep{cooke1978,ricker1978}  and the 
first AGN known to host a warm absorber \citep{halpern1984}. The quasar is located on the 
outskirts of a cluster of $\sim50$ galaxies \citep{phillips1980} and is surrounded by an 
extended nebula of diffuse gas out to 10--20\,kpc, 
which gives rise to [O\,\textsc{ii}], [O\,\textsc{iii}] and H$\alpha$ 
emission at optical wavelengths \citep{macchetto1990,phillips1980}. 
The source has a central black hole mass of 
$\sim2.4\times10^{8}$\Msun\ \citep{dunn2008}, is observed to be a weak radio emitter 
(with a radio loudness parameter, $R_{\rm L}=F_{\rm 5 GHz}/F_{\rm
  4400\AA}=-0.43$; \citealt{reeves&turner2000}),  
and has a Fanaroff-Riley type I (FR\,I) 
radio morphology \citep{macchetto1990}.

The first detailed study of MR\,2251-178 in the X-ray regime was conducted by \cite{halpern1984} 
who, using spectra from the Einstein X-ray observatory, noticed soft X-ray variability on 
time-scales of $\sim1$\,yr caused by changes in both the column density of photoionized material 
along the line of sight and an associated change in the materials
ionization state. 
The ionization state of the absorbing material was also later found to be 
strongly correlated with the source luminosity, with the absorber appearing to become more 
ionized when the source was at a larger luminosity, which thus strongly suggested the presence 
of partially-ionized `warm' material along the line of sight \citep{mineo1993}. 
Subsequent observations with \exosat, \ginga\ and \sax\ found the broad-band X-ray spectrum 
could be well described by a power-law of photon-index $\Gamma\sim1.6-1.7$ which is absorbed by 
a column density of around a few\,$\times10^{22}$\,cm$^{-2}$ \citep{pan1990,mineo1993}, 
and a high-energy roll-over at around $100$\,keV \citep{Orr2001}. In the UV, \cite{monier2001} 
found absorption lines due to Ly$\alpha$, N\,\textsc{v} and
C\,\textsc{iv} with a systematic blueshift  
of $\sim300$\vunit; the C\,\textsc{iv} absorption in particular showed variability over a period of 
roughly 4 years which constrained the absorption clouds to within $r\lesssim$2.4\,kpc of the 
continuum source \citep{ganguly2001}.

\cite{kaspi2004} performed a detailed spectral and temporal study of MR\,2251-178 
using a series of \asca, \fuse, \sax\ and \xmm\ observations which spanned a period of 
$\sim8.5$\,yrs. In confirmation of previous studies \cite{kaspi2004} also found the continuum 
to be described by an absorbed power-law of photon index $\Gamma\sim1.6$, but also found that 
the continuum required a supplementary soft-excess at $E<2$\,keV to achieve an acceptable fit 
to the soft X-ray data. The grating spectrum from the \xmmrgs\ revealed the warm absorber in 
MR\,2251-178 to be multi-phase, consisting of at least two or three ionised absorption 
components with column densities in the range $10^{20-22}$\,cm$^{-2}$, all of which had physical 
properties which appeared to vary between observations in accord to what was reported by 
\cite{halpern1984}. This led \cite{kaspi2004} to propose a scenario where absorption clouds 
were moving across the line of sight over a time-scale of `several months'. In the \fuse\ 
spectrum further UV absorption lines from C\,\textsc{iii},
H\,\textsc{i} and O\,\textsc{vi} were detected  
with velocity shifts similar to those found by \cite{monier2001}. A 2002 Chandra/HETG observation of 
MR\,2251-178 was published by \cite{gibson2005}. There the authors 
found evidence for a highly-ionised Fe\,\textsc{xxvi}~Ly$\alpha$ absorption line with a substantial 
blueshifted velocity, $\vout=-12700\pm2400$\vunit. By considering the kinematics of the 
absorber \cite{gibson2005} inferred that unless the absorber is of a low global covering fraction 
(in terms of the total fraction of $4\pi$\,sr covered by the absorber)
the mass-loss rate in MR\,2251-178 is at least an order of magnitude larger than the source 
accretion rate. 

 A recent analysis over the $0.6-180.0$\,keV broad-band X-ray
  spectrum has been performed by \cite{gofford2011} combining 
 a \textit{Suzaku} observation of MR\,2251-178 performed in May 2009 and
  \textit{Swift}/BAT data as part of the 58-months all-sky-survey
  \citep{baumgartner2010}. In accordance with previous observations,
  the authors found that the general continuum can be well described
  by a power-law with $\Gamma=1.6$, an apparent soft-excess below
  1\,keV and considerable curvature above around $\sim
  10$\,keV. However, the authors found that a good fit can also be found with a
softer $\Gamma\sim2.0$ power-law absorbed by a column of $N_{\rm
  H}\sim10^{23}$\,cm$^{-2}$ which covers $\sim30$\% of the source
flux. This softer photon index value is more consistent with that found 
generally in radio-quiet quasars 
\citep[e.g.,][]{reeves&turner2000,porquet2004,piconcelli2005,scott2011}.   
In addition, numerous significant warm absorption lines were detected (at
the $>$99$\%$ confidence level from Monte Carlo simulations) and associated to  
Fe\,UTA, Fe\,L shell (blend of 2s$\rightarrow$3p transitions from
Fe\,\textsc{xxiii-xxiv}), S\,\textsc{xv}, S\,\textsc{xvi} and
Fe\,\textsc{xxv-xxvi} lines. 
\cite{gofford2011} found at least 5 ionised absorption components with $10^{20} \la N_{\rm H}
\la 10^{23}$\,cm$^{-2}$ and $0 \la \log\xi/{\rm erg\,cm\,s^{-1}} \la
4$ are required to achieve an adequate spectral fit of all these
absorption features.  

In this paper the analysis of an  unprecedented deep follow-up
campaign of MR\,2251-178 in 2011 with \xmm\ and \chandra, is
presented. The \xmm\ and \chandra\ observations were both performed as
a large observing program, with the observations within about a month
of each other. The exposure times of these observations, of $\sim
400$\,ks, is significantly greater than obtained in the previous 2002
Chandra/HETG and XMM-Newton observations (net exposures of
$\sim140$\,ks and 60\,ks respectively). The increased exposure times
make it possible to study the warm absorber in this quasar in
unprecedented detail and resolution, with the RGS and HETG gratings
on-board XMM-Newton and Chandra respectively. Thus the overall goal of
this campaign was to obtain high signal-to-noise and high resolution
spectroscopy of MR\,2251-178 in order to measure the properties of the
primary continuum emission and in particular the ionized absorption
and outflow along the line of sight.\\ 
 
This paper is organized as follows. In section~\ref{sec:obs}, we
  describe the data reduction of both RGS and HETG
  observations. Section~\ref{sec:initial} is devoted to the initial
  spectral fitting of the HETG data, to atomic line detections and
  identifications as well to the initial kinematics of the absorption lines. 
Section~\ref{sec:absorption} presents
  photoionization modeling of the X-ray absorption in the RGS and HETG
  spectra combining fully and partial covering warm absorber
  components; in addition the variability of the X-ray
  absorption components and the possible presence of a highly ionized
  absorber are examined. 
  Section~\ref{sec:emission} focuses on the modeling of the emission
  line spectrum, especially the O\,\textsc{vii} line complex. In
  Section~\ref{sec:discussion}, we discuss about the origins and
  infer some physical properties of the absorption and emissions
  media observed in MR\,2251-178 and compare them to those found in other AGN. 

Values of $H_0=70$\,km\,s$^{-1}$\,Mpc$^{-1}$,
and $\Omega_{\Lambda_{\rm 0}}=0.73$ are assumed throughout and errors are quoted at 90\% 
confidence ($\Delta\chi^{2}=2.7$), for 1 parameter of interest. All spectral parameters
are quoted in the rest-frame of the quasar, at $z=0.06398$ 
 \citep{bergeron1983}, unless otherwise stated.

\section{Observations and Data Reduction}\label{sec:obs}
\subsection{XMM-Newton Observations of MR\,2251-178}

XMM-Newton observed MR\,2251-178 three times from 11-17 November 2011, over 3 consecutive 
satellite orbits. Each observation was approximately 130\,ks in length, with the details 
of the 3 observations listed in Table\,1. First order dispersed spectra were obtained  
with the Reflection Grating Spectrometer \citep{denherder01} and were reduced using the 
\textsc{rgsproc} script as part of the XMM-Newton SAS software v11.0. The spectra from 
each of the orbits were found to be consistent with each other, with the only variation being due to a 10\% 
change in the count rate of the source over the 3 observations. Therefore spectra and response files 
for each RGS were combined to give a single spectrum with a total net exposure of 389.1\,ks. 
There were no periods of strong background flares during the 
observations, the background rate in each RGS being only 7-8\% of the total source rate. Prior to spectra 
analysis, channels due to bad pixels on the RGS CCDs were ignored as well as the two malfunctioning 
CCDs for RGS\,1 and RGS\,2 respectively. 

The net background subtracted count rates were $0.496\pm0.001$\,s$^{-1}$, $0.535\pm0.001$\,s$^{-1}$ 
for RGS\,1 and RGS\,2 respectively, yielding a total of over $4\times10^{5}$ counts 
for the two RGS spectra together. Spectra were binned into 
$\Delta \lambda=0.02$\,\AA\ bins, which over-samples the RGS spectral resolution by a factor of 
$\times 4$ compared to the FWHM resolution. Due to the high count rate statistics, $\chi^{2}$ 
minimization was employed in the subsequent spectral fitting. An additional $\pm3$\% systematic error 
was added in quadrature to each combined RGS spectrum, in order to allow for systematic differences 
between the two grating spectra. A constant multiplicative offset was subsequently allowed between 
the RGS\,1 and RGS\,2 in all the spectral fits, which was found to be within $\pm3$\%. Data were 
fitted over the 0.33-2.0\,keV energy range in the observed frame.

\subsection{Chandra HETG Observations of MR\,2251-178}

The High Energy Transmission Grating (HETG) onboard \chandra\ 
  \citep{weisskopf2000,canizares2005} also observed MR\,2251-178  
from 26 September to 2 October 2011, occurring approximately 40\,days 
before the \xmm\ observations. As per the \xmm\ observations, the \chandra\ observations 
occurred over 3 consecutive orbits, with the last sequence somewhat shorter than the first 
two - see Table\,1 for details. Spectra were extracted with the
\textsc{ciao} package v4.3. 
Only the first order dispersed spectra were considered for both the MEG (Medium Energy Grating) and 
HEG (High Energy Grating) and the $\pm1$ orders for each grating were subsequently combined 
for each sequence. No significant spectral variability was observed between the 3 sequences and the 
spectra were consistent, with only modest $\sim 10$\% variations in source flux. 
Therefore the spectra were combined from all three sequences to yield a single 1st order spectrum 
for each of the MEG and HEG, yielding respective net source count rates of $0.485\pm0.001$\,s$^{-1}$ 
and $0.245\pm0.001$\,s$^{-1}$ respectively for a total exposure time of 392.9\,ks. 
Thus the total counts obtained exceeded $1.9\times10^{5}$ and $9.5\times10^{4}$ counts
for MEG and HEG respectively. Note that the background contribution towards the count rate was negligible. 

The resultant 2011 
source spectra were subsequently binned to $\Delta \lambda = 0.02$\,\AA\ and 
$\Delta \lambda = 0.01$\,\AA\ bins for 
MEG and HEG respectively, which samples their respective FWHM spectral resolutions. 
The MEG and HEG spectra were analyzed over the energy ranges of 0.5--5.0\,keV and 
1.0--9.0\,keV respectively. The C-statistic was 
employed in the subsequent spectral fits to the HETG, as although the overall count rate is high, towards 
the lower energy (longer wavelength) 
end of each grating spectrum the total source counts per bin drops below $N<20$ 
in some bins. In the case of $\chi^{2}$ minimization, this would lead to the continuum level being 
somewhat underestimated at soft X-ray energies.

An archived \chandra\ HETG observation of MR\,2251-178 also took place 11 September 2002, with a total 
net exposure of 146.3\,ks. First order spectra for MEG and HEG were re-extracted as above, yielding count 
rates of $0.317\pm0.001$\,s$^{-1}$ and $0.164\pm0.001$\,s$^{-1}$ respectively. Thus the 2011 observation 
was approximately 50\% higher in count rate or flux than the earlier 2002 observation and therefore the 
2002 dataset provides a lower flux comparison spectrum. The data were binned and 
analyzed over the same energy ranges as per the 2011 observation and the C-statistic was employed 
in all subsequent spectral fits.

\section{Initial Spectral Fitting}\label{sec:initial}

\subsection{The Overall Spectral Form}

\begin{figure}
\begin{center}
\rotatebox{-90}{\includegraphics[width=12cm]{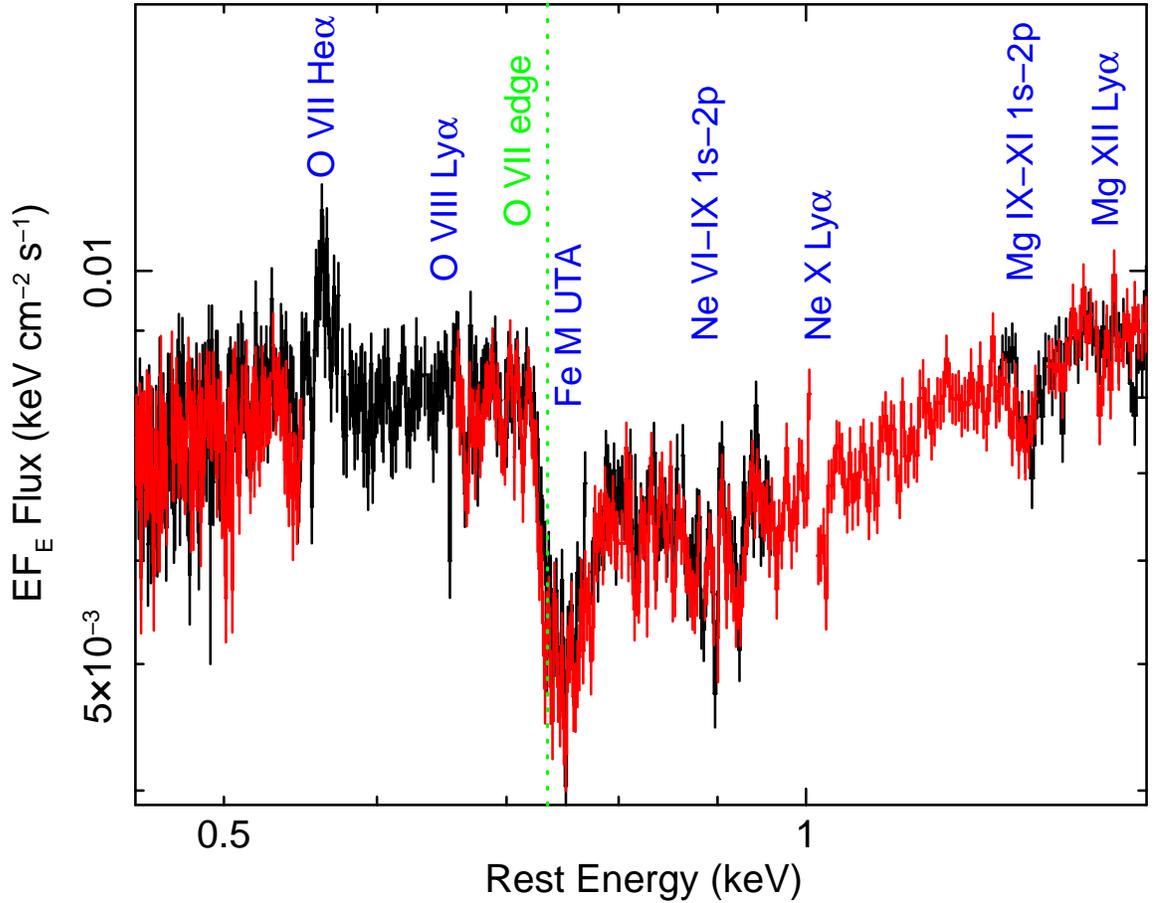}}
\end{center}
\caption{Fluxed 2011 XMM-Newton RGS spectra of MR\,2251-178 between 0.4--1.5\,keV. 
RGS\,1 is shown in black, RGS\,2 in red. 
The spectrum shows a clear imprint of a warm absorber, with the main features 
in the spectrum labelled. The absorption due to the UTA of M-shell iron is particularly 
prominent above 0.7 keV, as well as absorption due to Ne (and iron L-shell) between 0.9-1.0 keV, 
as well as an absorption trough due to Mg near 1.3 keV. Note the strong O\,\textsc{vii} emission 
at 0.56--0.57\,keV. Energy is plotted in the quasar rest frame.} 
\label{rgs-eeuf}
\end{figure}

Initially we concentrated on the 2011 RGS and HETG observations. All parameters are given in the 
rest frame of the quasar at $z=0.06398$, 
unless otherwise stated and spectral parameters are quoted in energy units (thus 1\,keV is 
equivalent to 12.3984\,\AA).
In all the fits, a Galactic absorption of hydrogen column density of
$N_{\rm H}=2.4\times10^{20}$\,cm$^{-2}$ \citep{kalberla2005} was adopted, modeled with the 
``Tuebingen--Boulder'' absorption model ({\sc tbabs} in {\sc xspec}) using 
the cross--sections and abundances of \citet{wilms2000}. 
Figure\,\ref{rgs-eeuf} shows the overall 
2011 fluxed RGS spectrum of MR\,2251-178, plotted against a powerlaw of $\Gamma=2$ in 
the soft X-ray band and 
in the quasar rest frame at $z=0.06398$.  
The spectrum shows several clear signatures of a warm absorber and emitter. A deep absorption trough 
is present between $0.7-0.8$\,keV which is most likely identified with
an unresolved transition array (UTA), due  
to $2p \to 3d$ transitions from lower ionization M-shell iron (i.e. Fe less ionized than Fe\,\textsc{xvii}) 
\citep{behar2001}. The iron M-shell UTA has been commonly observed in high resolution grating 
spectra of many AGN  \citep{mckernan2007}, 
e.g., IRAS 13349+2438: \citep{sako2001}, 
NGC\,3783: \citep{kaspi2000,kaspi2001,krongold2003}, 
NGC\,5548: \citep{kaastra2002,andrade2010}, 
Mrk\,509: \citep{pounds2001,yaqoob2003,smith2007}, 
NGC\,7469: \citep{blustin2007}, 
Mrk\,841: \citep{longinotti2010}, 
IC\,4239A: \citep{steenbrugge2005b},
NGC\,3516: \citep{holczer&behar2012},
Ark\,564: \citep{papadakis2007}, 
MCG-6-30-15: \citep{lee2001,turner2004},
NGC\,4051: \citep{pounds2004a},
Mrk\,279: \citep{costantini2007}, 
I Zw1: \citep{gallo2004},
1H\,0419-577: \citep{pounds2004b},
PG\,1114+445: \citep{ashton2004}.

Several narrow absorption lines appear to be present between 
$0.85-1.0$\,keV, likely due to K-shell $1s \to 2p$ lines of Neon as well as higher ionization L-shell 
($2p \to 3d$) lines of iron (i.e. Fe\,\textsc{xvii-xxii}). A
broad absorption trough appears to be present near 1.3\,keV  
in the rest frame, close to the expected K-shell lines of Mg, the origin of which is discussed 
in Section 4. Strong and resolved line emission is also especially prominent in the RGS\,1 spectrum 
between $0.56-0.58$\,keV, at the expected energy of the O\,\textsc{vii} triplet.

\begin{figure}
\begin{center}
\rotatebox{-90}{\includegraphics[width=12cm]{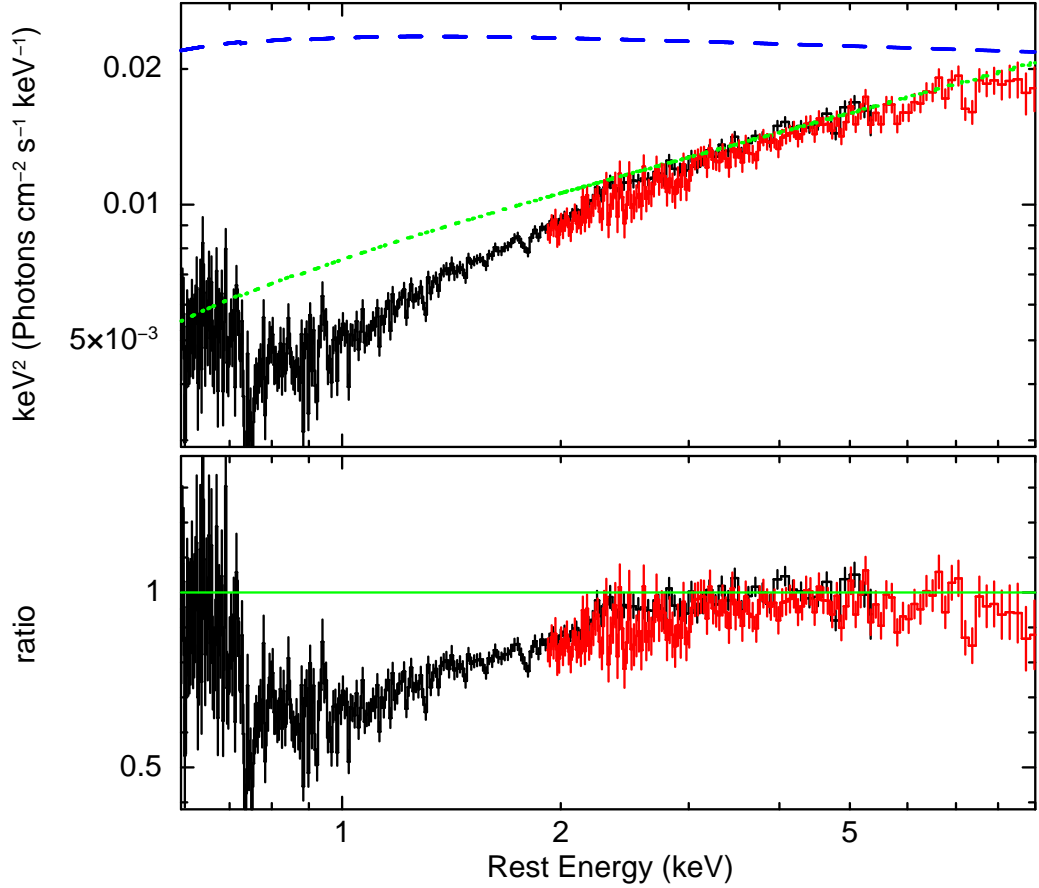}}
\end{center}
\caption{Fluxed 2011 Chandra HETG spectra of MR\,2251-178, MEG is shown in black, HEG is red. 
The top panel shows the spectra, while the dotted green line is a representative power-law 
continuum with $\Gamma=1.6$. The upper dashed blue line shows the actual intrinsic level of the continuum, 
once the absorption is modeled. The lower panel shows data/model ratio to the $\Gamma=1.6$ power-law, 
the downwards continuum curvature due to the warm absorber 
is clearly present. Note that data are binned at $4\times$ the HWHM resolution for clarity, while the 
HEG spectrum is only plotted above 2 keV.} 
\label{hetg-pl}
\end{figure}

For comparison, the fluxed 2011 HETG spectrum of MR\,2251-178 is shown in Figure\,\ref{hetg-pl}. 
The spectrum plotted is against 
a power-law of photon index $\Gamma=1.6$ for comparison purposes only; as is discussed later 
in Section 4.2, the likely 
intrinsic photon index of the source is perhaps much steeper ($\Gamma\gs2$) once all the layers of 
absorption in MR\,2251-178 are accounted for. 
Although the power-law provides a good representation of the HETG spectrum above 3\,keV, 
the data/model ratio residuals show pronounced curvature due to the presence of the known 
warm absorber in this AGN. Indeed fitting a single power-law (modified by Galactic absorption only) 
provided a very poor representation of the whole HETG spectrum fitted from 0.5--9.0\,keV, with 
a very hard photon index of $\Gamma=1.33\pm0.02$ and an unacceptable fit statistic of 
$C=3806.4$ for 2360 degrees of freedom (dof). Note a multiplicative cross-normalization constant 
was included between the MEG and HEG spectra, the HEG normalization was found to be slightly lower
($0.97\pm0.01$) than the MEG (which was normalized to 1.00). 

In order to investigate and identify the atomic lines present in the HETG spectra, a 
more complex continuum shape was adopted in order to better account for the clear spectral curvature. 
A power-law continuum was adopted, modified by a neutral partial covering absorber (the \textsc{pcfabs} 
model in \textsc{xspec}). While this simple partial coverer is not meant to provide a physical 
description of the spectrum, its advantages are that it provides a better parameterization of the 
spectral curvature, while not imparting any discrete atomic lines on the spectrum, thus providing a 
reference continuum from which lines can be identified against. A similar approach 
was also taken to provide an initial parameterization of the broad-band Suzaku spectrum of 
MR\,2251-178 \citep{gofford2011}. In addition to the partial covering absorption, a phenomenological 
absorption edge component was initially included to account for the pronounced spectral drop above 0.7\,keV, due to 
a possible combination of the Fe M-shell UTA and O\,\textsc{vii} edge. Again this was not meant to provide a 
physical fit to the spectrum. The edge energy was $E=730.6\pm2.1$\,eV with an optical depth of 
$\tau=0.36\pm0.05$. The partial coverer had a column density 
$N_{\rm H}=(2.9\pm0.2) \times 10^{22}$\,cm$^{-2}$ and a covering fraction of $0.35\pm0.03$, while the 
photon index was $\Gamma=1.69\pm0.03$. The overall fit statistic was much improved compared to the 
power-law only case, with $C=3047$ for 2356 degrees of freedom.

\subsection{Atomic Lines in HETG Spectrum}

\begin{figure}
\centering
\rotatebox{0}{\includegraphics[width=10cm]{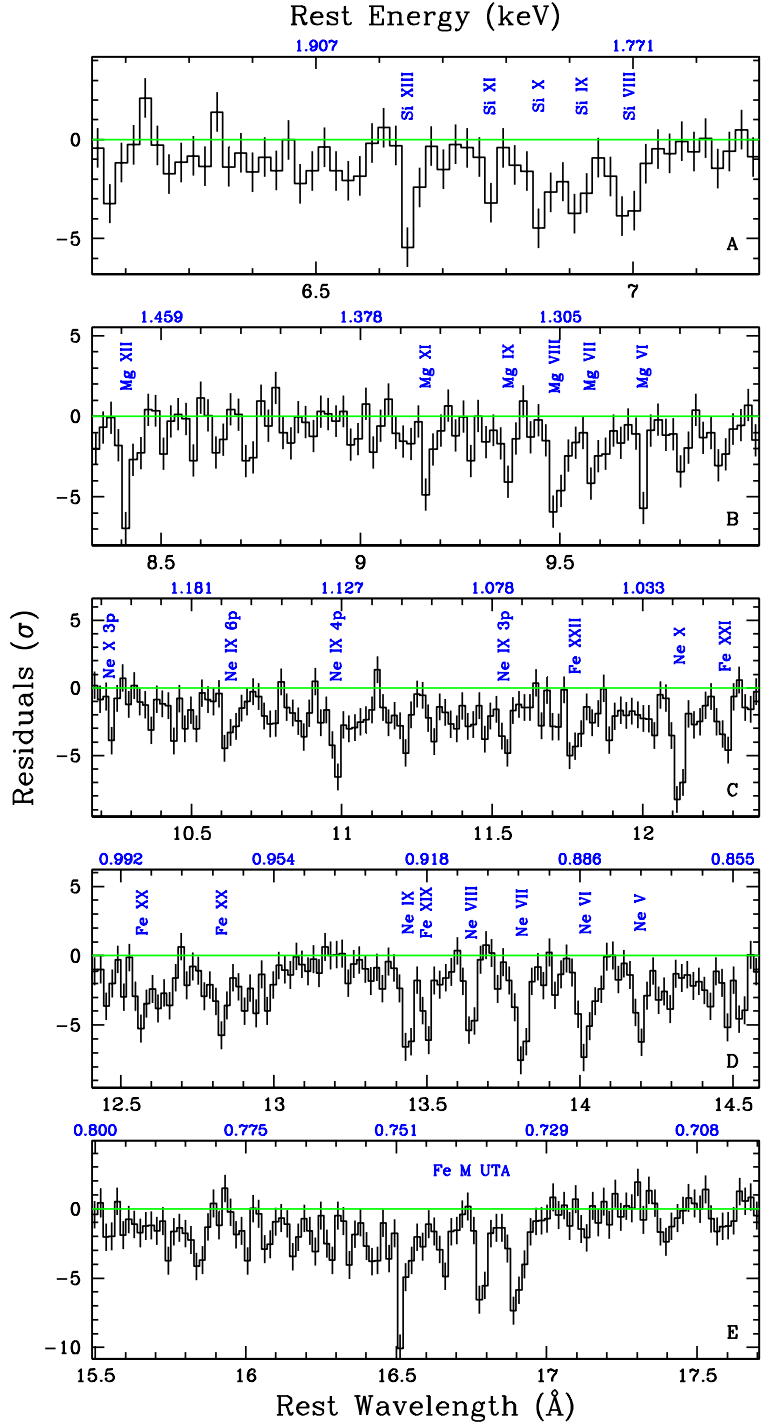}}
\caption{2011 Chandra MEG spectrum of MR\,2251-178, showing the wealth of absorption lines below 2\,keV. 
The data are shown as residuals (in $\sigma$) against the baseline continuum and are plotted 
in the quasar rest frame in wavelength (with energy marked along the upper axis).
Panel (a) shows the Si K band, including inner shell absorption from Si\,\textsc{viii}--\textsc{xiii};
(b) the Mg K band, including inner shell absorption from Mg\,\textsc{vi}--\textsc{xi};
(c) absorption from Ne\,\textsc{ix}, Ne\,\textsc{x} and L-shell Fe; 
(d) inner shell absorption due to Neon ions from Ne\,\textsc{v}--Ne\,\textsc{ix} and 
(e) the iron M-shell UTA band.}
\label{meg-panels}
\end{figure}

\begin{figure}
\centering
\rotatebox{0}{\includegraphics[width=10cm]{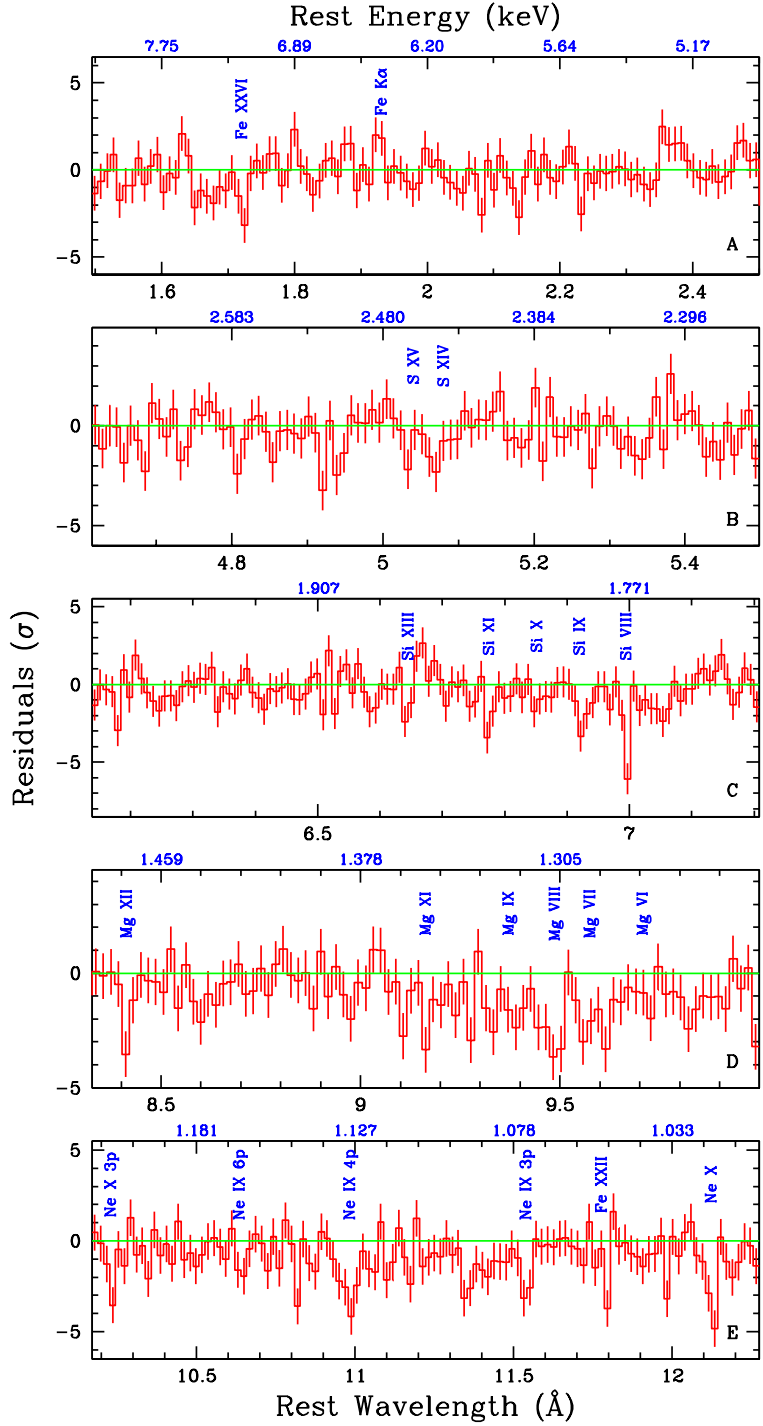}}
\caption{As per Figure~\ref{meg-panels}, but showing the Chandra HEG spectrum of MR\,2251-178, 
Panels (c), (d) and 
(e) show the absorption present in the Si, Mg and Ne bands respectively. Panels (a) 
and (b) also show the spectrum in the Fe and S K-shell bands.}
\label{heg-panels}
\end{figure}

Figure\,\ref{meg-panels} (MEG) and Figure\,\ref{heg-panels} (HEG) show the residuals
against the neutral partial covering model in  
the soft X-ray band below 2\,keV. A wealth of absorption lines are
clearly present in the HETG spectrum against  
the continuum model over the 0.7--2.0\,keV energy range (or 6--18\AA). 
In order to parameterize the lines, 
successive narrow Gaussian absorption lines were included in the continuum model; an individual line was 
deemed to be statistically significant if its addition to the model 
resulted in an improvement of the fit statistic of 
$\Delta C>9.2$, corresponding to 99\% significance for 2 interesting parameters. The width of the 
absorption lines was initially assumed to be less than the
instrumental resolution. The parameters of all 31 of  
the statistically significant absorption lines detected in the soft X-ray HETG spectrum are shown in 
Table\,2. A narrow structure is also clearly present in the UTA region around $0.73-0.76$\,keV, these are 
parameterized by 2 lines in Table\,2, which by comparison with the blends of transitions noted in 
\cite{behar2001} may be due to iron in the ionization states Fe\,\textsc{vii-x}. 

Further low ionization gas appears to be present in the form of a multitude of inner K-shell lines 
of Ne, Mg and Si. These are $1s \to 2p$ absorption lines whereby the L-shell is partially occupied, 
i.e. due to charge states 
corresponding to Li, Be, B, C, N, O-like etc ions. We refer to
\cite{behar2002} for a compilation of 
these inner shell lines, adopting the known energies (wavelengths) of these lines from this paper 
in Table\,2. 
Indeed such lines have been detected in other high signal to noise
grating spectra of Seyfert 1 AGN, 
such as in NGC\,3783 \citep{kaspi2002,blustin2002},
  NGC 4151 \citep{kraemer2005}, Mrk 509 \citep{kaastra2011b}, NGC 3516
  \citep{holczer&behar2012}, NGC\,4051 \citep{lobban2011} and
  NGC 5548 \citep{steenbrugge2005}. In the MR\,2251-178 HETG spectrum, 
absorption lines due to Ne\,\textsc{v-viii} (i.e. C-like through to Li-like ions) are detected from 
0.87--0.91\,keV (13.6--14.3\AA) in the rest frame (Figure\,\ref{meg-panels}). Similarly $1s \to 2p$ inner shell lines from Mg 
are detected due to Mg\,\textsc{vi-ix} (N-like through to Be-like ions), from $1.26-1.33$\,keV (9.3--9.8\AA). Likewise 
inner shell absorption is also detected from Si, from Si\,\textsc{viii-xi} (N-like to Be-like) 
around 1.8\,keV (6.5--7.0\AA). The inner shell absorption is also independently
detected in the HEG (Figure\,\ref{heg-panels}) as  
well as the MEG (Figure\,\ref{meg-panels}) spectra. Thus the detection of the strong Fe M-shell UTA, plus the 
inner-shell absorption due to Ne, Mg and Si suggests the imprint of a significant amount of absorption 
due to both low and high ionization gas in MR\,2251-178.

Absorption lines due to more highly ionized gas are also significantly detected in the HETG spectrum. 
He and H-like lines of O, Ne, Mg and Si are all detected (with the exception of the O\,\textsc{vii} 
$1s \to 2p$ line due to the lack of S/N below 0.6 keV in the MEG spectrum). In some cases, higher order 
$1s \to np$ lines are detected, especially in the case of Ne\,\textsc{ix} where the series of 
resonance lines up to $1s \to 6p$ are seen. Higher ionization L-shell lines of iron are also present, e.g. 
from Fe\,\textsc{xix-xxii}. The spectra over the S and Fe K band are also shown Figure\,\ref{heg-panels}, although note that neither strong emission nor absorption features appear to be present in 
these parts of the spectrum. In the S band, weak absorption may be present at the expected 
energies of the He and Li-like lines of S, although they are below the formal detection threshold. 
The details of the iron K band spectrum will be discussed further in Section 4.4.

It is also apparent from Table\,2 that most of the measured rest frame energies of the absorption lines 
are close to the known atomic energies. This suggests that the outflow velocity of the soft X-ray absorbing 
gas is relatively small. We discuss below some of the velocity profiles of the strongest detected 
H and He-like lines.

\subsection{Atomic Lines in RGS Spectrum}

The RGS provides an energy coverage of $0.3-2.0$\,keV with high throughput 
and therefore provides a high quality view of the soft X-ray warm absorber, with a lower energy 
bandpass than Chandra HETG. 
The initial analysis of the absorption line spectrum suggests 
that multiple absorption components may be required in order to model the 
wide range in the ionization state of the gas, e.g. covering for instance 
Fe\,\textsc{vii-xxii}, Ne\,\textsc{v-x} or Mg\,\textsc{vi-xii}. 

\begin{figure}
\centering
\includegraphics[width=11cm]{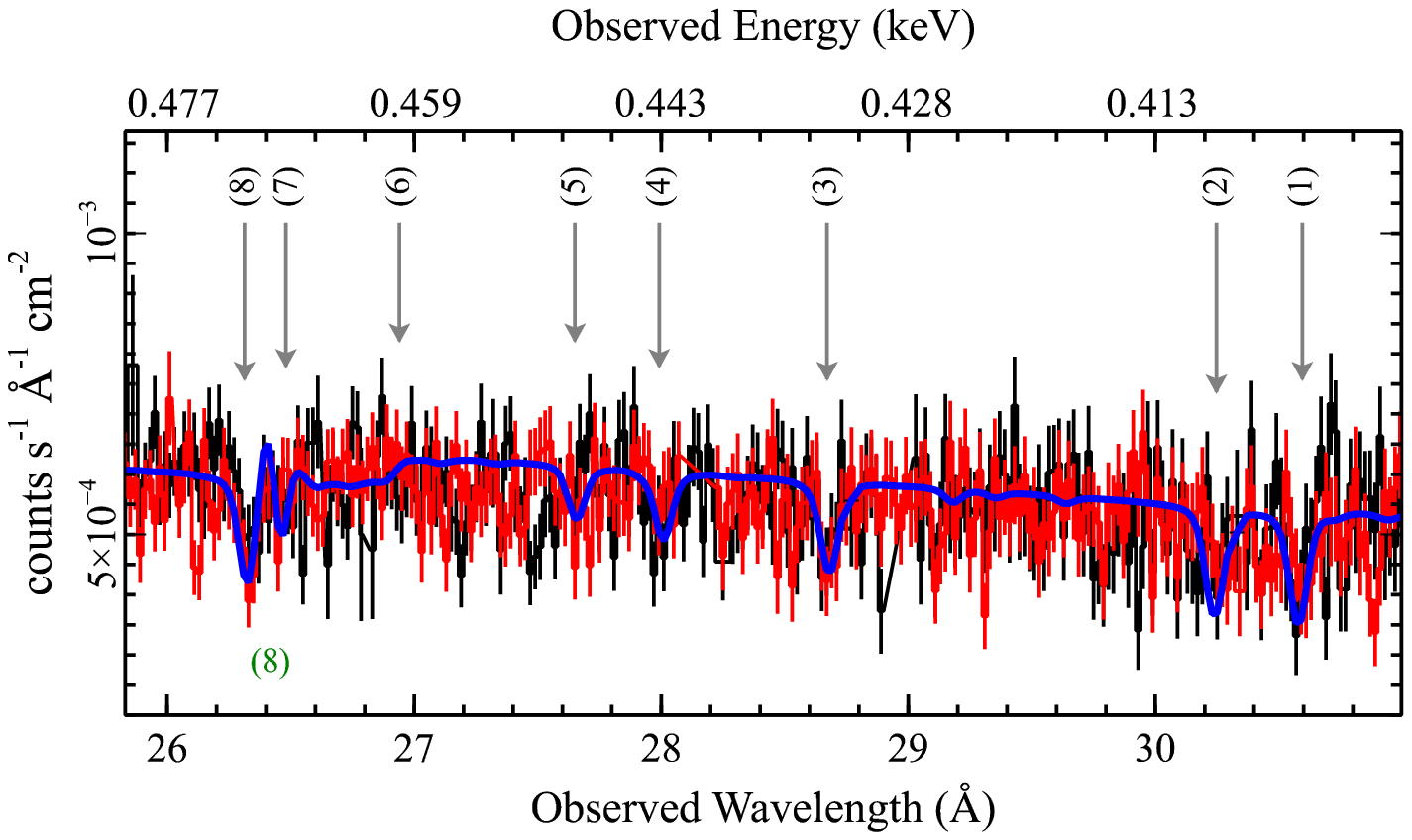}
\includegraphics[width=11cm]{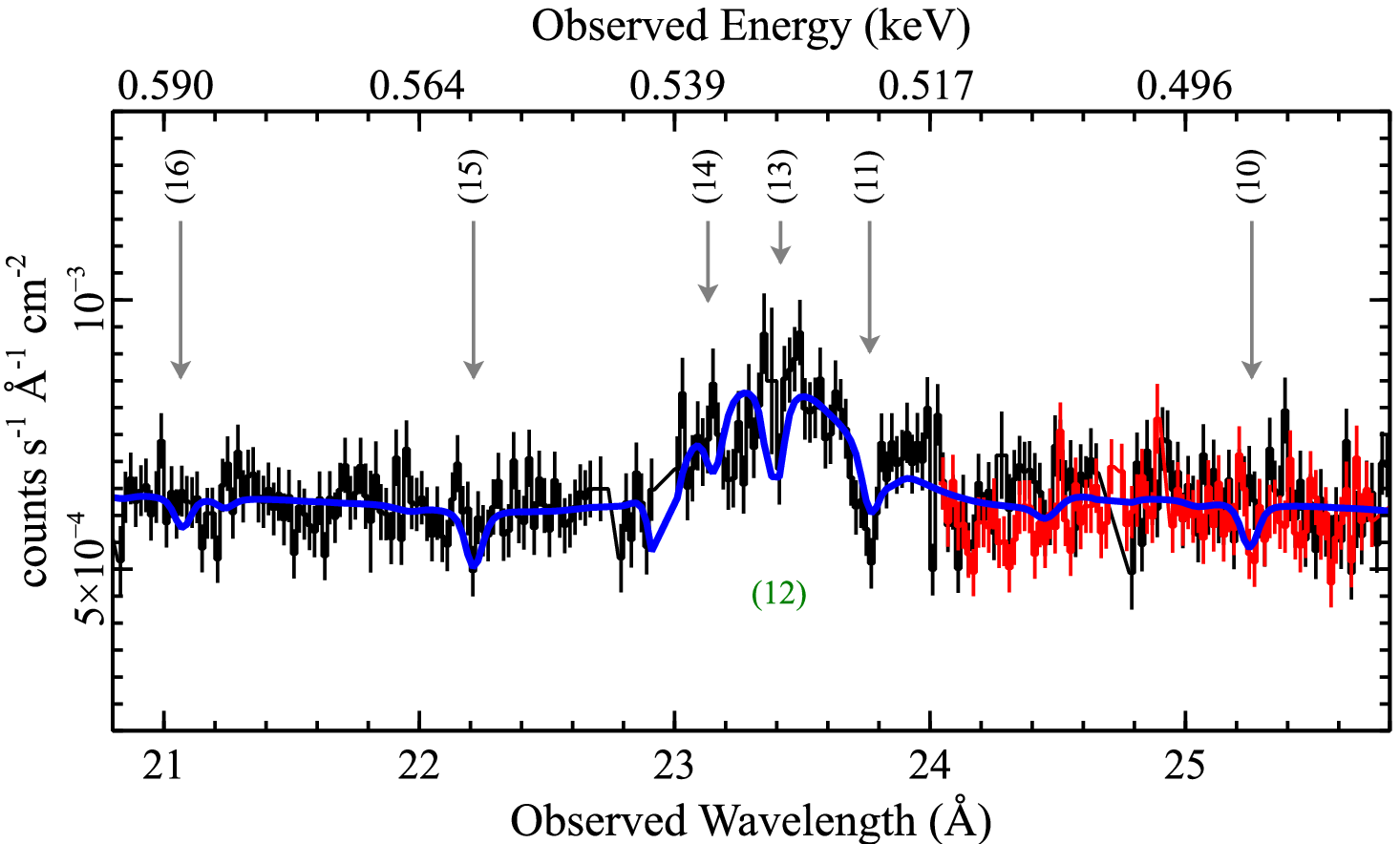}
\includegraphics[width=11cm]{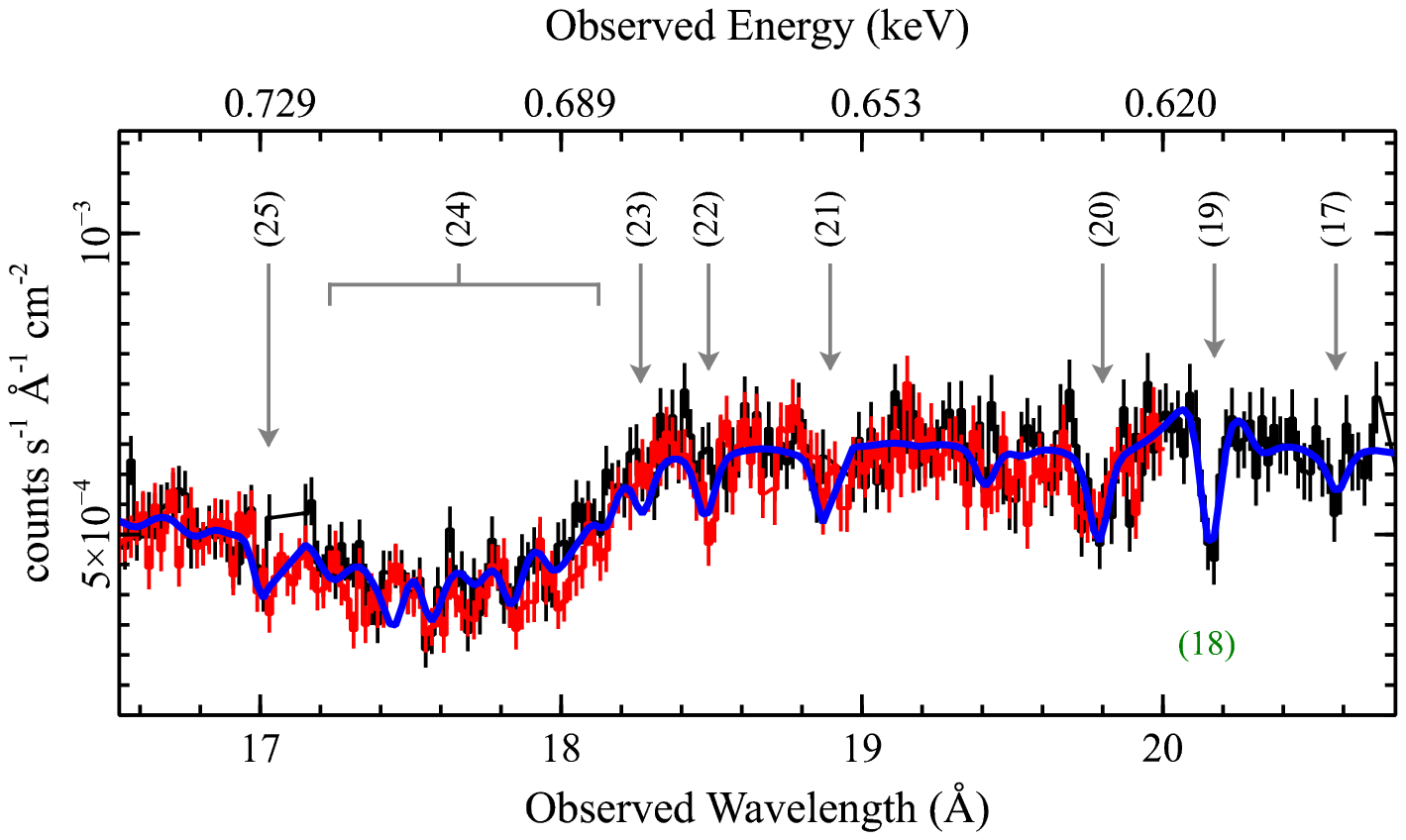}
\caption{Enlarged view of the \xmmrgs\ data (RGS1: black, RGS2: red), showing the count rate 
spectra normalized to the instrument effective area.
The best-fit model for the soft X-ray absorber, comprising three absorption components, is shown by the blue line. There are a plethora of ionized lines present originating from C, N, O and the Fe\,M-shell. There is also some complex interplay between the absorption and underlying emission components, which are further discussed in the text. The likely identification of the numbered lines are presented in Table~\ref{rgs-lines}. Green numbers denote the position of emission components.}
\label{rgs-panels1}
\end{figure}

\begin{figure}
\centering
\includegraphics[width=11cm]{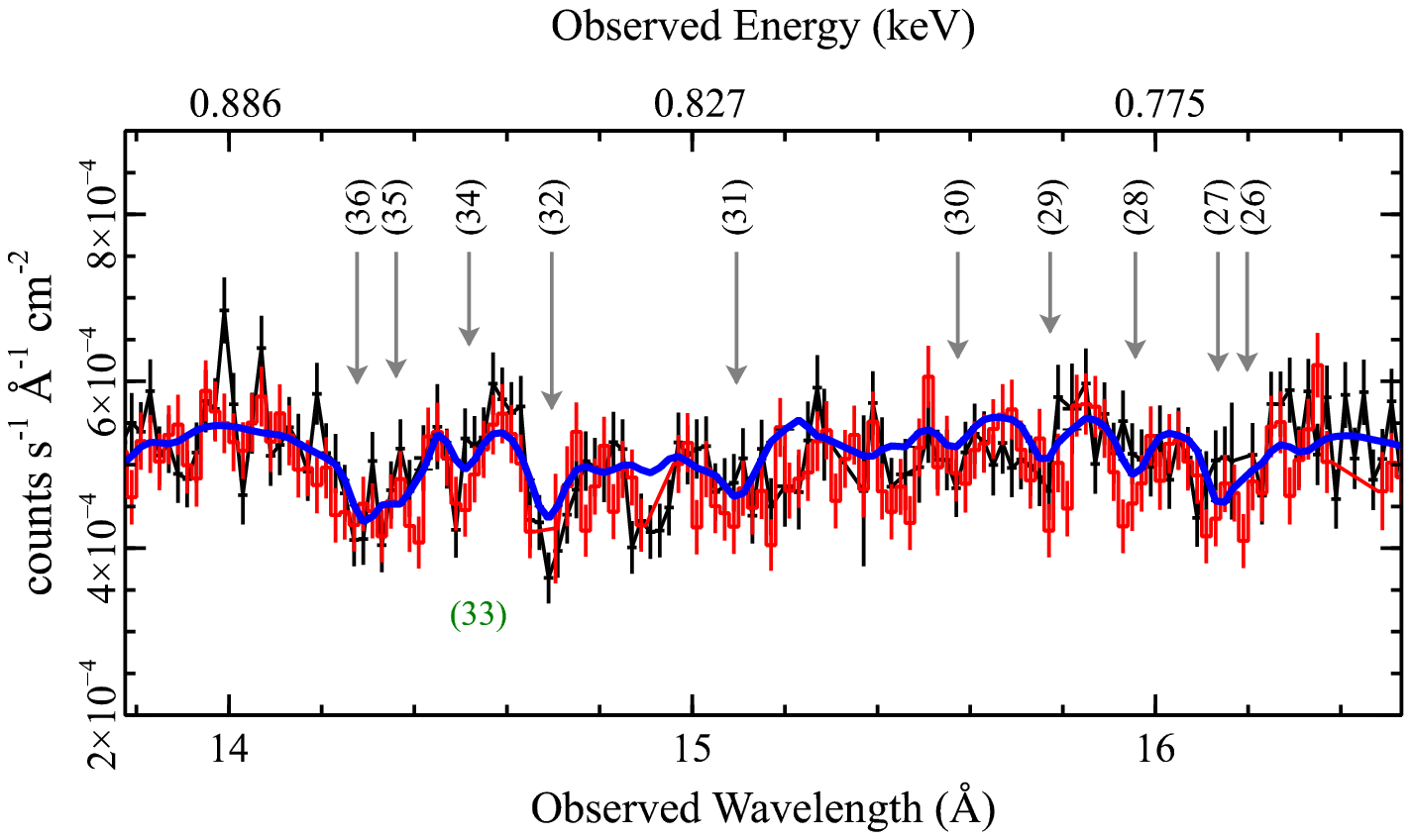}
\includegraphics[width=11cm]{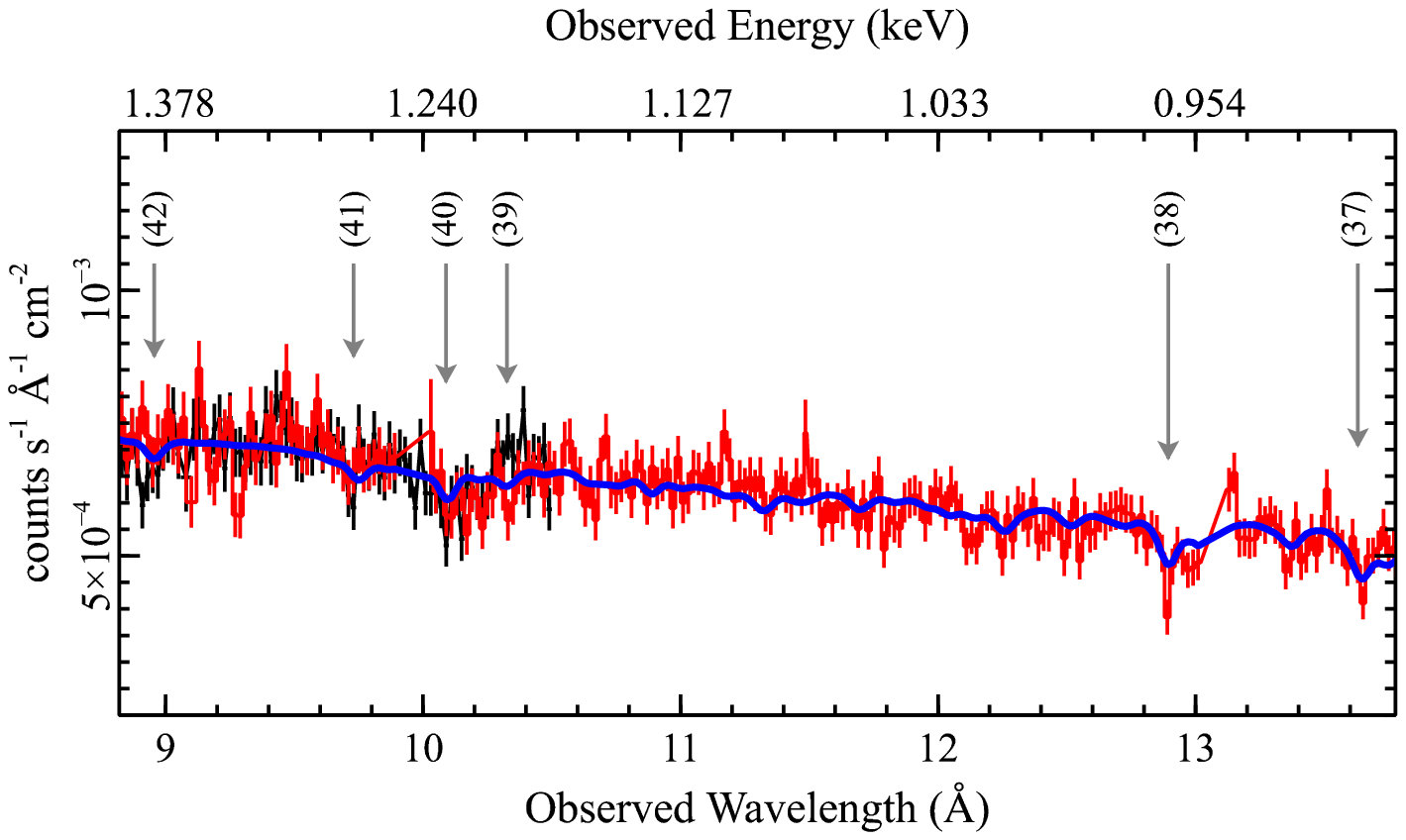}
\caption{Enlarged view of the \xmmrgs\ data showing the Ne and Mg energy regimes. The blue line again shows the best-fit absorption model. Both Ne and Mg have a number of inner-shell lines (i.e., the B-, Be-, Li-like charge states) present in the spectrum. As in the Figure\,\ref{rgs-panels1}, the likely identification of the numbered lines are presented in Table~\ref{rgs-lines}.}
\label{rgs-panels2}
\end{figure}

Indeed enlarged portions of the RGS spectrum of MR\,2251-178 are 
shown in Figures~\ref{rgs-panels1} and \ref{rgs-panels2}. Note these are plotted in the observed frame and 
not the rest frame.
The warm absorber is clearly complex, comprising a wealth of atomic 
features. Notably, inner-shell (Li-like and below) and higher-order 
(i.e., the $1s \to np$ transitions, where $n\geq3$) absorption lines are detected throughout the spectrum, 
due to C, N, O, Ne and Mg. 
Figure~\ref{rgs-panels1} shows that the higher-order line series of C\,\textsc{vi} is 
particularly prominent, while N\,\textsc{vi}, N\,\textsc{vii}, O\,\textsc{vii} and O\,\textsc{viii} also have higher-order 
line series, with each ion reaching at least the 
$1s \to 4p$ transition. 

Complementing the array of absorption lines there is also some interesting 
interplay between emission and absorption components; e.g. 
see the O\,\textsc{vii} line at $517-539$\,eV (23--24\AA) observed frame in Figure~\ref{rgs-panels1}. 
The O\,\textsc{vii} ($1s2p \to 1s^{2}$) emission line complex is superimposed on by three 
narrow absorption lines corresponding to inner-shell absorption due to O\,\textsc{v} (line 11, Figure 5) and the two lines 
which make up the O\,\textsc{vi} ($1s^{2}2s \to 1s2s2p$) doublet (lines 13, 14, Figure 5). Again, similar structures are 
present at other energies, with N\,\textsc{vii}, O\,\textsc{viii} and Ne\,\textsc{ix} all showing emission 
superposed by absorption. The nature of the emission line spectrum will be discussed further in Section 5.

From panels (a) and (b) of Figure~\ref{rgs-panels2}, both Neon and Magnesium also show evidence for 
inner-shell absorption from at least their Be-like ionisation states \citep{behar2002}. 
Indeed, the inner-shell lines for Mg in particular occur throughout the $\sim1.2-1.3$\,keV energy range, 
as per the HETG. 
This appears to be the origin of the absorption trough visible in
Figure\,\ref{rgs-eeuf} and first noted in the  
lower resolution \suzaku\ spectrum of MR\,2251-178 published by \cite{gofford2011}.
The complete list of atomic lines identified in the RGS data
 -- including details such as the responsible ion, the electron transition and the 
centroid energies in the source rest-frame -- is given in Table~\ref{rgs-lines}.

\subsection{Velocity Profiles}

\begin{figure}
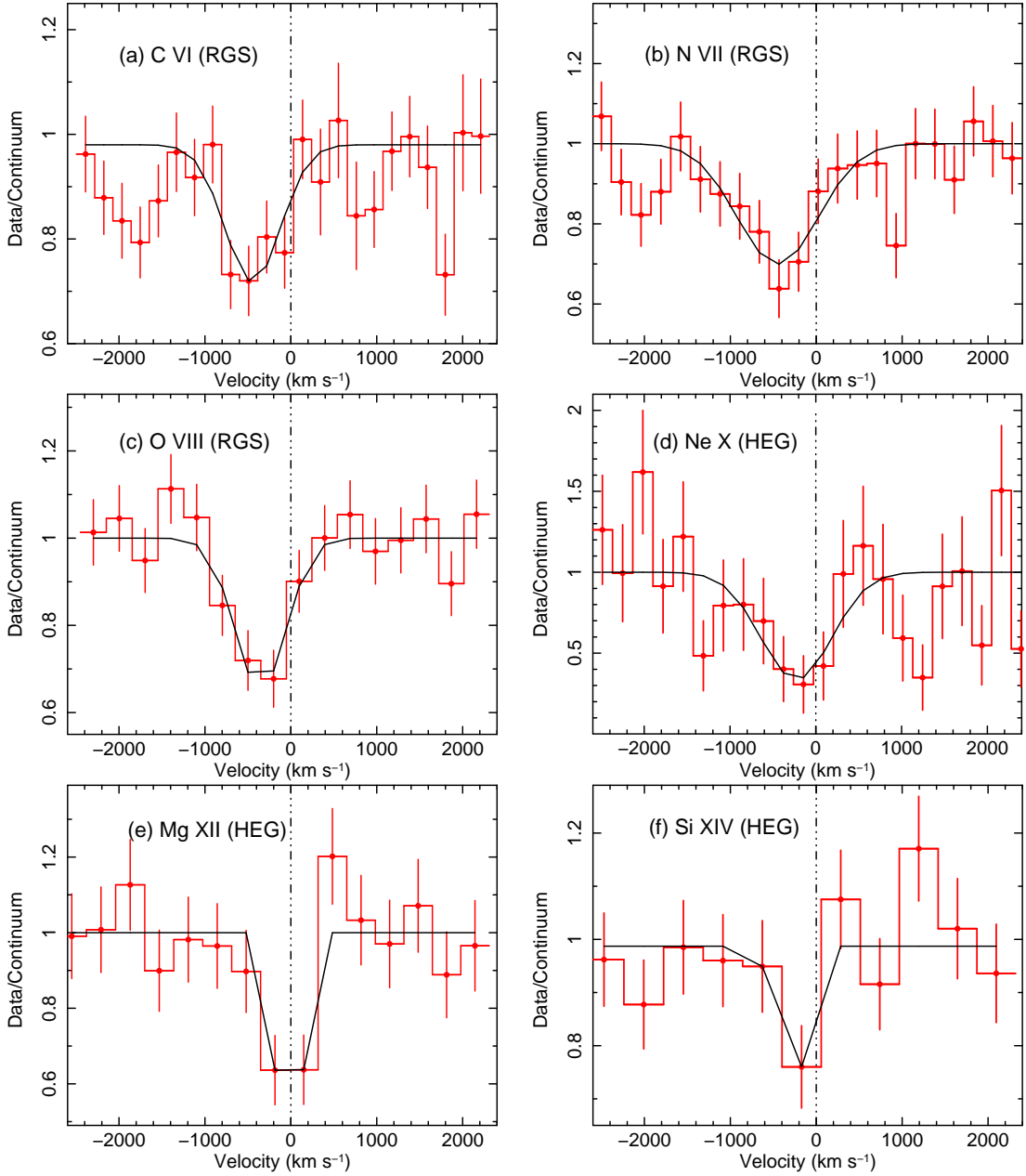

\begin{center}
\includegraphics[angle=-90,width=0.45\textwidth]{f7a.eps}
\includegraphics[angle=-90,width=0.45\textwidth]{f7b.eps}
\includegraphics[angle=-90,width=0.45\textwidth]{f7c.eps}
\includegraphics[angle=-90,width=0.45\textwidth]{f7d.eps}
\includegraphics[angle=-90,width=0.45\textwidth]{f7e.eps}
\includegraphics[angle=-90,width=0.45\textwidth]{f7f.eps}
\end{center}
\caption{Velocity profiles of the main H-like lines, as measured by XMM-Newton RGS (for C\,\textsc{vi}, 
N\,\textsc{vii}, O\,\textsc{viii}) and Chandra HEG (for Ne\,\textsc{x}, Mg\,\textsc{xii}, 
Si\,\textsc{xiv}), see Section 3.4 for details. The data points show the data divided by continuum 
model for each line and negative velocities correspond to blue-shifts. 
The solid line indicates the simple single Gaussian absorption 
profile fitted to each line profile. In the case of the C, N, O (and to a lesser extent Ne) lines, 
a clear blue-shift of the Gaussian centroid is observed, while the higher energy Mg and Si lines 
do not require any net blue-shift and appear unresolved. The subsequent 
best-fit values of the Gaussian profiles are reported in Table\,\ref{line-widths}.}
\label{Hprofiles}
\end{figure}

\begin{figure}
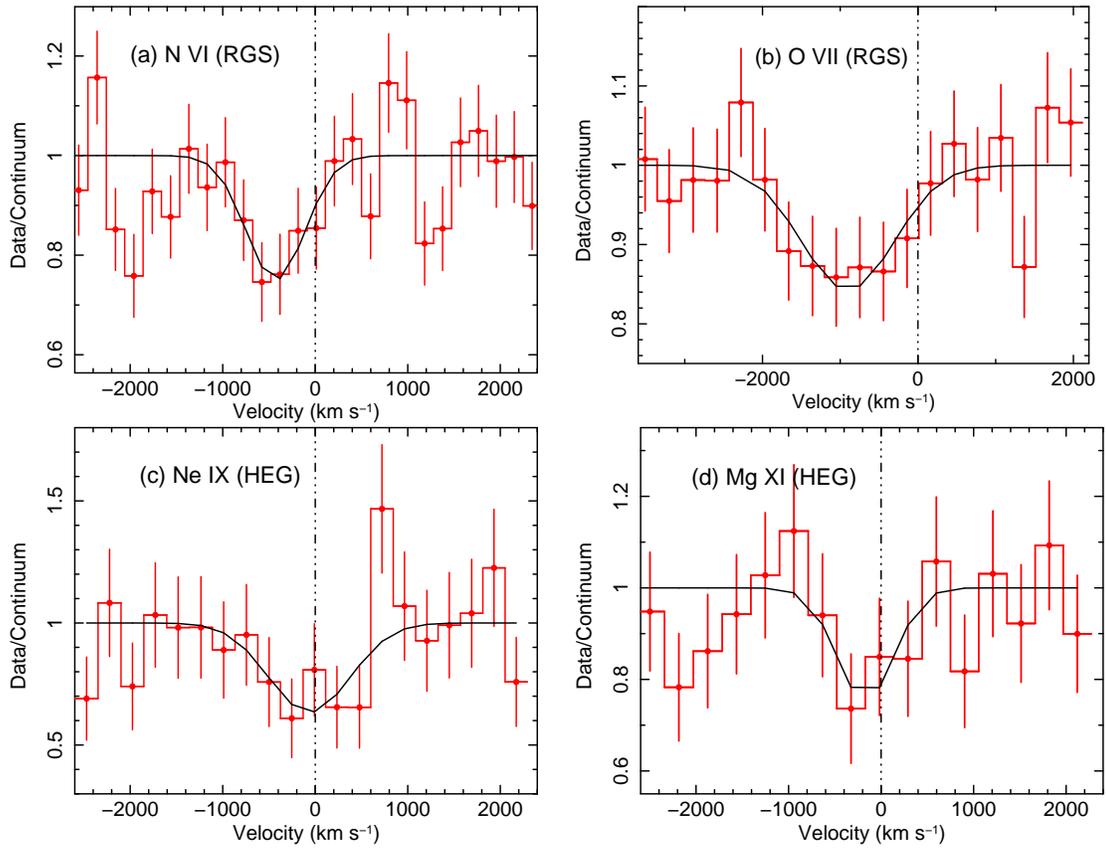

\begin{center}
\includegraphics[angle=-90,width=0.45\textwidth]{f8a.eps}
\includegraphics[angle=-90,width=0.45\textwidth]{f8b.eps}
\includegraphics[angle=-90,width=0.45\textwidth]{f8c.eps}
\includegraphics[angle=-90,width=0.45\textwidth]{f8d.eps}
\end{center}
\caption{As per Figure\,\ref{Hprofiles}, except the velocity profiles correspond to the 
He-like lines of N\,\textsc{vi}, O\,\textsc{vii} (RGS) and Ne\,\textsc{ix}, Mg\,\textsc{xi} (HEG).}
\label{Heprofiles}
\end{figure}

We constructed velocity profiles of the strongest H and He-like absorption lines identified in 
the above HETG and RGS spectra. In each case the 
profiles were constructed by taking the ratio of the data to the 
best fit parameterization of the continuum model described above and transposing them into velocity space 
around the known lab frame energy (wavelength) of each line.
For the H-like ions, the C\,\textsc{vi}, N\,\textsc{vii}, 
O\,\textsc{viii}, Ne\,\textsc{x}, Mg\,\textsc{xii} and Si\,\textsc{xiv} profiles have been produced, 
with the profiles plotted in Figure\,\ref{Hprofiles}. Note that the C\,\textsc{vi} line corresponds 
to the $1s-3p$ absorption line (as the $1s-2p$ line at the redshift of MR\,2251-178 
is close to the edge of the RGS bandpass), while 
the other profiles correspond to the $1s-2p$ lines. Similarily, profiles were also 
constructed for the He-like resonance lines of N\,\textsc{vi}, 
O\,\textsc{vii}, Ne\,\textsc{ix}, Mg\,\textsc{xi} and Si\,\textsc{xiii} and are shown in 
Figure\,\ref{Heprofiles} (note only the first 4 profiles are actually plotted here). 
In the case of the He-like ions, the $1s-3p$ lines of O\,\textsc{vii} and Ne\,\textsc{ix},
are used instead of the $1s-2p$ lines, due to contamination with other lines present 
in the spectrum. Overall the profiles from the C, N and O lines were derived from the RGS data  
in the soft X-ray part of the spectrum (taking the mean of RGS\,1 and 2 where both were available), 
while the Ne, Mg and Si profiles were derived from 
the HEG data at higher energies. Note that negative velocities indicate blue-shift throughout this 
paper\footnote{Note that any upper limits on outflow 
velocities are expressed as absolute values for clarity.}. The profiles are as 
measured from the data, without correcting for the spectral resolution of the instrument.

The subsequent lines were fitted with Gaussian profiles and the results of the fits 
are shown in Table\,\ref{line-widths}, which gives both the overall velocity shift ($v_{\rm out}$) 
of the line profile (as determined from the centroid of the Gaussian profile) as well as the 
observed $1\sigma$ velocity width of the profile ($\sigma_{\rm obs}$). 
Firstly it can be seen both from the profiles themselves and 
the fits that the outflow velocities of the lines tend to decrease in magnitude with increasing 
ionization state, e.g. from C through to Si. 
For instance for the H-like ions, C\,\textsc{vi} and N\,\textsc{vii} profiles have a velocity shift of 
$v_{\rm out} \sim -450$\,km\,s$^{-1}$, with the Mg\,\textsc{xii} profile having a 
formal upper limit on the outflow velocity of only
$v_{\rm out}<40$\,km\,s$^{-1}$, while the 
velocity centroids for O\,\textsc{viii} and Ne\,\textsc{x} are somewhat intermediate in value.
We note that a similar possible trend was found in emission in the Seyfert 2 NGC\,1068 
\citep{Kinkhabwala2002}, whereby the higher energy (excitation) lines had somewhat 
lower velocities. 

The velocity profiles and fits also indicate that a second higher velocity component may be 
present in the lower energy lines of C\,\textsc{vi}, N\,\textsc{vi} and 
N\,\textsc{vii}, with an outflow velocity of $v_{\rm out}\sim-2000$\,km\,s$^{-1}$. 
Such a component is not present in the higher energy lines. The outflow velocities of the 
possible higher velocity components are also given in Table\,\ref{line-widths}, noting that 
the line width of this component was assumed to be the same as for the respective lower 
outflow velocity lines. Thus while we note the possible presence of a higher velocity 
component to some of the lines, we do not discuss this further here, 
as the improvement in fit statistic upon adding this second velocity component 
(see Table\,\ref{line-widths}) was generally less 
than the more robust low velocity component which is always present.

The observed velocity widths of the Gaussian profiles ($\sigma_{\rm obs}$) 
are also given in Table\,\ref{line-widths}. 
These are not corrected for instrument resolution, however for comparison the $\sigma$ widths 
of the RGS (RGS\,1+2 combined) varies between $\sigma=300-380$\,km\,s$^{-1}$ for C\,\textsc{vi} 
to O\,\textsc{viii} and for the HEG between $\sigma=120-230$\,km\,s$^{-1}$ for 
Ne\,\textsc{x} to Si\,\textsc{xiv}.
The intrinsic line widths corrected for instrument resolution 
($\sigma_{\rm int}$) are also given in Table\,\ref{line-widths}. Thus some of the line profiles 
appear resolved, with typical widths of $\sigma_{\rm int}=300-400$\,km\,s$^{-1}$, 
while the higher energy lines (e.g. Mg\,\textsc{xii} and Si\,\textsc{xiii}) appear to be 
unresolved, similar to the possible above trend in outflow velocity.


\section{Photoionization Modeling of the X-ray Absorption Spectrum}\label{sec:absorption}

Given the substantial presence of partially ionized gas in the X-ray spectrum of MR\,2251-178, 
we attempted to model the 
absorption spectrum with photoionized grids of models using the \textsc{xstar} code v2.2 
\citep{kallman2004}. 
Absorption grids were generated in the form of \textsc{xspec} multiplicative tables (or mtables). 
The absorption spectra within each grid were computed between 0.1--20\,keV with $N=10000$ spectral bins. 
The photoionizing X-ray continuum between 1--1000\,Rydberg was assumed to be a power-law of a 
photon index $\Gamma=2$, except for the grid which covered the lowest range in ionization, which 
we discuss further below. Given the narrow (or unresolved) widths of the absorption lines detected in the 
\chandra\ HETG, grid turbulence velocities of either $\sigma=100$\,km\,s$^{-1}$ or $\sigma=300$\,km\,s$^{-1}$ 
were generated; grids with higher turbulences all gave substantially
worse fits in the models considered below. 
An electron density of $n_{\rm e}=10^{10}$\,cm$^{-3}$ was assumed for the absorption grids, although we note 
that the absorption spectra are largely insensitive to the density over a wide range of values. 
Solar abundances were adopted for all the abundant elements, using the
values of \cite{grevesse1998}, 
except for Ni which is set to zero (the default option within \textsc{xstar}).

We generated one generic grid of models that covered a wide range in ionization and column density 
parameter space, from $N_{\rm H}=1\times10^{18}$\,cm$^{-2}$ to $N_{\rm H}=3\times10^{24}$\,cm$^{-2}$ 
and $\logxi = 0-5$ in logarithmic steps of $\Delta (\log N_{\rm H}) = 0.5$ and 
$\Delta (\log \xi) = 0.5$ respectively. A turbulence velocity of $\sigma=100$\,km\,s$^{-1}$ was used. 
This grid was used to fit the high ionization absorption components, as well as the possible partial covering 
absorption which we discuss further below. A separate more finely tuned grid (covering 
a narrower range of parameters) was generated with the 
specific purpose of modeling the low ionization absorption in the MR\,2251-178 spectrum, especially the 
Fe M-shell UTA and the inner-shell lines. The column density of this low ionization grid 
covered the range from $N_{\rm H}=0.5-5.0\times10^{21}$\,cm$^{-2}$ in steps of 
$\Delta N_{\rm H} = 1\times10^{20}$\,cm$^{-2}$, with the ionization range extending from 
$\logxi = 0-3$ in 15 steps of $\Delta (\log \xi)=0.2$. 
A fine spectral resolution of $N=10^{5}$ points over an energy range of $0.1-20$\,keV was 
also employed. A turbulence velocity of $\sigma=100$\,km\,s$^{-1}$ was also adopted. 
The other significant 
difference with this absorption grid was that a steeper photoionizing X-ray continuum of $\Gamma=2.5$ 
was employed, the requirement for this is discussed further in Section 4.2.

\subsection{XMM-Newton RGS}

We first considered the RGS spectrum. 
The initial analysis of the absorption line spectrum from the HETG and RGS observations 
in Section 3 suggests 
that multiple absorption components may be required in order to model the 
wide range in the ionization state of the gas, e.g. covering for instance 
Fe\,\textsc{vii-xxii}, Ne\,\textsc{v-x} or Mg\,\textsc{vi-xii}. 

In order to model the absorption spectrum we successively added individual 
components of absorbing gas, fully covering the line of sight to the source, until the fit statistic was  
no longer improved at the 99.9\% confidence level. Three components of
fully covering gas are formally required  
in the RGS model, which are listed as components 1--3 in Table\,5. The lowest ionization absorber 
(component 1) was modelled by the low ionization \textsc{xstar} grid as described above and components 2-3 
by the higher ionization grid. We note that the continuum itself was assumed to be a power-law 
of variable photon index, absorbed by the Galactic column, while we no longer retain either the ad-hoc 
absorption edge or the simple neutral partial coverer in the models. However we do allow 
for at least one additional component of partially ionized 
absorbing gas (as modeled by an \textsc{xstar} grid) to partially cover the X-ray source, in addition 
to the three fully covering components of gas described above, which appears to be required statistically 
to achieve a good fit. Soft X-ray emission lines are also added to the model as Gaussians 
when statistically required by the data at $>99$\% and will be discussed in detail later. 
Thus the phenomenological form of the spectral model fitted to the RGS data is:-

\begin{equation}
F(E) = {\rm tbabs} \times {\rm comp1} \times {\rm comp2} \times {\rm comp3} \times 
[{\rm pow}_{\rm uncov} + {\rm Gauss} + ({\rm pc_{1}} \times {\rm pow}_{\rm cov})]
\end{equation}

\noindent where here comp\,1-3 represent the 3 fully covering warm absorber components, {\rm Gauss} represents the 
Gaussian emission lines and tbabs the Galactic absorption. The partial covering absorber is 
represented by ${\rm pc}_1$ which covers a fraction $f_{\rm cov}$ of the line-of-sight to the X-ray 
source, while $1-f_{\rm cov}$ is subsequently unattenuated by the partial covering 
component. Thus the fraction of the continuum that is absorbed 
simply given by the respective ratio of the power-law normalizations, i.e:-
$f=N_{\rm cov}/(N_{\rm cov} + N_{\rm uncov})$. The spectral parameters of the RGS fit are listed in 
Table~\ref{absorbers}.

Overall the three warm absorber components that are required to model the 
RGS spectrum cover the range in column from $N_{\rm H}=1.5-3.6\times10^{21}$\,cm$^{-2}$ 
and ionization parameter from $\logxi = 1.27-2.80$. Consistent outflow velocities 
are found for the low and medium ionization components 1 and 2, 
with $v_{\rm out}=-480\pm40$\,km\,s$^{-1}$ and 
$v_{\rm out}=-460\pm60$\,km\,s$^{-1}$ respectively. However the
highest ionization component 3 does not require  
an outflow velocity (formally consistent with zero) 
and only a limit can be placed with $v_{\rm out}<130$\,km\,s$^{-1}$. 
We note that the lack of any outflow velocity of component 3 also appears consistent 
with the velocity profile analysis in Section\,3.4, where the velocities of the higher excitation 
lines appear to be lower.
The column density 
($6\times10^{22}$\,cm$^{-2}$) and ionization ($\logxi=1$) of the partial 
covering component are not well constrained in the RGS fit, mainly because of the limited higher 
energy bandpass of the RGS makes it difficult to constrain multiple continuum components, while the 
partial coverer itself does not impart discrete detectable lines upon the soft X-ray spectrum 
(but it does impart continuum curvature).
Thus its column and ionization have been fixed in the model, while we note that these values are consistent 
with those obtained with the HETG in Section 4.2. 

Nonetheless the partial coverer is 
certainly required in the model, the fit statistic is increased by $\Delta \chi^{2} = 192.4$ upon 
removing the partial coverer from the model and refitting; its exclusion leads to systematic broad residuals 
in the data/model ratio suggesting the continuum is inadequately modeled. 
The covering fraction of the partial coverer is $f_{\rm cov}=0.61\pm0.05$. 
Overall the fit statistic for the best-fit warm absorber model is $\chi^{2}/{\rm dof}=2991.7/2562$, 
while the continuum photon index upon modeling all the three required components of warm absorption is 
steeper, with $\Gamma=2.32\pm0.08$. The warm absorber model reproduces well the absorption 
lines observed in the RGS spectrum, as shown by the solid line in 
Figures~\ref{rgs-panels1} and \ref{rgs-panels2}.
We also note that addition to the warm absorption,
an additional neutral component of absorption is required in the rest frame of MR\,2251-178. 
However its column density is quite small, $N_{\rm H}=(2.8\pm0.3)\times10^{20}$\,cm$^{-2}$, and it may 
plausibly be associated with absorption in the quasar host galaxy rather than the AGN.

The relatively low turbulence velocity (of $\sigma=100$\,km\,s$^{-1}$) 
of the warm absorber components aides 
in the modeling of the higher order lines, as some of the $1s \to 2p$ 
lines may lie on the saturated part of the 
curve of growth. This means the some of the higher order lines can be of comparable strength as 
the $1s \to 2p$ lines, while some of the line series are detected up to $1s \to 6p$. Indeed the 
warm absorber model matches well the profiles of the higher order lines, as can be seen 
in Figures~\ref{rgs-panels1} and \ref{rgs-panels2}.

To correctly account for the intensity of the low ionisation lines the absorbing 
grid requires a much softer (steeper) input continuum than the other
higher ionization absorption components  
(which have $\Gamma_{\rm input}=2.0$), in order not to over-ionize
the gas and reduce their depth in the model. The necessary power-law continuum required by 
the \textsc{xstar} grid in order to model the low ionization lines is $\Gamma_{\rm input}=2.5$. 
This is much softer than what has typically been found for MR\,2251-178 assuming a fully-covering 
absorption model, which is of the order of $\Gamma=1.6-1.7$
\citep{pan1990,mineo1993,kaspi2004,gibson2005}. However the underlying soft X-ray 
photon index recovered in the RGS spectrum ($\Gamma=2.32\pm0.08$), after the required absorbing layers of 
gas are accounted for, is in reasonable agreement with 
the required photon index to reproduce the soft X-ray lines. 
This lends weight to the notion that 
MR\,2251-178 may, indeed, have an intrinsically soft continuum which is partially-covered by a complex 
and stratified absorber. We discuss this further in Section 4.2.

\subsection{Chandra HETG}

The above best-fit model was then applied to the 2011 HETG spectrum, 
allowing the continuum and warm absorber parameters to vary between
the datasets. A second partial covering  component of higher column
density of $\sim7\times10^{23}$\,cm$^{-2}$ was added to the model, as
the direct application of the RGS model gave a slight excess at higher energies in the HETG spectrum.
Otherwise the model construction applied to the HETG data is identical to the RGS.

The absorber fit parameters applied to the 2011 HETG spectrum are also listed in Table~\ref{absorbers}. 
The parameters of the 3 warm absorber components are rather similar to those obtained from the RGS data, 
with most of the values consistent within the errors between the observations. Similar to the RGS, the 
warm absorber column densities cover the narrow range $N_{\rm H}=1.5-2.1\times10^{21}$\,cm$^{-2}$, 
while the ionization 
spans a range from $\logxi=1.15-2.9$. There is evidence for a small change in the 
ionization of the warm absorber of the low ionization component 1, increasing from 
$\logxi=1.15\pm0.05$ to $\logxi=1.27\pm0.02$ 
between the HETG and RGS, following the same direction as the 0.4--2.0\,keV continuum flux which 
also increased from the HETG to the RGS, we discuss this further in Section 4.3 below.
The column density of component 1 is consistent between observations, 
with $N_{\rm H}=2\times10^{21}$\,cm$^{-2}$, although the 
outflow velocity is slightly smaller\footnote{The differences are likely within the absolute wavelength 
scales of the HETG and RGS.}, with $v_{\rm out}=-315\pm40$\,km\,s$^{-1}$. The ionization and 
columns of components 2 and 3 are consistent within the errors, while as per the RGS, the highest ionization 
component 3 does not require any outflow, as noted above.

\begin{figure}
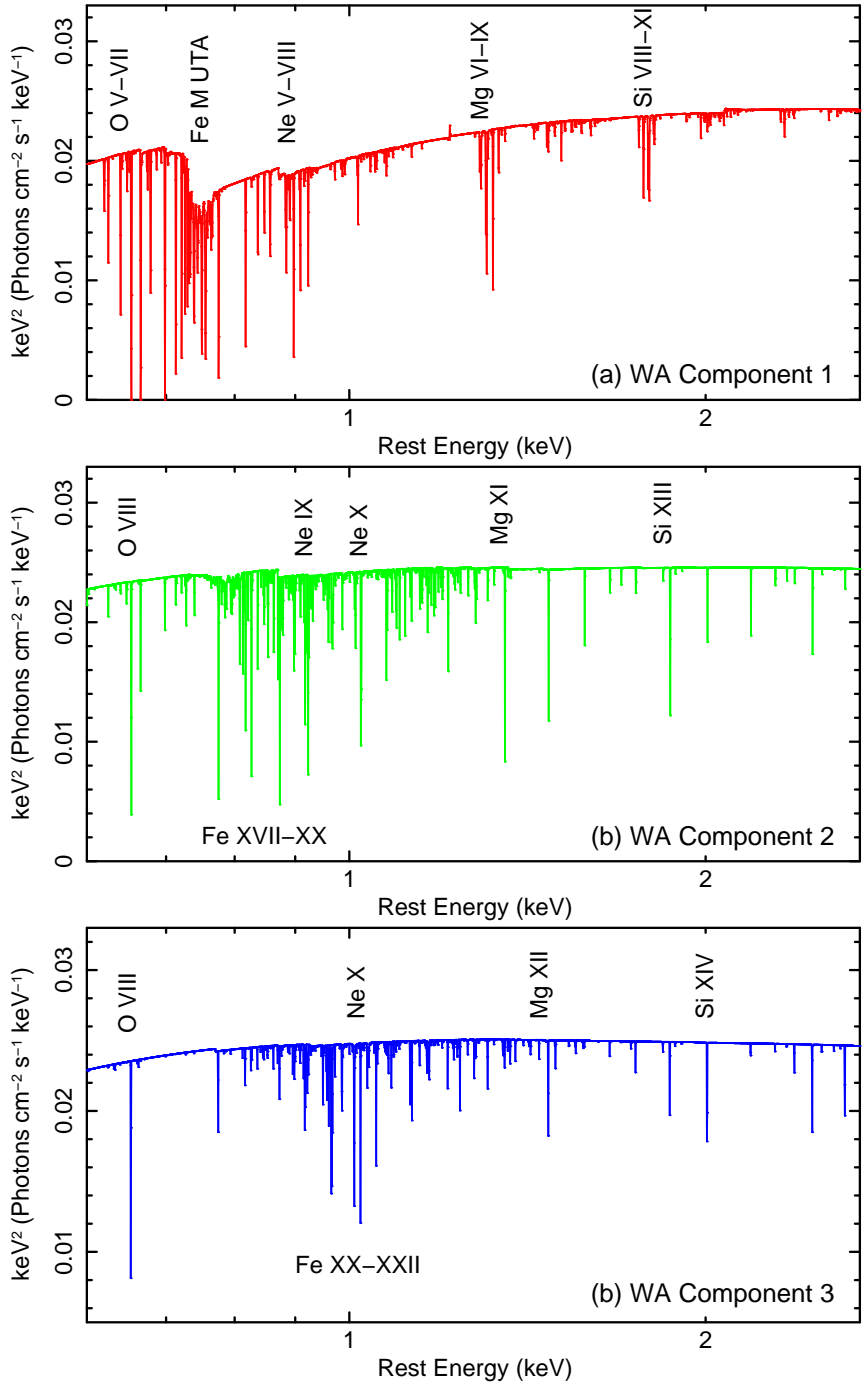

\centering
\rotatebox{-90}{\includegraphics[height=12cm]{f9a.eps}}
\rotatebox{-90}{\includegraphics[height=12cm]{f9b.eps}}
\rotatebox{-90}{\includegraphics[height=12cm]{f9c.eps}}
\caption{Contribution of respective warm absorption components towards the X-ray spectrum. The low ionization component 1 (panel a)
carries the largest opacity with absorption due to inner shell O, Ne, Mg, Si and M-shell Fe; component 2 (panel b) 
contains absorption due to He-like ions and moderately ionized Fe and component 3 (panel c) contributes absorption 
due to H-like ions and highly ionized Fe.}
\label{components}
\end{figure}

Figure\,\ref{components} shows the relative contributions of each of the 3 warm absorbers components against a power-law 
continuum. The lowest ionization component 1 (top panel) contributes the lower ionization ions, i.e. 
O\,\textsc{v-vii}, Ne\,\textsc{v-viii}, Mg\,\textsc{vi-ix}, Si\,\textsc{viii-xi} as well as M-shell 
iron, as expected. The higher ionization components produce most of the He and H-like ions, as well as 
the higher ionization (L-shell) iron ions (see lower panels).

\subsubsection{The Nature of the Photoionizing Continuum}

\begin{figure}
\centering
\includegraphics[width=14.2cm]{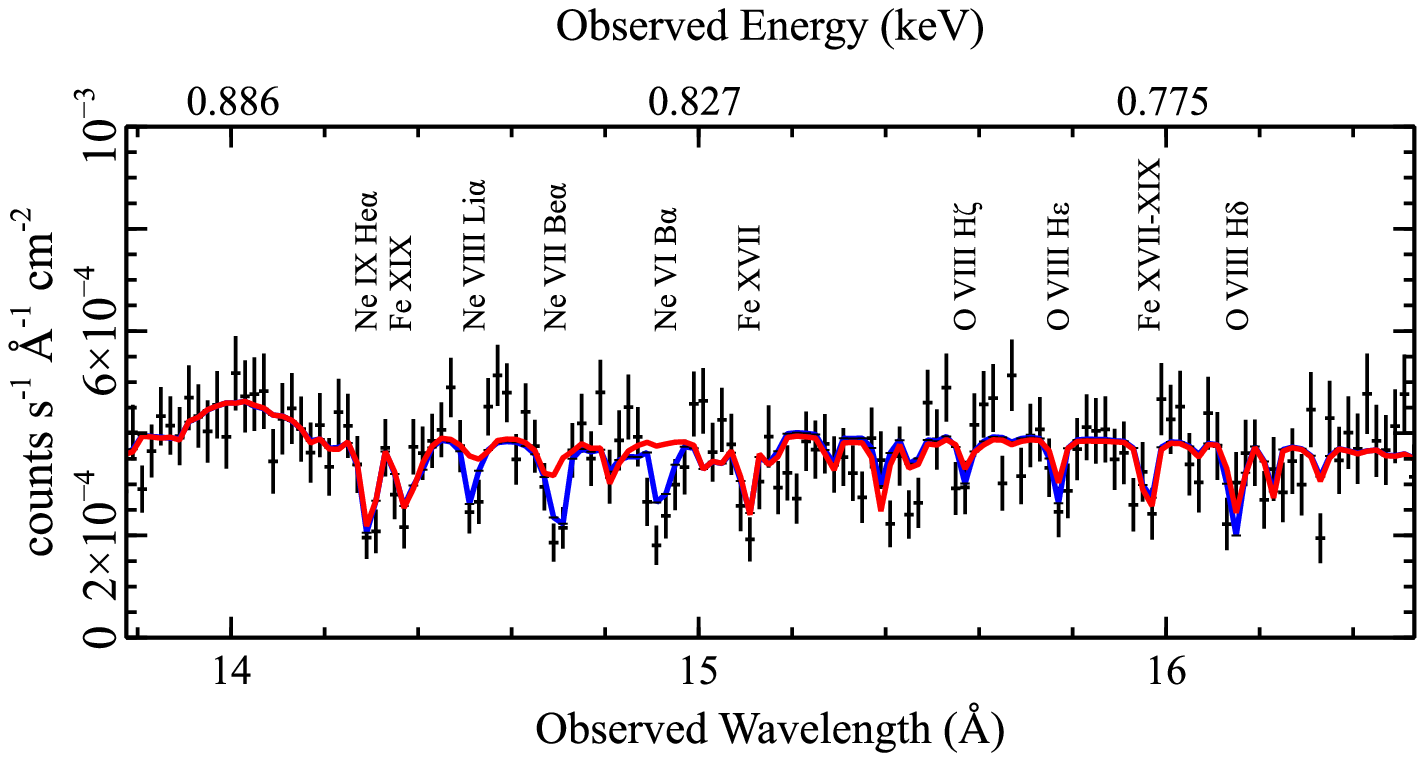}
\includegraphics[width=14cm]{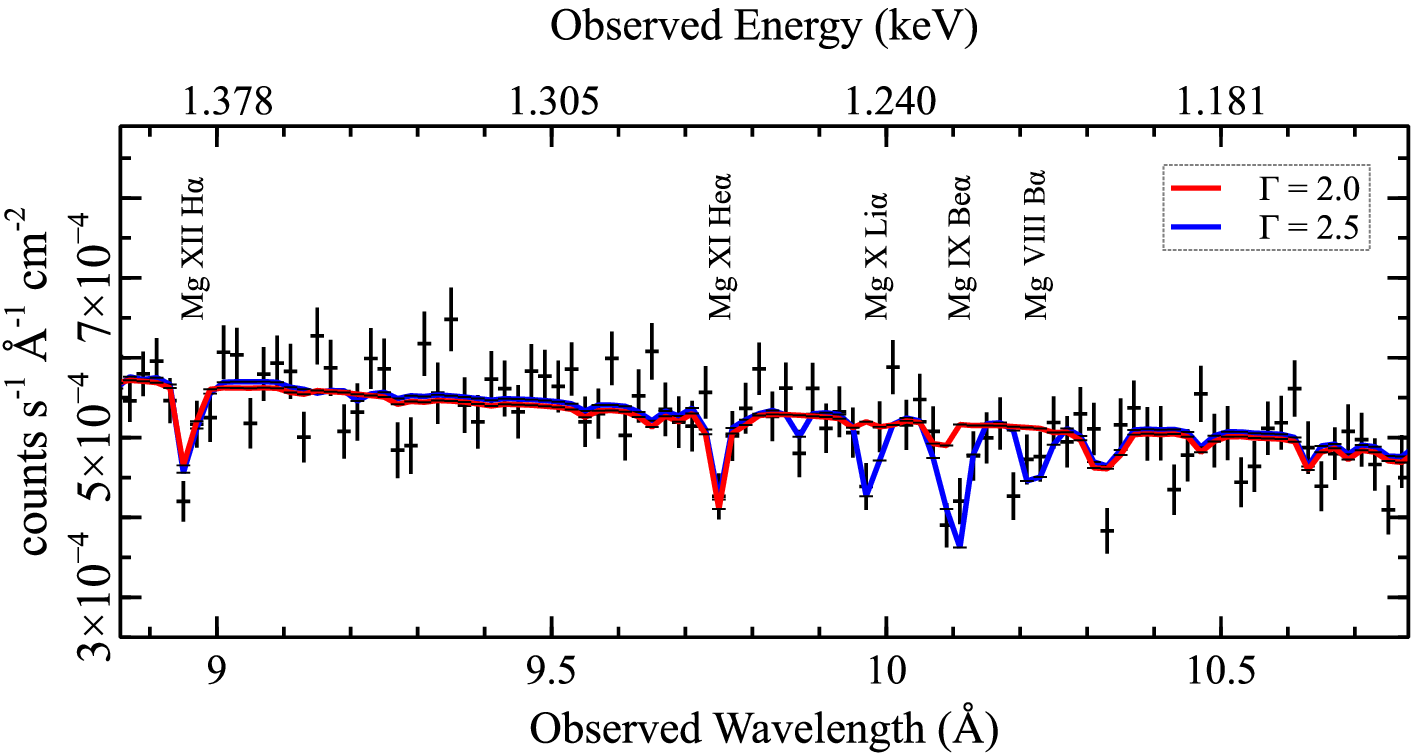}
\caption{Enlarged portions of the 2011 \chandrahetg\ observation of MR\,2251-178, focusing on the 
Ne and Mg energy bands. The HETG data give a much clearer view of the inner-shell Ne and Mg lines 
than was possible with the \xmmrgs. Both elements have lines due to their B-, Be- and Li-like 
charge states. The solid lines correspond to the fit that is obtained when the 
low ionization xstar absorber (component 1, Table 5) 
has an input photon continuum of $\Gamma_{\rm input}=2.0$ (red) and 
$\Gamma_{\rm input}=2.5$ (blue). Importantly, the inner-shell lines cannot be accounted for 
without an intrinsically soft X-ray continuum, which in turn provides evidence for a 
partially-covered X-ray spectrum. See text for further details.}
\label{hetg-inner}
\end{figure}

The HETG has a wider bandpass and higher resolution than the RGS, which enables some additional tests 
to be applied to the inner-shell lines in particular. 
Figure~\ref{hetg-inner} shows a comparison between the fit to the warm absorber when 
the low ionization component (component 1) of \textsc{xstar} absorption has a 
$\Gamma_{\rm input}=2.5$ input photoionizing continuum (blue line) or $\Gamma_{\rm input}=2.0$ (red line). 
For the case of the harder $\Gamma=2$ input continuum, the model is clearly unable to account for the 
depth of the inner-shell (Li-like and below) 
charge states of Ne or Mg, whereas the $\Gamma_{\rm input}=2.5$ absorber  
is able to model the low ionization absorption lines. This suggests that the softer input continuum 
is strongly required to model the absorption. The absorption grid with the steeper continuum also provides a better fit to the Fe M-shell UTA and also the Silicon inner-shell lines. 
These differences are reflected in the fit statistic, which for
the $\Gamma=2$ grid is $C=2665.9$ for 2335 degrees of freedom, whereas for the $\Gamma=2.5$ grid 
the fit statistic is $C=2542.4$ for the same number of degrees of freedom, corresponding to a 
difference of $\Delta C = 123.5$. 

Overall the photon index of the continuum recovered 
after modeling all the layers of absorption is $\Gamma=2.13\pm0.10$. 
Thus the index is somewhat flatter than in the RGS ($\Gamma=2.32$), 
but this may reflect the fact that the RGS is more sensitive at soft X-ray energies than the HETG, 
especially if the intrinsic continuum has subtle curvature, becoming slightly steeper towards lower energies. 
Note that Figure\,\ref{hetg-pl} also shows the level of the intrinsic continuum (the dashed blue line) after 
correcting for all the absorbing layers of gas. Thus the observed continuum without 
modeling the absorption (which would otherwise appear to have a very hard 
photon index of $\Gamma=1.3$) does not necessarily represent the intrinsic emission, where $\Gamma\gs2$, 
more typical of radio-quiet quasars
\citep[e.g.,][]{reeves&turner2000,porquet2004,scott2011}. 

The partial covering components also appear to be required by the data. 
The moderate column partial covering component (named pc\,1, Table~\ref{absorbers}) appears well 
constrained, with $N_{\rm H}=5.5\pm0.3 \times 10^{22}$\,cm$^{-2}$ and 
$\logxi = 1.04^{+0.08}_{-0.11}$, while its covering fraction 
is $f_{\rm cov}=0.4\pm0.1$. 
The highest column component (pc2, Table~\ref{absorbers}) is less well constrained, 
but the fit is still worse by $\Delta C=31.6$ if this component 
is removed from the model and the continuum refitted. The removal of the pc2 absorber 
results in the fitted photon index
hardening from $\Gamma=2.13\pm0.10$ to $\Gamma=1.77\pm0.05$. Furthermore if the more moderate column 
partial coverer (pc1) is also removed then the fit is considerably worse $\Delta C=213.1$ and the 
photon index then becomes an unphysical $\Gamma=1.49\pm0.03$. 

Such a hard continuum slope 
also poses a problem for the modeling of the warm absorber components, as the low ionization (inner shell) 
absorption requires a soft input photoionizing continuum of $\Gamma\sim2.5$ as above, which cannot be recovered 
in the model without applying the partial covering absorption. The other possibility is that the intrinsic 
continuum shape and high energy SED 
are unusual, consisting of a rather hard powerlaw component $\Gamma\sim1.5$ (and much harder 
than usually observed in radio-quiet quasars), then softening 
to an index of $\Gamma\gs2.5$ at soft X-ray energies. The broad band continuum modeling will be 
explored in more detail in a forthcoming paper (Nardini et al. 2013, in prep), where the \xmm\ 
EPIC and Optical Monitor data will be considered, as well as archival Suzaku and Swift/BAT 
observations, thereby covering the optical/UV through to hard X-ray bandpass.

We note that although a softer continuum does provide a better fit to the inner shell
lines and some improvement to the Fe UTA, the model fits for these
inner shell features is dependent on the calculation of
the ionization balance for these elements. For example, in their analysis of the
900ksec HETG spectrum of NGC 3783, \citet{netzer2003} noted that their
best warm absorber model did not accurately reproduce the Fe UTA due to the
predicted iron being too highly ionized. \citet{netzer2003} suggested that the 
problem was the lack of accurate low-temperature ($\Delta {\rm n} = 0$) dielectronic
recombination (DR) rates for the M-shell sequence of iron (Fe~\,\textsc{ix} -- Fe\,\textsc{xvi}). 
Following this, \citet{netzer2004} and \citet{kraemer2004} 
incorporated estimated $\Delta {\rm n} = 0$ DR rates into the
codes ION \citep{Netzer1996} and Cloudy \citep{ferland1998}, respectively,
and demonstrated that such rates would shift the overall ionization balance
of M-shell iron downward, hence solving the problem described by \citet{netzer2003}. 

More recently, DR rates have been computed \citep{badnell2006} for the M-shell states of iron, 
which are included within \textsc{xstar}. These are an order of magnitude greater
than the radiative recombination rates for these ions and several times greater
than the estimated DR rates from \citet{netzer2004} and \citet{kraemer2004}.
Furthermore, these rates have been confirmed in storage-ring experiments
\citep{Schmidt2006}. However while for the same physical parameters as those
used in \citet{netzer2003}, Cloudy models using the new DR rates predict 
similar C, N, and O column densities, the predicted Fe ionization is now too low
to fit the UTA. Although it may be possible to recover the fit by changing model
parameters (e.g., the continuum slope), these results may also indicate that some process which mitigates the
effects of the new DR rates is not being accurately treated. 
One possibility is (multi-electron) autoionization following inner-shell ionization (D. Savin,
private communication). In any event, given such
sensitivity to the accuracy and availability of atomic data, the exact parameterization of the 
low ionization absorber could differ, with the ionization perhaps somewhat lower than currently 
inferred by \textsc{xstar}.

\subsection{Variability of the X-ray Absorption}

The best fit absorption model to the 2011 HETG spectrum was also applied to the earlier 2002 HETG 
observation. The signal to noise of the 2002 observation is substantially lower, due to the overall lower 
flux level (and count rate) and shorter exposure of this observation (see Table\,1), which means that 
most of the individual absorption lines were not detected (see \citealt{gibson2005} for a description 
of this dataset). However the same spectral model can still be applied to the 2002 data, allowing the 
continuum and warm absorber parameters to vary between the observations. For ease of comparison 
the photon index of the 2002 observation was tied to that of the 2011 observation, i.e. $\Gamma=2.13$. 
The column and ionization of the partial covering components were also fixed to the 2011 values, as otherwise 
they are less well determined, although the covering fractions were allowed to vary. The warm 
absorber parameters (column, ionization, outflow velocity) were allowed to vary between the observations. 

The absorber parameters of the 2002 observation are shown in Table~\ref{absorbers}. 
Again the absorption values are largely consistent between the 2002 and 2011 HETG observations, 
as well as with the 2011 RGS observations, suggesting that the absorber components appear stable over time.
The main parameter that does appear to change is the ionization of the low ionization component 1 absorber. 
Indeed if the 2011 RGS observation is also considered, the ionization of component 1 appears to increase from 
$\xi=8.1^{+3.5}_{-2.5}$\,erg\,cm\,s$^{-1}$ (Sept 2002/HETG) to $\xi=14.1\pm1.6$\,erg\,cm\,s$^{-1}$ 
(Sept 2011/HETG) to 
$\xi=18.6\pm0.8$\,erg\,cm\,s$^{-1}$ (Nov 2011/RGS). 
Indeed the changes in $\xi$ appear increase in direct proportion 
to the observed 0.5--2.0\,keV band flux, varying from 
$0.75\pm0.01\times10^{-11}$\,erg\,cm$^{-2}$\,s$^{-1}$ (Sept 2002/HETG) to 
$1.33\pm0.01\times10^{-11}$\,erg\,cm$^{-2}$\,s$^{-1}$ (Sept 2011/HETG) to 
$1.80\pm0.01\times10^{-11}$\,erg\,cm$^{-2}$\,s$^{-1}$ (Nov 2011/RGS). 
Thus from the lowest ionization to highest, $\xi$ increases by a factor $\times2.3$, 
while the soft X-ray flux increase by the same factor. This would appear to suggest that the 
low ionization absorber is in photoionization equilibrium with the continuum. In contrast
there appears to be no change in the higher ionization components 2 and 3, within the errors.
Note that this behavior is also consistent with a December 2002 (80\,ks) Chandra LETG 
observation (not analyzed here), which was at about a 35\% lower flux than the 2002 HETG observation, but 
observed the low ionization absorber to have an even lower ionization, 
of $\logxi=0.63\pm0.06$ \citep{ramirez2008}.

\begin{figure}
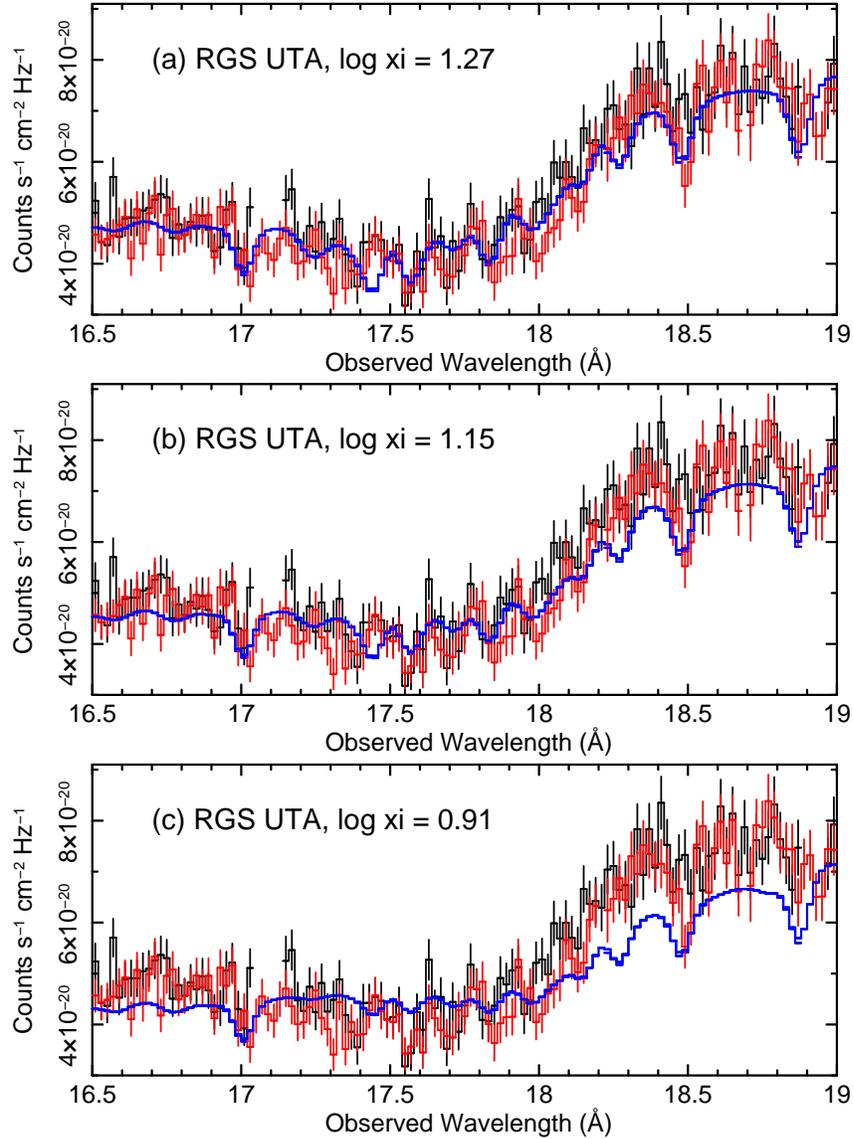

\centering
\rotatebox{-90}{\includegraphics[height=12cm]{f11a.eps}}
\rotatebox{-90}{\includegraphics[height=12cm]{f11b.eps}}
\rotatebox{-90}{\includegraphics[height=12cm]{f11c.eps}}
\caption{Zoom-in of the 2011 RGS spectrum in the region of the iron M-shell UTA, showing the 
effect of the change in the ionization state of component 1 in the warm absorber model (solid line). 
Panel (a) shows the best fit case to the RGS, 
whereby the ionization parameter of component 1 is $\logxi=1.27$. Panel (b) shows the model 
fitted when the ionization is lowered to $\logxi=1.15$, as found in the 2011 Chandra HETG 
spectral fits. In panel (c) the ionization parameter is $\logxi=0.91$, as found in the spectral 
fits to the 2002 Chandra HETG data. Thus the fits to the UTA region are 
sensitive to the ionization parameter 
in the xstar absorber model and models with a substantially lower ionization, as found in the 
Chandra datasets, can be ruled out by the RGS data.}
\label{uta}
\end{figure}

We also illustrate the apparent change in ionization further in Figure\,\ref{uta}, which 
plots the change in the xstar model from varying the ionization of warm absorber component\,1, 
against the 2011 RGS data in the Fe M-shell UTA band. The upper panel of Figure\,\ref{uta} plots 
the best fit model obtained, with an ionization parameter of $\logxi=1.27$ for component 1, 
as reported in Table\,\ref{absorbers}. Then the ionization parameter was lowered (and fixed) to 
$\logxi=1.15$, equal to the value found for component\,1 in the 2011 HETG spectrum. 
This results in a worse fit, as seen in panel (b) of Figure\,\ref{uta}, indeed even allowing the 
other warm absorber and continuum parameters in the fit to vary resulted in a worse fit 
by $\Delta \chi^{2}=24.6$. Similarily if the ionization parameter is lowered still further, 
to $\logxi=0.91$ as obtained from the 2002 Chandra HETG data, the fit is substantially worse 
by $\Delta \chi^{2}=125.4$, compared to the best fit case shown in panel (a). Indeed 
this can be seen in panel (c) of Figure\,\ref{uta}, whereby the drop in the Fe M-shell UTA region 
observed at 17.5--18.5\AA\ is too 
shallow compared to the data, while the spectrum is then too absorbed red-wards of this 
feature. Thus overall the Fe M-shell UTA region appears to be quite sensitive to the ionization state 
of the spectrum.

The other possible change in the spectra is in the partial covering absorption. Considering all 
three grating observations, the uncovered fraction (or $1-f$) of the power-law (in other words the 
fraction that is not obscured by the partial covering absorption) appears to increase 
as the flux increases from the 2002 through to the 2011 observations, from $(1-f)=0.18\pm0.02$ 
to $(1-f)=0.39\pm0.03$. This may suggest that the AGN 
is more obscured when it is in a lower flux state, which has been claimed in several Seyferts 
to date (e.g., NGC\,3516: \citealt{turner2005,turner2008};
  PG\,1211+143: \citealt{bachev2009,PR2009}; H\,0557-385:
  \citealt{longinotti2009}; NGC\,4051: \citealt{Terashima2009, lobban2011}),  
and indeed variable X-ray absorption
was first suggested from soft X-ray band  
variations in MR\,2251-178 itself \citep{halpern1984}. 
This variability behaviour will be investigated further in a subsequent paper (Porquet et al. 2013, 
in preparation), 
considering a broad-band X-ray analysis of all the contemporary and archival observations of 
MR\,2251-178.

\subsection{Is there a very highly ionized absorber?}\label{sec:zonehigh}

Previous studies of MR\,2251-178, with a 2009 Suzaku observation \citep{gofford2011} and the 
2002 HETG observation \citep{gibson2005}, have suggested the presence of a highly ionized and 
possibly strongly outflowing, absorption component in the iron K band. Such absorption 
could be similar to the very highly ionized outflows (or ``ultra fast outflows'') 
detected in about 40\% of local type I AGN with \xmm\ \citep{tombesi2011a} and 
\suzaku\ \citep{gofford2013}. Thus we have analyzed 
the higher energy 2011 Chandra HETG observation above 2\,keV, using the High Energy Grating (HEG) 
spectrum, to assess whether such a component is present in the new data.  
The 2002 HETG spectrum was also re-analyzed for comparison, while the results are also compared to the 
Suzaku analysis in \cite{gofford2011}. 

\begin{figure}
\begin{center}
\rotatebox{-90}{\includegraphics[width=10cm]{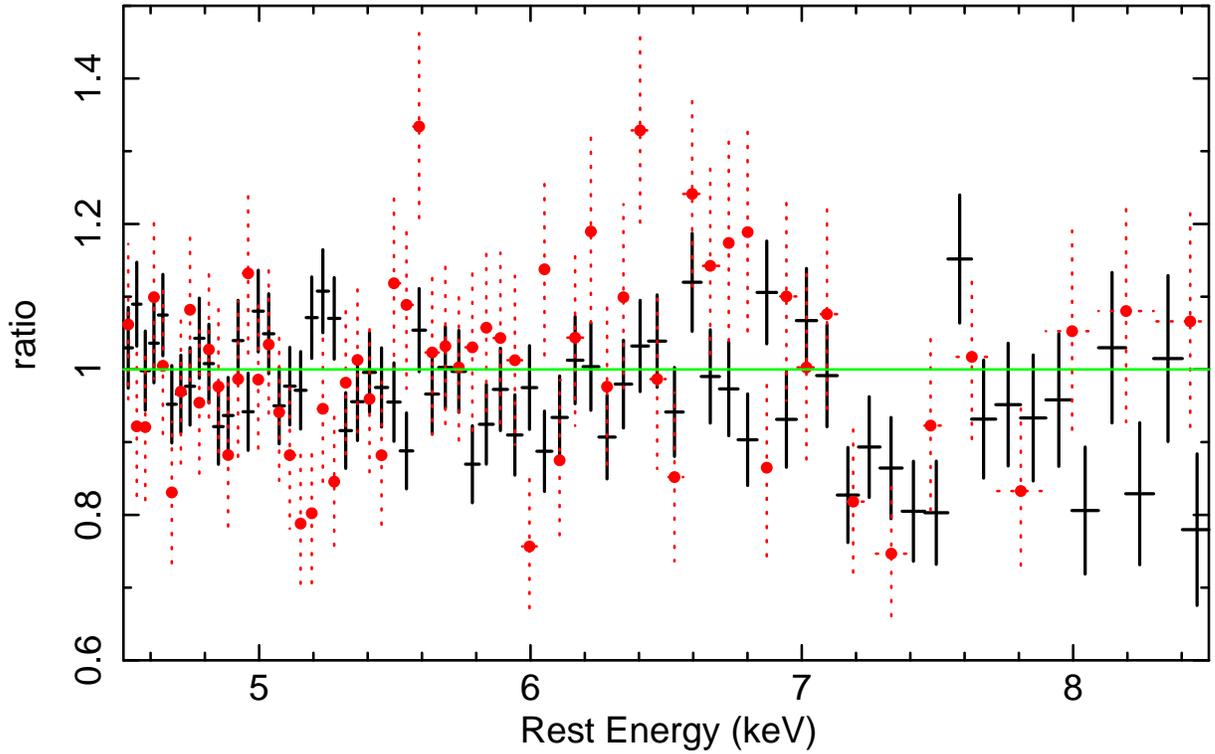}}
\end{center}
\caption{Data/model ratio residuals to the HEG spectrum of MR\,2251-178, in the iron K band, 
to the best fit continuum model. The 2011 data are shown as black crosses, the 2002 HEG data shown 
in red circles (with dashed errors). The datasets have also been binned to have a minimum of 20 
counts per bin, in addition to the instrument resolution binning.
Energy is plotted in the quasar rest frame at $z=0.064$. 
Both datasets appear to show a weak, but statistically required, 
absorption feature near 7.3\,keV, which if identified with H-like iron would require a 
blueshift of $\sim-15000$\,km\,s$^{-1}$. Note the lack of a strong K$\alpha$ emission line 
at 6.4\,keV. } 
\label{ratio-fe}
\end{figure}

Figure~\ref{ratio-fe} shows the data/model residuals of the 2011 HEG spectrum to the best-fit 
absorption model discussed above, plotted over the Fe K band in the quasar rest frame {\and 
further binning the spectrum to 20 counts per bin to increase the signal to noise.}
First we consider the iron K band emission.
The lack of any strong iron K$\alpha$ emission is quite apparent in the residuals. 
Indeed the limit on the equivalent width of a narrow 6.4\,keV line is $11\pm6$\,eV and 
is only very marginally required at $\sim95$\% confidence in the fit, with $\Delta C=6.3$. 
The limit on the width of the line is $\sigma<28$\,eV or $\sigma<1300$\,km\,s$^{-1}$. 
No other iron K emission component is required in the spectrum, either narrow or broad. 
The weakness of the iron K$\alpha$ line in MR\,2251-178 has also been noticed previously 
(\citealt{gofford2011} and references therein), and is much weaker that 
the typical narrow iron line equivalent width of $\sim50-100$\,eV observed 
in most Seyfert 1s \cite[e.g.,][]{nandra1997,patrick2012,tatum2013}.
The weakness of the iron K line may be accounted for by the X-ray Baldwin effect, whereby 
the equivalent width of the iron K$\alpha$ line appears to decrease with AGN X-ray luminosity 
\cite[e.g.,][]{iwasawa1993,nandra1997,reeves&turner2000,page2004,bianchi2007,shu2010}. The 2-10\,keV X-ray 
luminosity of MR\,2251-178 in this observation is $3.7\times10^{44}$\,erg\,s$^{-1}$ 
(or absorption corrected, $5.8\times10^{44}$\,erg\,s$^{-1}$), higher than most local Seyfert 1\,s.

There does appear to be a broad but shallow absorption trough in the 2011 data at 7.3\,keV. Fitting the 
trough with a Gaussian absorption profile gives a rest frame centroid energy of $E=7.34\pm0.08$\,keV with an 
equivalent width of ${\rm EW}=-58\pm24$\,eV and the fit statistic improves by $\Delta C=15.2$. 
Note this appears to be consistent with the high energy absorption line that was previously claimed 
in the 2002 HETG observation by \cite{gibson2005}; there the line centroid was at $E=7.26\pm0.04$\,keV. 
Furthermore \cite{gofford2011} claimed an absorption trough in the Suzaku observation 
at an energy of $E=7.57^{+0.19}_{-0.12}$\,keV, 
which is only marginally inconsistent at 90\% confidence 
with the line energy measured by the 2011 Chandra data, while the equivalent width of $-26^{+18}_{-12}$\,eV is 
consistent.  
In Figure~\ref{ratio-fe} the 2002 HEG spectrum has been overlayed on the 2011 data, with 
the normalization of the 2002 spectrum allowed to vary to account for the overall lower flux level 
in the 2002 observation, it appears that the trough in the 2002 data has a consistent profile in both 
energy and depth with the 2011 data. With the 2002 and 2011 fitted together with a single Gaussian profile, 
then consistent parameters were obtained, with a line energy of $E=7.32\pm0.06$\,keV and an equivalent 
width of $-60\pm18$\,eV. The fit statistic was improved by $\Delta C=26.2$ with respect to a model 
without the absorption line.
The profile appears to be resolved compared to the HETG resolution, 
with a width of $\sigma=120^{+50}_{-40}$\,eV or $\sigma=4900^{+2100}_{-1600}$\,km\,s$^{-1}$. 
Note if the absorption line is associated with the Fe\,\textsc{xxvi} (H-like) $1s \to 2p$ 
doublet at 6.97\,keV, then the velocity shift implied is $-15000\pm2600$\,km\,s$^{-1}$.
We also note that no significant iron K$\alpha$ emission was required from refitting the 2002 
HEG spectrum, although the upper limit to its equivalent width is less well determined ($<40$\,eV) and 
is consistent with the 2011 measurement. 

We attempted to model the Fe K band absorption with a highly ionized \textsc{xstar} grid. 
Unlike for the warm absorber components, a high turbulence velocity grid was used, with 
$\sigma=5000$\,km\,s$^{-1}$, consistent with the observed line width and an 
illuminating hard X-ray continuum of $\Gamma=2$. The ionization parameter 
is not so well constrained, with $\logxi=4.8^{+1.0}_{-0.8}$, but suggests 
that either H-like or He-like iron contributes to the absorption. The column was found to be largely 
degenerate upon the ionization parameter (i.e. as the ionization increases the column increases to compensate) 
and only a lower-limit can be placed of $N_{\rm H}>1.5\times10^{23}$\,cm$^{-2}$. The outflow velocity 
derived was consistent with the line analysis, with $v_{\rm out}=-15600\pm2400$\,km\,s$^{-1}$ and 
is consistent with the \citet{gibson2005} value of $v_{\rm out}=-12700\pm2400$\,km\,s$^{-1}$.
However we also note that at this velocity, the absorption is only marginally excluded at 90\% 
confidence from being associated from a local $z=0$ absorber.

We also tested whether the iron K-shell region could instead be fitted with a photoelectric edge, 
from neutral or mildly ionized iron, without any velocity shift as was implied from the highly ionized 
absorption model. Indeed fitting the Chandra data with a simple edge model results in a 
equally good fit statistically, with a best fit edge energy of $E=7.15\pm0.05$\,keV and 
optical depth $\tau=0.15\pm0.05$. Such an edge component could plausibly result from a partial covering 
absorber with column density typically exceeding $N_{\rm H}>10^{23}$\,cm$^{-2}$ and as has 
been discussed, this may also
be required from fitting the broader band HETG spectrum. Thus it is not possible to distinguish 
here between the high velocity absorber and possible partial covering cases in MR\,2251-178 and 
higher resolution data in the Fe K bandpass, such as with the calorimeter to be flown 
on Astro-H, would be required to differentiate between these cases.

Thus the detection of the Fe K band absorption trough appears to be confirmed from the two 
Chandra observations, with the parameters consistent in both and at the same rest frame energy, 
although its exact origin remains uncertain. 
\cite{gofford2011} also claimed further blueshifted absorption features at lower energies 
from the Suzaku data; 
in particular absorption lines at $E=2.52\pm0.02$\,keV and $E=2.79\pm0.03$\,keV in the quasar rest 
frame, which were identified 
with blueshifted S\,\textsc{xv} and S\,\textsc{xvi} $1s \to 2p$ respectively. A 1.3\,keV 
absorption trough was present in the \suzaku\ data near 1.3\,keV and tentatively identified 
with blue-shifted iron L-shell transitions. In the latter case, the much higher resolution HETG and RGS 
spectra resolve the 1.3\,keV absorption into a series of lower ionization lines of inner shell Mg from 
Mg\,\textsc{vi-ix}, with only a modest outflow velocity of $\sim-400$\,km\,s$^{-1}$. However the 
absorption line at 2.52\,keV appears to be only marginally detected in the 2011 Chandra 
spectrum at $\sim 99$\% confidence 
($\Delta C = 9.3$) at $E=2.521\pm0.002$\,keV in the quasar rest frame (or 4.92\AA) 
with an equivalent width of $-2.0\pm1.2$\,eV; these 
parameters are entirely consistent with those measured by \suzaku. An absorption line is not 
detected at 2.79\,keV, however the limit on the equivalent width of ${\rm EW}<4$\,eV from Chandra 
is consistent with the Suzaku measurement of $-5\pm2$\,eV. Thus the presence of this 
possible higher velocity component appears uncertain based on the current data and such a 
component does not appear to be present in line profiles of C through to Si. 

\section{Modeling the Emission Line Spectrum}\label{sec:emission}

As we have noted previously, the 2011 RGS and HETG observations contain several soft X-ray emission 
lines, which have been fitted with simple Gaussian emission line profiles. The parameters 
of these emission lines are listed in Table~\ref{emission-lines}. Most of the lines were detected 
in the RGS rather than the HETG, as the RGS has a higher effective area below 1\,keV. 
Many of the lines detected are substantially broadened, with typical widths of several 
thousand km\,$^{-1}$, from C\,\textsc{vi} Lyman-$\alpha$, N\,\textsc{vi}, O\,\textsc{vii}, 
and Ne\,\textsc{ix}. Two weaker narrow components are also present from N\,\textsc{vii} Lyman-$\alpha$, and 
Ne\,\textsc{ix}, with velocity widths typically $\ls1000$\,km\,s$^{-1}$ (FWHM). 
The latter line is detected at an energy of $905\pm1$\,eV in both the RGS and HETG 
and would appear to be consistent with expected energy of the
forbidden line of the Ne\,\textsc{ix} triplet. As we discuss 
below, a weak narrow component of the O\,\textsc{vii} forbidden line cannot be ruled out in the RGS 
spectrum. Thus it may be plausible that the broad lines originate from BLR type gas, while the 
narrow (and forbidden) lines originate from gas associated with the NLR.

The O\,\textsc{vii} line complex is by far the strongest and most statistically significant 
emission feature detected (with $\Delta \chi^{2}=345.1$ upon its addition 
to the model), while it also appears be detected with consistent parameters in the HETG spectrum (albeit 
less well constrained). We therefore concentrate on the analysis of
the O\,\textsc{vii} line complex, using the 
high signal to noise RGS spectrum. The line complex width is
certainly broadened, with a FWHM velocity width of  
$10200^{+1200}_{-1400}$\,km\,s$^{-1}$. Note that the width of the
C\,\textsc{vi} line complex is poorly constrained, as  
it lies at the low energy end of the RGS bandpass, and so has been set equal to the O\,\textsc{vii} 
line complex width, which is the best determined broad line. 

\begin{figure}
\begin{center}
\rotatebox{-90}{\includegraphics[width=12cm]{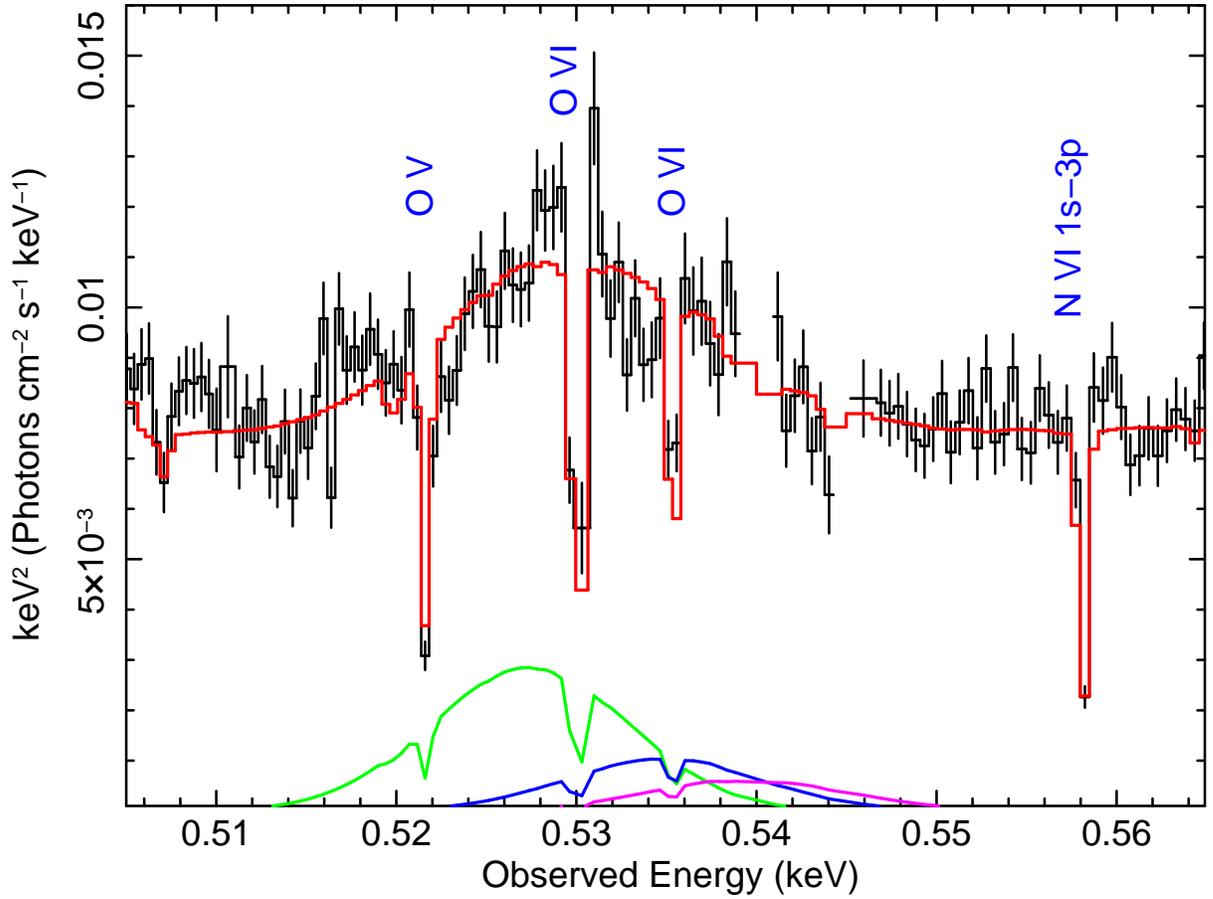}}
\end{center}
\caption{Broad O\,\textsc{vii} emission line complex as observed
  by RGS\,1, plotted against {\it observed} frame   
energy. Black crosses show the data, while the solid (red) line is the best fit model. 
The line profile has been fitted by a blend of forbidden, intercombination and resonance emission 
components of equal velocity width (FWHM 7300\,km\,s$^{-1}$), 
as shown by the solid lines below from left to right (green, blue and magenta respectively). 
The forbidden line dominates the profile, implying densities of $n_{e}<10^{11}$\,cm$^{-2}$. 
Note that narrow absorption lines of O\,\textsc{v-vi} are superimposed upon the broad emission profile.} 
\label{ovii}
\end{figure}

An enlarged view of the O\,\textsc{vii} RGS line complex profile is plotted in Figure~\ref{ovii}. 
Note that this portion of the spectrum only contains data from RGS\,1, due to the malfunctioning 
RGS\,2 chip over this energy range. The fit with a single broad line profile is good, 
with an overall fit statistic of $\chi^{2}/{\rm dof}=3007.7/2564$. It is also apparent that 
three narrow absorption lines are superimposed upon the emission line profile, which have been identified 
with inner shell O\,\textsc{v-vi}, e.g. see Table~\ref{rgs-lines}. We tested whether a narrow 
($\sigma<1$\,eV) 
component due to the O\,\textsc{vii} forbidden line at 561.0\,eV could also be added to the profile and indeed 
such a component cannot be excluded, with an equivalent width of ${\rm EW} = 0.9\pm0.4$\,eV and an improvement 
in fit statistic of $\Delta \chi^{2}=16.0$. The equivalent width of the narrow component  
is much weaker than that of the broad line, which has $EW = 8.3^{+0.9}_{-1.1}$\,eV and thus its overall 
contribution towards the profile is negligible.

Note the energy of the broad O\,\textsc{vii} emission line is $E=564.5\pm0.9$\,eV, which is somewhat 
blue-shifted compared to the expected energy of the forbidden line at 561.0\,eV. If the broad emission 
is purely associated with the forbidden emission, this would suggest an overall blue-shift of
$-1900\pm500$\,km\,s$^{-1}$. Alternatively it may be that the profile consists of a blend of 
forbidden (561.0\,eV), intercombination (568.6\,eV) and resonance (573.9\,eV) emission.
A blend of narrow lines can be ruled out at high confidence, as the fit statistic is substantially 
worse ($\chi^{2}/{\rm dof}=3135.5/2565$) and the majority of the O\,\textsc{vii} flux is not accounted for. 
However the profile can be fitted by a blend of velocity broadened lines. In order to test this, 
the forbidden, intercombination 
and resonance lines were fitted with line energies fixed at their expected values, with a common 
velocity width for all 3 line components allowed to vary. This provides an excellent fit to the line 
profile, with $\chi^{2}/{\rm dof}=2974.4/2563$, while the FWHM width of the 3 lines is now 
$7300^{+1000}_{-1500}$\,km\,s$^{-1}$. The parameters of the three line components are listed in 
Table~\ref{emission-lines}, while the line model is the one overlaid on the O\,\textsc{vii} profile in 
Figure~\ref{ovii}. From the line fluxes listed in Table~\ref{emission-lines}, it is apparent 
that the flux of the forbidden line component dominates over the intercombination emission, 
while only an upper-limit is placed on the resonance line emission. The dominance of the forbidden 
line emission over the other components is perhaps expected, as the centroid of the broad 
O\,\textsc{vii} profile is closest to the expected forbidden line energy. In section 6.2.1 we attempt 
to place constraints on the density and location of the emitter given these O\,\textsc{vii} parameters. 

\section{Discussion}\label{sec:discussion}

\subsection{Main observational results}

The exposure time of both the HETG and RGS observations allow us to
perform an unprecedented high signal-to-noise and high resolution
spectroscopy study of the properties of both the primary continuum
and the ionized absorption and emission features in the quasar MR\,2251-178. 
The main observational results are the following:

\noindent In the soft X-ray range, numerous absorption features are
  clearly detected: \\ 
$-$ A deep absorption trough  between
  $0.7-0.8$\,keV most likely identified with an unresolved
  transition array (UTA), due  to $2p \to 3d$ transitions from low
  ionization M-shell iron, i.e. Fe\,\textsc{vii-x}; \\ 
$-$ A multitude of inner K-shell lines of O, Ne, Mg and Si, 
due to charge states corresponding to Li, Be, B, C, N, O-like etc ions.\\
$-$ Several higher ionization L-shell ($2p \to 3d$) lines of iron
(i.e.\ Fe\,\textsc{xvii-xxiv}). \\
$-$ Resonance ($1s \to 2p$) lines from He and H-like ions of C, N, O, Ne, Mg and
Si, and in some cases, higher order $1s \to np$ lines up to $n$=6.\\
\indent In most cases the (strongest) absorption line profiles 
are narrow or not resolved, with velocity widths typically
$\sigma\ls300$\,km$\,s^{-1}$. 
Similarly the outflow velocities inferred from the measured rest frame
energies of the absorption lines are small or consistent with zero, of the order 
$v_{\rm out}\ls 400$\,km\,s$^{-1}$.  

The spectral fit using photo-ionised {\sc xstar} model grids shows that
3 fully-covering WA components are required in order to model the wide range in the ionization state of
  the gas, with $N_{\rm H}=1.5-3.6\times10^{21}$\,cm$^{-2}$ 
and $\logxi = 1.27-2.80$. 
The small outflow velocities found for the low and medium
ionization components 1 and 2 are consistent with each other, with $v_{\rm out}=-480\pm40$\,km\,s$^{-1}$
and $v_{\rm out}=-460\pm60$\,km\,s$^{-1}$ respectively, while the
highest ionization component 3 does not require an outflow velocity with
$v_{\rm out}<130$\,km\,s$^{-1}$. Notably, the necessary power-law continuum required by 
the \textsc{xstar} grid in order to model the low ionization lines
(component 1) is $\Gamma_{\rm input}=2.5$, which is softer than that required for the
higher ionized lines (i.e., $\Gamma_{\rm input}=2.0$). 
Moreover, one additional component of partially (covering factor $\sim$ 61\%) 
ionized absorbing gas with $N_{\rm H}\sim6\times10^{22}$\,cm$^{-2}$ and $\logxi\sim1$ 
is required to achieve a good fit. 
Interestingly after the required absorbing layers of gas are
accounted for, the soft X-ray photon index found
($\Gamma=2.32\pm0.08$) is in good agreement with what is required 
to reproduce the soft X-ray inner shell absorption lines (i.e. $\Gamma\sim2.5$). 
Therefore, MR\,2251-178 may have an intrinsically soft continuum, at least below 2\,keV, which
is partially-covered by a complex and stratified absorber. 

For the 2011 HETG spectrum the parameters of the three fully-covering 
WA are rather similar to those obtained from the RGS spectra with $N_{\rm
  H}=1.5-2.1\times10^{21}$\,cm$^{-2}$, and $\logxi=1.15-2.9$,
 but a second partial covering component (covering factor $\sim$ 40\%) of higher
 column density $\sim7\times10^{23}$\,cm$^{-2}$ seems to be required, from 
the spectral curvature above 2\,keV. 
As for the RGS spectrum, a softer input continuum is strongly required
to model the low ionization warm absorber component 1.  However,
there is evidence for a small, but significant change in its ionization parameter that
appears to be correlated with the soft X-ray flux. Applying this model
to the 2002 HETG spectrum we confirm that the change of the ionization parameter
is in direct proportion to the soft X-ray flux suggesting that this
component is in photoionization equilibrium with the continuum. 
The other possible change in the spectra is in the partial covering absorption. Considering all 
three grating observations, the uncovered fraction of the power-law appears to increase 
as the flux increases from the 2002 through to the 2011 observations, from $0.18\pm0.02$ 
to $0.39\pm0.03$, suggesting that this AGN 
is more obscured at lower flux states. 

The soft X-ray spectra also display several emission lines
  from a photoionized emitter from He- and H-like ions of C, N, O, and Ne. 
Notably, a strong and broad emission line near 0.56\,keV is clearly
detected in the RGS\,1 spectrum at the expected energy of the
O\,\textsc{vii} triplet, and is well represented by a blend of the 
forbidden (dominant), intercombination and resonance emission lines
with a common velocity of $\sim$7300\,km\,s$^{-1}$ (FWHM). 
This broad O\,\textsc{vii} triplet profile is superimposed on by three 
narrow absorption lines corresponding to inner-shell absorption due to O\,\textsc{v} and the two lines 
which make up the O\,\textsc{vi} ($1s^{2}2s \to 1s2s2p$) doublet. Similar structures are 
present at other energies, with N\,\textsc{vii}, O\,\textsc{viii} and Ne\,\textsc{ix} all showing emission 
superimposed by absorption.

In the hard X-ray energy band of the HETG spectrum, 
there is a lack of any strong iron
K$\alpha$ emission, with EW=11$\pm$6\,eV. This could be accounted
for by the X-ray Baldwin effect, since MR2251-178 has a much
higher 2-10\,keV luminosity than most local Seyfert 1s. However, we
found the presence of a significant absorption feature at 7.3\,keV consistent
with what was previously reported from the 2002 HETG observation by
\cite{gibson2005}, but only marginally inconsistent at 90\% confidence with
the line energy measured in the 2009 Suzaku observation by
\cite{gofford2011}. This Fe K band absorption is well modelled by a highly
ionized {\sc xstar} grid with a high turbulence velocity of
5000\,km\,s$^{-1}$ and an outflow velocity of $\sim$ -15600\,km\,s$^{-1}$.
However an alternative origin from a low ionization partial covering absorber 
without requiring any velocity shift cannot be excluded. 
The much higher spectral resolution of both HETG and RGS data allows us
to resolve the 1.3\,keV absorption feature -- first observed in the lower resolution 2009
Suzaku XIS spectrum \citep{gofford2011} and tentatively identified with blue-shifted iron
L-shell transitions --  into a series of lower ionization lines of inner shell Mg from 
Mg\,\textsc{vi-ix}, with only a modest outflow velocity of
$\sim-400$\,km\,s$^{-1}$. 

\subsection{The Origins of the Warm Absorption and Emission in MR\,2251-178}

\subsubsection{Constraints from the O\,\textsc{vii} line triplet}
Given the constraints on the O\,\textsc{vii} line triplet, we can attempt to estimate the 
density and likely radial location of the emitting gas. The line ratios $G = (x + y + z) / w$ and 
$R = z / (x + y)$ give a measure of the temperature and density of the gas, where $z$ corresponds 
to the forbidden line, $(x + y)$ to the intercombination emission and $w$ to the resonance line 
\citep{Porquet2000}. From the line ratios in Table~\ref{emission-lines}, this yields 
$G>3.9$ and $R=2.9\pm1.4$. Thus from the calculations in \cite{Porquet2000}, the high G ratio 
corresponds to the gas being photoionized rather than collisionally ionized, with 
a temperature $T<10^6$\,K. However, photo-excitation of the resonance
lines can be important in X-ray photoionized sources as AGN (e.g.,
\citealt{Kinkhabwala2002,Porquet2010}), thus other complementary
temperature diagnostics should be used such as those based on the
width measurement of the recombination continuum (RRC) features 
\citep{Liedahl1996}. Unfortunately in the spectrum of MR\,2251-178, 
no RRC emission is detected, so it is not possible to determine the temperature 
by this method. 

The $R$ values suggests a density of $n_{\rm e}\sim10^{10}$\,cm$^{-3}$, 
while the fact that the forbidden line is required to be stronger than the intercombination 
emission (i.e the lower limit is $R>1.4$) implies that the maximum possible density is $<10^{11}$\,cm$^{-3}$. Thus a density of $n_{\rm e}=10^{10} - 10^{11}$\,cm$^{-3}$ would seem to imply an origin of the broad 
line emission consistent with the optical Broad Line Region \citep{davidson1979}. 
The ionization of the emitter can also be constrained, 
given that a line flux ratio of O\,\textsc{vii}/O\,\textsc{viii}\,$\sim6$, 
e.g. Table~\ref{emission-lines}. From running an \textsc{xstar} simulation with a density of 
$n_{\rm e}=10^{10}$\,cm$^{-3}$, the line ratio 
implies an ionization parameter of $\logxi=1.25$. 
Thus an estimate of the radial distance can be obtained via the definition of the ionization parameter, 
i.e. $r = (L_{\rm ion} / \xi n_{\rm H})^{1/2}$, where $L_{\rm ion}$ is the $1-1000$\,Rydberg luminosity  
and $n_{\rm H}$ is the hydrogen number density. From extrapolating 
the best-fit spectrum from above, the ionizing luminosity of MR\,2251-178 is 
$L_{\rm ion}=2\times10^{45}$\,erg\,s$^{-1}$. Thus for a density in the range 
$n_{\rm e}=10^{10} - 10^{11}$\,cm$^{-3}$, then the radius is $r=0.3-1.0\times10^{17}$\,cm 
(or 0.01-0.03\,pc), again 
consistent with typical BLR radii \cite[e.g.,][]{kaspi2005}.  
The radius of the emission can also be estimated from the O\,\textsc{vii} width of 
$\sigma=3200$\,km\,s$^{-1}$. Assuming 
a virial relation between the black hole mass and the radius $r$, of $3\sigma^{2}=GM/R$ 
\citep{peterson2004} and adopting a black hole mass of $2.4\times10^{8}$\Msun\ 
for MR\,2251-178 \citep{dunn2008}, gives a radius of $r\sim 10^{17}$\,cm, consistent with the 
above estimate. 

Given the estimate of the ionization parameter of the soft X-ray emitter of $\log \xi=1.25$, it 
can plausibly be associated with one of two absorption components, either the low ionization warm absorber 
component 1 or the partial covering component (pc1), as summarised in
Table~\ref{absorbers}. We have calculated  
the total (global) covering fraction as a fraction of $4\pi$\,sr ($f_{tot}$) 
for either absorbing layer in order to produce the total luminosity of 
the broad O\,\textsc{vii} emission, of $3.1\times10^{42}$\,erg\,s$^{-1}$. The \textsc{xstar} code is used 
to calculate the line luminosity from a spherical shell of gas, covering a full $4\pi$\,sr around the 
AGN, illuminating by the above ionizing luminosity. The component 1 absorber has a column density of 
$N_{\rm H}=2\times10^{21}$\,cm$^{-2}$ and produces an O\,\textsc{vii} luminosity over $4\pi$\,sr of 
$3.7\times10^{42}$\,erg\,s$^{-1}$, while the partial covering component has 
$N_{\rm H}=5\times10^{22}$\,cm$^{-2}$ and produces an O\,\textsc{vii} luminosity of 
$4.5\times10^{43}$\,erg\,s$^{-1}$. Thus in order to reproduce the O\,\textsc{vii} luminosity, the
component 1 absorber would require a high covering fraction of $f_{\rm tot}=0.84$ of $4\pi$\,sr$^{-1}$, 
while the partial coverer only requires a fraction of $f_{\rm tot}=0.07$.

However some of the narrow absorption lines that are produced from the 
component 1 warm absorber itself are 
superimposed upon the O\,\textsc{vii} broad emission profile. This would appear to require the component 1 
absorber to be physically placed outside the line emitting region, making it less likely to 
be the origin of the broad soft X-ray lines. Furthermore the kinematics of component 1, with a low outflow 
velocity ($\sim-400$\,km\,s$^{-1}$) and small or unresolved line widths/turbulences, would also suggest 
it is placed at larger distances perhaps coincident with the NLR. Therefore one possibility is that 
the broad line region clouds themselves not only produce the broad soft X-ray lines, but are also 
responsible for the partial covering of the X-ray continuum itself. 
 Such broad X-ray emission ionized lines have been detected in several
  other AGN thanks to high-resolution X-ray data suggesting that such a BLR origin
  for the X-ray emission may be common in AGN  (e.g., Mrk 279: \citealt{costantini2007},
  Mrk 841: \citealt{longinotti2010}, NGC 4051: \citealt{ogle2004},
  Mrk\,509: \citealt{detmers2011}, 3C\,445: \citealt{Reeves2010}).

Note that the estimate of the total covering fraction, of $\sim7$\% for the partial coverer/emitter, 
may be substantially higher if some of the broad line 
emission is itself obscured, depending on the exact spatial distribution of emitting and absorbing 
clouds. We note that in the RGS data about 60\% of the intrinsic X-ray continuum is obscured by the 
$\sim5\times10^{22}$\,cm$^{-2}$ partial coverer (40\% remains unobscured). If this obscuration is 
also applied to the 
broad O\,\textsc{vii} emission, that may imply a total covering fraction of the emitting clouds 
closer to $f_{\rm tot}\sim0.2$. Furthermore the X-ray BLR emission can be further obscured by the 
warm absorber which fully covers the line of sight to the AGN, which obscures the continuum 
level by a factor of about 
30-40\% at the energy of the O\,\textsc{vii} emission line. Thus the total covering is likely 
to be consistent with typical estimate of the overall covering fraction of optical BLR clouds, 
of the order $5-30$\% \cite[e.g.,][]{netzer&laor1993}. 
If the BLR clouds do partially cover the X-ray source, then this can give an approximate estimate of a 
size of a cloud. Thus for X-ray absorption of the order $\sim10^{23}$\,cm$^{-2}$ and for a density 
of $n_{\rm e}\sim10^{10}$\,cm$^{-3}$, then that implies a size of $\Delta r \sim 10^{13}$\,cm, likely 
smaller than the size of the X-ray emission region (e.g. $10R_{\rm g}$ here would correspond to 
a few $\times10^{14}$\,cm). Thus it seems plausible for such clouds to only partially cover the 
line of sight to the continuum X-ray emission.

The low ionization component 1 warm absorber could instead plausibly
reproduce some of the weak narrow emission  
lines in the spectrum, e.g. the narrow forbidden components, which have line widths of 
$\sigma\ls 500$\,km\,s$^{-1}$ typically. This would correspond to radial distances of a few pc or greater. The distance to the component 1 absorber is estimated below, via its response to the 
soft X-ray continuum.

\subsubsection{Constraints from the variability of the warm absorber
  components}\label{sec:recomb}
The component 1 warm absorber appears to respond to the overall increase in
the continuum between the 2002 and 2011  
observations, but also in the $\sim40$\,days timescale between the 2011 \xmm\ and \chandra\ 
observations (Section 4.2) and thus would appear to be in photoionization equilibrium. We can 
therefore attempt to place a lower limit on the density of this absorber via the recombination 
timescale. For this we use the recombination time-scale formula from
  \cite{Bottorff00} that does account simultaneously for  
the cascade into the population of $X_{i}$ ions from
the population of $X_{i+1}$ ions, and the cascade out of the population
of $X_{i}$ ions into the population of $X_{i-1}$ ions: \\
\begin{equation} 
t(X_{i})=\frac{1}{\alpha(X_{i})n_{e}[\frac{f(X_{i+1})}{f(X_{i})}-
\frac{\alpha(X_{i-1})}{\alpha(X_{i})}]}.
\end{equation}

\noindent where $f(X_{i})$ the ionic fraction of the $X_{i}$ ion,
$\alpha(X_i,T_e)$ is the recombination coefficient of
the $X_{i}$ ion at the electronic temperature $T_e$, and $n_{e}$ is
the electron density ($\sim$ 1.2 $n_{H}$ for cosmic
abundance). We apply this formula to O\,\textsc{vii}. At log~$\xi$=1.27,
the ratio O\,\textsc{vii}/O\,\textsc{viii} 
is 6.0 and $T_e$ is 4$\times$10$^{4}$\,K.
Using the  recombination coefficient from \cite{Nahar03} and a recombination time of 
$t\lesssim40$\,days between observations, we find a
lower limit for the hydrogen density of 3.8$\times$10$^{4}$\,cm$^{-3}$. Hence, this 
implies an upper limit for the radial distance ($R_{\rm var}$) of 5.3$\times$10$^{19}$\,cm
(i.e. $\lesssim$ 17\,pc) or a few pc. Moreover as discussed below in
\S\ref{sec:energetics} the minimum radius for component 1 is about $2.8\times10^{19}$\,cm
(i.e. $\gtrsim$ 9\,pc). Therefore the location of component 1 is well
constrained between 9\,pc and 17\,pc. 
For comparison, the expected distances of the
torus and of the NLR that are about 7\,pc and
about 140\,pc, using the following formula of \cite{Krolik01}
and \cite{Mor09} respectively:

\begin{equation} 
R_{\rm torus}\sim L_{\rm ion,44}^{1/2} ~~~~~~~~ (pc)
\end{equation}

\begin{equation} 
R_{\rm NLR} = 295 \times L_{46}^{0.47\pm 0.13} ~~~~ (pc).
\end{equation}

\noindent For MR\,2251-178 the ionizing (1-1000\,Rydberg) luminosity was taken as 
$L_{\rm ion, 44}=20$ (in units of $10^{44}$\,erg\,s$^{-1}$) and 
$L_{46}=0.43$\footnote{This is is likely to be a somewhat conservative estimate of the bolometric 
luminosity. Applying a bolometric correction of a factor of 30 for the 2-10\,keV X-ray luminosity 
\citep{VF2009}, would result in $L_{\rm bol}=10^{46}$\,erg\,s$^{-1}$.}  
is assumed as the bolometric luminosity \citep{dunn2008}, 
in units of $10^{46}$\,erg\,s$^{-1}$. 

Therefore component 1 appears to be located consistent with the pc scale torus and/or inner NLR
radius-scales. We note that $R_{\rm var}$ is much greater than the BLR distance that is of only
about 75 light-days (using the recent R$_{\rm BLR}$--$\lambda L_{\lambda}$(5100\AA) relationship from \citealt{Bentz13}
and the average 5100\,\AA\, flux from \citealt{Lira11}), i.e. 0.06\,pc.
The lack of response from $2002-2011$ 
of the higher ionization (components 2,3) absorbers may place this gas at greater distances.
However a more intense monitoring campaign 
(over weeks to months) would be needed to place a firmer constraint on the density and therefore 
radial location of the absorbers.

\subsubsection{Warm absorber properties: radii, outflows rates and energetics}\label{sec:energetics}

We estimate the lower and upper limits of the distance, mass outflow
rate and kinetic power of the WAs following the assumptions and
definitions outlined in \cite{Tombesi2013} for the fully covering warm
absorbers (components 1, 2, and 3) and for the highly ionized absorber
discussed in $\S$\ref{sec:zonehigh}.  
An upper limit for the radial location of an absorber can be derived
from the definition of the ionization parameter and the requirement
that the thickness of the absorber does not exceed its distance to the
supermassive black hole, i.e., $N_\mathrm{H} \simeq n_\mathrm{H} \Delta R <
n_\mathrm{H} R$, then:  
\begin{equation}
r_{\mathrm{max}} \equiv L_{\mathrm{ion}}/\xi N_\mathrm{H}. 
\end{equation}\label{eq:rmax}
\noindent Note the material can not be farther away than this given the observed
ionization and column. 
An estimate of the minimum distance can be derived from the
radius at which the observed velocity corresponds to the escape
velocity: 
\begin{equation}
 r_{\mathrm{min}} \equiv 2 G M_{\mathrm{BH}}/ v_{\mathrm{out}}^{2}. 
\end{equation}
\noindent Here the black hole mass estimate for MR\,2251-178 is taken as 
$2.4\times10^{8}M_{\odot}$ \citep{dunn2008}. 

For the calculation of the mass outflow rate we use the expression
derived by \cite{krongold2007} that is appropriate for a
biconical wind-like geometry and that does not rely on the estimate of
the covering and filling factors (see \citealt{Tombesi2013} for details):   
\begin{equation}
\dot{M}_{\mathrm{out}} \equiv f(\delta, \phi) \pi \frac{n_\mathrm{H}}{n_\mathrm{e}} m_\mathrm{p} N_\mathrm{H} v_\mathrm{out} r, 
\end{equation}
where $f(\delta, \phi)$ is a function that depends on the angle
between the line of sight to the central source and the accretion disc
plane, $\delta$, and the angle formed by the wind with the accretion
disc, $\phi$ (see Fig.~12 of \citealt{krongold2007}). As in
\cite{krongold2007} and \cite{Tombesi2013}, we assume $f(\delta,\phi)$$\simeq$1.5 that corresponds to a
roughly vertical disc wind ($\phi$$\simeq$$\pi/2$) and an average line
of sight angle of $\delta$$\simeq$$30^{\circ}$ for a type-I
AGN, while $n_\mathrm{H}$/$n_\mathrm{e}$ is about 1/1.2 for Solar
elemental abundances, so:  
\begin{equation}
\dot{M}_{\mathrm{out}} \simeq 6.6\times10^{-24}~ N_\mathrm{H} ~v_\mathrm{out}~ r ~~~~~[g/s]. 
\end{equation}

\noindent To determine the $\dot{M}_{\mathrm{out}}$ interval range, we use the
values of $r_{\mathrm{max}}$ and $r_{\mathrm{min}}$ inferred from equations
(5) and (6), except for component 1 for which with use as $r_{\mathrm{max}}$
the value found above due to the recombination time-scale of
O\,\textsc{vii}, i.e. $r_{\mathrm{var}}$ (see values reported in Table\,7). 

Neglecting additional acceleration of the outflow, i.e. assuming that
it has reached a constant terminal velocity, the kinetic (or
mechanical) power can consequently be derived as:  
\begin{equation}
\dot{E}_\mathrm{K} \equiv \frac{1}{2} \dot{M}_{\mathrm{out}} v_\mathrm{out}^2. 
\end{equation}

\noindent We also calculated the outflow momentum rate as $\dot{P}_\mathrm{out} \equiv
\dot{M}_\mathrm{out} v_\mathrm{out}$ and subsequently compared it to
the momentum flux of the radiation field, $\dot{P}_\mathrm{rad} \equiv
L_\mathrm{bol}/c$.
All values are reported in Table~\ref{energetics}. 

The inner and outer radii of component 1 are the best determined, between $9-17$\,pc, 
with the upper bound being set by the 
40 day timescale response of the absorber to the continuum. 
The higher ionization component 3 is constrained between 
between $\sim120-290$\,pc; this is consistent with being placed outside component 1, noting that 
no response of this absorber was detected to continuum variations, consistent with a lower 
density. Component 2 is the least well determined and is consistent with the radial 
estimates for components 1 and 3 (Table 7). 
Thus the locations of components 1 and 3 are consistent with the torus and NLR respectively, 
as estimated above. The possible highly ionized (iron K band) absorber (Table 7, component high), 
with an outflow velocity of 
$\sim 15000$\,km\,s$^{-1}$, would appear to be located much closer to the black hole ($\lesssim 
0.01$\,pc) 
with a location perhaps consistent with an accretion disk wind \citep{Tombesi2013}. 

The kinetic power of the three warm absorbers (components\,1--3) appear to be $\lesssim 0.01$\% of the
bolometric luminosity; while for the highly ionized absorber we found
a minimum value of 1\% of the bolometric luminosity, hence its 
mechanical power can potentially affect the host galaxy via feedback \citep{hopkins&elvis2010}. 
Nonetheless the mass outflow rates of all components\,1--3 as well as the 
highly ionized absorber are rather similar, the lower limits on $\dot{M}_{\rm out}$ 
vary between $0.2-1.9 M_{\odot}$\,yr$^{-1}$ for components\,1--3, while for the highly ionized 
absorber, $\dot{M}_{\rm out}\gtrsim 0.6 M_{\odot}$\,yr$^{-1}$. In comparison for a bolometric 
luminosity of $4.3\times10^{45}$\,erg\,s$^{-1}$ and assuming an accretion efficiency of 
$\eta=0.06$, the expected mass accretion rate of MR\,2251-178 is 
$\dot{M}_{\rm acc}\sim 1.3 M_{\odot}$\,yr$^{-1}$; thus the combined mass outflow rate from 
MR\,2251-178 is likely to be at least equal to (or somewhat exceeding) the accretion rate onto the 
black hole. Finally the outward momentum rate of the putative highly ionized absorber is estimated 
to be at least $\sim50\%$ of $L_{\rm bol}/c$, which suggests efficient ($\tau\sim1$) scattering 
between photons and electrons in a Thomson scattering driven outflow, as may be expected in a highly ionized 
accretion disk wind \citep{king2003}. 

\subsection{Comparisons with UV observations}

Ultra-violet absorption has also been found previously in the spectrum of
MR\,2251-178. Using HST/FOS data obtained in 1996, 
\cite{monier2001} found absorption lines due to Ly$\alpha$, N\,\textsc{v} and
C\,\textsc{iv} with a systematic blueshift  
of $\sim300$\vunit\ with a total hydrogen column density of about
5$\times$10$^{21}$\,cm$^{-2}$.
From the comparison between HST data taken with FOS in 1996 and STIS in 2000, 
the C\,\textsc{iv} absorption in particular showed variability -- both
in terms of the velocity and column density -- over a period of 
roughly 4 years. This relatively short timescale variability 
showed that this UV absorption is truly intrinsic
and constrained the absorption clouds to within $r\lesssim$2.4\,kpc of the 
continuum source \citep{ganguly2001} consistent with the estimate of
\cite{monier2001}.  
\cite{kaspi2004} reported for the first time the entire FUSE spectrum
of MR\,2251-178 and detected at least four blueshifted absorption
systems of C\,\textsc{iii}, H\,\textsc{i}, and O\,\textsc{vi}; one at
$-$580\vunit, and at least three others blended components with centroid
velocities at about $-$150, $-$300, and $-$430\vunit. 

We note that the velocity profiles obtained here from the 
X-ray data, e.g. from C\,\textsc{vi} and O\,\textsc{viii}, appear to be 
consistent with these UV profiles, with the X-ray absorption line profiles having 
typical velocity shifts of the order $\sim-300--400$\,km\,s$^{-1}$, as shown in 
Figure\,\ref{Hprofiles}. The only exception may be from the highest ionization 
lines, such as Mg\,\textsc{xii} and Si\,\textsc{xiv}, which do not require any 
net blue-shift, but this very highly ionized gas may be more apparent in the X-ray spectrum than in 
the UV.
The total depth of the FUV and UV
absorption lines appeared larger than the underlying continuum, which indicates
that the broad UV emission lines are absorbed by the UV absorber and
therefore the UV absorber lies outside the BLR. This is similar to what 
is found in the X-ray spectrum presented here, 
whereby the broad soft X-ray lines (BLR) are absorbed 
by the narrow lines from the X-ray warm absorber.

The above UV absorber properties are similar to that found here for
the fully-covered soft X-ray warm absorber components (namely components 1, 2, and
3), indeed both the column densities and the outflow velocities appear consistent. 
Moreover, the
relatively tight constraint for the location of component 1 shows that it
lies outside the BLR region too, with the components 2 and 3 consistent with 
being further out due to their lack of variability. 
In conclusion the UV and soft X-ray warm absorption components 
cover a similar range of column densities and appear to be kinematically consistent 
with each over in terms of their outflow velocities, although the X-ray absorption 
likely originates from more highly ionized gas.
A more detailed comparison with the 2011 
Chandra and XMM-Newton observations and 
a contemporaneous 2011 HST/COS observation will be deferred until future work. 
We note however that from a preliminary analysis of the HST/COS spectrum as 
well as optical spectroscopy (M. Crenshaw, private communication), the FWHM widths 
of the C\,\textsc{iv} and H$\beta$ emission lines appear in the range $3200-3600$\,km\,s$^{-1}$. 
This is similar to, if somewhat smaller than the widths of the X-ray emission lines 
such as O\,\textsc{vii}, with FWHM $\sim7300$\,km\,s$^{-1}$. 
This could suggest that the broad X-ray lines originate from the innermost part of the 
BLR, which would likely be more highly ionized. 

\subsection{Comparison with the recent observations of Markarian 509}

The Seyfert 1 galaxy Mrk\,509 (z=0.03450) -- which  has a similar black hole mass (1-3 $\times$
10$^{8}$\, M$_{\odot}$) and a bolometric luminosity of only a factor
of about 2-3
smaller than that of MR\,2251-178 \citep{Raimundo2012} -- 
has also been recently monitored (in 2009) from UV to hard
X-rays (HST/COS, XMM-Newton, Chandra, Swift and Integral) to constrain
the location of the outflow components \citep{Kaastra2011a}. 
Thus a comparison between MR\,2251-178 and Mrk\,509 may be informative, 
given their similar properties at the higher luminosity end of the Seyfert 
population, while both AGN have long XMM-Newton or Chandra exposures.
The deep (600\,ks) XMM-Newton/RGS spectrum of Mrk\,509 revealed the
presence of a multitude of blueshifted absorption lines from  
three slow velocity absorber components ($\sim$$-$13\,km\,s$^{-1}$,
$\sim$$-$320\,km\,s$^{-1}$, and $-$770\,km\,s$^{-1}$), with two strong and discrete ionization
parameter peaks in the $\logxi=1-3$ range at about $\logxi=2.0$ and $\logxi=2.8$
\citep{detmers2011}. 
The ionization parameters of the UV components with similar outflow
velocities are much lower than those
found in X-rays, which could indicate that the UV and X-ray absorbers 
 are co-spatial but have different densities, as also inferred from
 the LETG spectrum \citep{Ebrero2011}. 
The presence of a possible fast outflow with $v_{\rm out}\simeq$ -14000\,km\,s$^{-1}$
was claimed using the summed spectrum of previous XMM-Newton observations
\citep{Ponti2009}, and was only marginally detected in the LETG 2009
observation and the XMM-Newton/pn spectrum \citep{Ponti2013}.

The location of the outflowing components in Mrk\,509 are claimed to be consistent with a torus
wind or NLR origin \citep{Kaastra2012}. 
While the kinetic luminosity of the outflow is
small in Mrk 509, the mass carried away is larger than the likely $0.5 M_{\odot}$\,yr$^{-1}$
accreting onto the black hole. These properties appear to be similar 
to those presented here for MR\,2251-178. Observationally 
the X-ray column densities, outflow velocities 
and ionization parameters cover a very similar range, while 
in terms of radial location, the warm absorbers of both AGN appear 
commensurate with a pc scale wind consistent with the outermost torus or inner NLR.

\section{Conclusions}

This paper has presented deep (400\,ks) Chandra HETG and XMM-Newton RGS 
observations of the nearby quasar, MR\,2251-178. The high 
resolution spectra have revealed the presence of a three ionization component 
warm absorber, with the ionization parameter covering the range from 
$\logxi=1-3$. The lowest ionization component is responsible 
for the absorption seem from the Fe M-shell UTA as well as the inner-shell 
lines of O, Ne, Mg and Si, while the higher ionization components produces the 
He and H-like lines as well as L-shell Fe ions. The lowest ionization gas 
tentatively appears to be in photoionization equilibrium with the continuum 
flux. From this and from the lower and upper-limits to the radial location of the gas, 
the low ionization absorber appears consistent with a pc scale location, coincident 
with either the torus or innermost NLR, while the highest ionization component may arise 
from more distant gas. The outflow velocities of the warm absorbing gas all appear 
within $\ls500$\,km\,s$^{-1}$, also consistent with the 
outflow velocities of the known UV absorber in this AGN \citep{ganguly2001,monier2001,kaspi2004}. 

Several broad emission lines also appeared to be present in the soft X-ray spectrum, 
most notably from O\,\textsc{vii}. The width derived for the broad O\,\textsc{vii} 
line complex, of FWHM $7300^{+1000}_{-1500}$\,km\,s$^{-1}$, is consistent with an origin on sub-pc 
scales from the optical BLR. In addition, we have suggested that the BLR clouds themselves, 
which are presumably responsible for the broad soft X-ray emission lines, may indeed 
partially cover the X-ray continuum, with a typical column density of $N_{\rm H}=10^{23}$\,cm$^{-2}$. 
The presence of such a partial coverer has also been recently invoked to account 
for the hard X-ray excesses observed towards several type I AGN \citep{tatum2013} and 
may be required here to explain the unusually hard X-ray continuum (with $\Gamma=1.5$) that is 
observed in MR\,2251-178. 
Overall the X-ray observations of MR\,2251-178 have revealed a 
complex and stratified absorption and emission region, which modify the overall X-ray spectrum. 
These appear to exist on several spatial scales; from a putative accretion disc wind responsible for 
the highly ionized Fe K band absorption; to the BLR clouds responsible 
for the broad soft X-ray emission lines and potentially the partial covering absorption; and 
to the more extended outflowing gas on parsec and NLR scales. The latter is the likely origin of the 
historical soft X-ray warm absorber observed towards this AGN.

\section{Acknowledgements}

J.N.\ Reeves acknowledges Chandra grant number GO1-12143X. 
D. Porquet acknowledges financial support from the French GDR PCHE. 
T.J.\ Turner acknowledges NASA grant number AR2-13006X. We
would also like to thank Margherita Giustini for helpful discussions. 
Based on observations obtained with the XMM-Newton, and ESA science
mission with instruments and contributions directly funded by ESA
member states and the USA (NASA).      
The scientific results reported in this article are based on observations made by the Chandra X-ray Observatory.
This research has made use of software provided by the Chandra X-ray Center (CXC) in the application packages CIAO, ChIPS.

\bibliographystyle{apj} 
\bibliography{newreferences}

\begin{thebibliography}{135}
\expandafter\ifx\csname natexlab\endcsname\relax\def\natexlab#1{#1}\fi

\bibitem[{{Andrade-Vel{\'a}zquez} {et~al.}(2010){Andrade-Vel{\'a}zquez},
  {Krongold}, {Elvis}, {Nicastro}, {Brickhouse}, {Binette}, {Mathur}, \&
  {Jim{\'e}nez-Bail{\'o}n}}]{andrade2010}
{Andrade-Vel{\'a}zquez}, M., {Krongold}, Y., {Elvis}, M., {Nicastro}, F.,
  {Brickhouse}, N., {Binette}, L., {Mathur}, S., \& {Jim{\'e}nez-Bail{\'o}n},
  E. 2010, \apj, 711, 888

\bibitem[{{Antonucci}(1993)}]{antonucci1993}
{Antonucci}, R. 1993, \araa, 31, 473

\bibitem[{{Ashton} {et~al.}(2004){Ashton}, {Page}, {Blustin}, {Puchnarewicz},
  {Branduardi-Raymont}, {Mason}, {C{\'o}rdova}, \& {Priedhorsky}}]{ashton2004}
{Ashton}, C.~E., {Page}, M.~J., {Blustin}, A.~J., {Puchnarewicz}, E.~M.,
  {Branduardi-Raymont}, G., {Mason}, K.~O., {C{\'o}rdova}, F.~A., \&
  {Priedhorsky}, W.~C. 2004, \mnras, 355, 73

\bibitem[{{Bachev} {et~al.}(2009){Bachev}, {Grupe}, {Boeva}, {Ovcharov},
  {Valcheva}, {Semkov}, {Georgiev}, \& {Gallo}}]{bachev2009}
{Bachev}, R., {Grupe}, D., {Boeva}, S., {Ovcharov}, E., {Valcheva}, A.,
  {Semkov}, E., {Georgiev}, T., \& {Gallo}, L.~C. 2009, \mnras, 399, 750

\bibitem[{{Badnell}(2006)}]{badnell2006}
{Badnell}, N.~R. 2006, \apjl, 651, L73

\bibitem[{{Baumgartner} {et~al.}(2010){Baumgartner}, {Tueller}, {Markwardt}, \&
  {Skinner}}]{baumgartner2010}
{Baumgartner}, W.~H., {Tueller}, J., {Markwardt}, C., \& {Skinner}, G. 2010, in
  Bulletin of the American Astronomical Society, Vol.~42, AAS/High Energy
  Astrophysics Division \#11, 675

\bibitem[{{Behar} \& {Netzer}(2002)}]{behar2002}
{Behar}, E. \& {Netzer}, H. 2002, \apj, 570, 165

\bibitem[{{Behar} {et~al.}(2003){Behar}, {Rasmussen}, {Blustin}, {Sako},
  {Kahn}, {Kaastra}, {Branduardi-Raymont}, \& {Steenbrugge}}]{Behar2003}
{Behar}, E., {Rasmussen}, A.~P., {Blustin}, A.~J., {Sako}, M., {Kahn}, S.~M.,
  {Kaastra}, J.~S., {Branduardi-Raymont}, G., \& {Steenbrugge}, K.~C. 2003,
  \apj, 598, 232

\bibitem[{{Behar} {et~al.}(2001){Behar}, {Sako}, \& {Kahn}}]{behar2001}
{Behar}, E., {Sako}, M., \& {Kahn}, S.~M. 2001, \apj, 563, 497

\bibitem[{{Bentz} {et~al.}(2013){Bentz}, {Denney}, {Grier}, {Barth},
  {Peterson}, {Vestergaard}, {Bennert}, {Canalizo}, {De Rosa}, {Filippenko},
  {Gates}, {Greene}, {Li}, {Malkan}, {Pogge}, {Stern}, {Treu}, \&
  {Woo}}]{Bentz13}
{Bentz}, M.~C., {Denney}, K.~D., {Grier}, C.~J., {Barth}, A.~J., {Peterson},
  B.~M., {Vestergaard}, M., {Bennert}, V.~N., {Canalizo}, G., {De Rosa}, G.,
  {Filippenko}, A.~V., {Gates}, E.~L., {Greene}, J.~E., {Li}, W., {Malkan},
  M.~A., {Pogge}, R.~W., {Stern}, D., {Treu}, T., \& {Woo}, J.-H. 2013, \apj,
  767, 149

\bibitem[{{Bergeron} {et~al.}(1983){Bergeron}, {Dennefeld}, {Boksenberg}, \&
  {Tarenghi}}]{bergeron1983}
{Bergeron}, J., {Dennefeld}, M., {Boksenberg}, A., \& {Tarenghi}, M. 1983,
  \mnras, 202, 125

\bibitem[{{Bianchi} {et~al.}(2007){Bianchi}, {Guainazzi}, {Matt}, \& {Fonseca
  Bonilla}}]{bianchi2007}
{Bianchi}, S., {Guainazzi}, M., {Matt}, G., \& {Fonseca Bonilla}, N. 2007,
  \aap, 467, L19

\bibitem[{{Blustin} {et~al.}(2002){Blustin}, {Branduardi-Raymont}, {Behar},
  {Kaastra}, {Kahn}, {Page}, {Sako}, \& {Steenbrugge}}]{blustin2002}
{Blustin}, A.~J., {Branduardi-Raymont}, G., {Behar}, E., {Kaastra}, J.~S.,
  {Kahn}, S.~M., {Page}, M.~J., {Sako}, M., \& {Steenbrugge}, K.~C. 2002, \aap,
  392, 453

\bibitem[{{Blustin} {et~al.}(2007){Blustin}, {Kriss}, {Holczer}, {Behar},
  {Kaastra}, {Page}, {Kaspi}, {Branduardi-Raymont}, \&
  {Steenbrugge}}]{blustin2007}
{Blustin}, A.~J., {Kriss}, G.~A., {Holczer}, T., {Behar}, E., {Kaastra}, J.~S.,
  {Page}, M.~J., {Kaspi}, S., {Branduardi-Raymont}, G., \& {Steenbrugge}, K.~C.
  2007, \aap, 466, 107

\bibitem[{{Blustin} {et~al.}(2005){Blustin}, {Page}, {Fuerst},
  {Branduardi-Raymont}, \& {Ashton}}]{blustin2005}
{Blustin}, A.~J., {Page}, M.~J., {Fuerst}, S.~V., {Branduardi-Raymont}, G., \&
  {Ashton}, C.~E. 2005, \aap, 431, 111

\bibitem[{{Bottorff} {et~al.}(2000){Bottorff}, {Korista}, \&
  {Shlosman}}]{Bottorff00}
{Bottorff}, M.~C., {Korista}, K.~T., \& {Shlosman}, I. 2000, \apj, 537, 134

\bibitem[{{Braito} {et~al.}(2007){Braito}, {Reeves}, {Dewangan}, {George},
  {Griffiths}, {Markowitz}, {Nandra}, {Porquet}, {Ptak}, {Turner}, {Yaqoob}, \&
  {Weaver}}]{braito2007}
{Braito}, V., {Reeves}, J.~N., {Dewangan}, G.~C., {George}, I., {Griffiths},
  R.~E., {Markowitz}, A., {Nandra}, K., {Porquet}, D., {Ptak}, A., {Turner},
  T.~J., {Yaqoob}, T., \& {Weaver}, K. 2007, \apj, 670, 978

\bibitem[{{Canizares} {et~al.}(2005){Canizares}, {Davis}, {Dewey}, {Flanagan},
  {Galton}, {Huenemoerder}, {Ishibashi}, {Markert}, {Marshall}, {McGuirk},
  {Schattenburg}, {Schulz}, {Smith}, \& {Wise}}]{canizares2005}
{Canizares}, C.~R., {Davis}, J.~E., {Dewey}, D., {Flanagan}, K.~A., {Galton},
  E.~B., {Huenemoerder}, D.~P., {Ishibashi}, K., {Markert}, T.~H., {Marshall},
  H.~L., {McGuirk}, M., {Schattenburg}, M.~L., {Schulz}, N.~S., {Smith}, H.~I.,
  \& {Wise}, M. 2005, \pasp, 117, 1144

\bibitem[{{Canizares} {et~al.}(1978){Canizares}, {McClintock}, \&
  {Ricker}}]{canizares1978}
{Canizares}, C.~R., {McClintock}, J.~E., \& {Ricker}, G.~R. 1978, \apjl, 226,
  L1

\bibitem[{{Chartas} {et~al.}(2002){Chartas}, {Brandt}, {Gallagher}, \&
  {Garmire}}]{chartas2002}
{Chartas}, G., {Brandt}, W.~N., {Gallagher}, S.~C., \& {Garmire}, G.~P. 2002,
  \apj, 579, 169

\bibitem[{{Cooke} {et~al.}(1978){Cooke}, {Ricketts}, {Maccacaro}, {Pye},
  {Elvis}, {Watson}, {Griffiths}, {Pounds}, {McHardy}, {Maccagni}, {Seward},
  {Page}, \& {Turner}}]{cooke1978}
{Cooke}, B.~A., {Ricketts}, M.~J., {Maccacaro}, T., {Pye}, J.~P., {Elvis}, M.,
  {Watson}, M.~G., {Griffiths}, R.~E., {Pounds}, K.~A., {McHardy}, I.,
  {Maccagni}, D., {Seward}, F.~D., {Page}, C.~G., \& {Turner}, M.~J.~L. 1978,
  \mnras, 182, 489

\bibitem[{{Costantini} {et~al.}(2007){Costantini}, {Kaastra}, {Arav}, {Kriss},
  {Steenbrugge}, {Gabel}, {Verbunt}, {Behar}, {Gaskell}, {Korista}, {Proga},
  {Quijano}, {Scott}, {Klimek}, \& {Hedrick}}]{costantini2007}
{Costantini}, E., {Kaastra}, J.~S., {Arav}, N., {Kriss}, G.~A., {Steenbrugge},
  K.~C., {Gabel}, J.~R., {Verbunt}, F., {Behar}, E., {Gaskell}, C.~M.,
  {Korista}, K.~T., {Proga}, D., {Quijano}, J.~K., {Scott}, J.~E., {Klimek},
  E.~S., \& {Hedrick}, C.~H. 2007, \aap, 461, 121

\bibitem[{{Crenshaw} \& {Kraemer}(2012)}]{crenshaw&kraemer2012}
{Crenshaw}, D.~M. \& {Kraemer}, S.~B. 2012, \apj, 753, 75

\bibitem[{{Crenshaw} {et~al.}(2003){Crenshaw}, {Kraemer}, \&
  {George}}]{crenshaw2003}
{Crenshaw}, D.~M., {Kraemer}, S.~B., \& {George}, I.~M. 2003, \araa, 41, 117

\bibitem[{{Davidson} \& {Netzer}(1979)}]{davidson1979}
{Davidson}, K. \& {Netzer}, H. 1979, Reviews of Modern Physics, 51, 715

\bibitem[{{den Herder} {et~al.}(2001){den Herder}, {Brinkman}, {Kahn},
  {Branduardi-Raymont}, {Thomsen}, {Aarts}, {Audard}, {Bixler}, {den Boggende},
  {Cottam}, {Decker}, {Dubbeldam}, {Erd}, {Goulooze}, {G{\"u}del}, {Guttridge},
  {Hailey}, {Janabi}, {Kaastra}, {de Korte}, {van Leeuwen}, {Mauche},
  {McCalden}, {Mewe}, {Naber}, {Paerels}, {Peterson}, {Rasmussen}, {Rees},
  {Sakelliou}, {Sako}, {Spodek}, {Stern}, {Tamura}, {Tandy}, {de Vries},
  {Welch}, \& {Zehnder}}]{denherder01}
{den Herder}, J.~W., {Brinkman}, A.~C., {Kahn}, S.~M., {Branduardi-Raymont},
  G., {Thomsen}, K., {Aarts}, H., {Audard}, M., {Bixler}, J.~V., {den
  Boggende}, A.~J., {Cottam}, J., {Decker}, T., {Dubbeldam}, L., {Erd}, C.,
  {Goulooze}, H., {G{\"u}del}, M., {Guttridge}, P., {Hailey}, C.~J., {Janabi},
  K.~A., {Kaastra}, J.~S., {de Korte}, P.~A.~J., {van Leeuwen}, B.~J.,
  {Mauche}, C., {McCalden}, A.~J., {Mewe}, R., {Naber}, A., {Paerels}, F.~B.,
  {Peterson}, J.~R., {Rasmussen}, A.~P., {Rees}, K., {Sakelliou}, I., {Sako},
  M., {Spodek}, J., {Stern}, M., {Tamura}, T., {Tandy}, J., {de Vries}, C.~P.,
  {Welch}, S., \& {Zehnder}, A. 2001, \aap, 365, L7

\bibitem[{{Detmers} {et~al.}(2011){Detmers}, {Kaastra}, {Steenbrugge},
  {Ebrero}, {Kriss}, {Arav}, {Behar}, {Costantini}, {Branduardi-Raymont},
  {Mehdipour}, {Bianchi}, {Cappi}, {Petrucci}, {Ponti}, {Pinto}, {Ratti}, \&
  {Holczer}}]{detmers2011}
{Detmers}, R.~G., {Kaastra}, J.~S., {Steenbrugge}, K.~C., {Ebrero}, J.,
  {Kriss}, G.~A., {Arav}, N., {Behar}, E., {Costantini}, E.,
  {Branduardi-Raymont}, G., {Mehdipour}, M., {Bianchi}, S., {Cappi}, M.,
  {Petrucci}, P., {Ponti}, G., {Pinto}, C., {Ratti}, E.~M., \& {Holczer}, T.
  2011, \aap, 534, A38

\bibitem[{{Dunn} {et~al.}(2008){Dunn}, {Crenshaw}, {Kraemer}, \&
  {Trippe}}]{dunn2008}
{Dunn}, J.~P., {Crenshaw}, D.~M., {Kraemer}, S.~B., \& {Trippe}, M.~L. 2008,
  \aj, 136, 1201

\bibitem[{{Ebrero} {et~al.}(2010){Ebrero}, {Costantini}, {Kaastra}, {Detmers},
  {Arav}, {Kriss}, {Korista}, \& {Steenbrugge}}]{Ebrero2010}
{Ebrero}, J., {Costantini}, E., {Kaastra}, J.~S., {Detmers}, R.~G., {Arav}, N.,
  {Kriss}, G.~A., {Korista}, K.~T., \& {Steenbrugge}, K.~C. 2010, \aap, 520,
  A36

\bibitem[{{Ebrero} {et~al.}(2011){Ebrero}, {Kriss}, {Kaastra}, {Detmers},
  {Steenbrugge}, {Costantini}, {Arav}, {Bianchi}, {Cappi},
  {Branduardi-Raymont}, {Mehdipour}, {Petrucci}, {Pinto}, \&
  {Ponti}}]{Ebrero2011}
{Ebrero}, J., {Kriss}, G.~A., {Kaastra}, J.~S., {Detmers}, R.~G.,
  {Steenbrugge}, K.~C., {Costantini}, E., {Arav}, N., {Bianchi}, S., {Cappi},
  M., {Branduardi-Raymont}, G., {Mehdipour}, M., {Petrucci}, P.~O., {Pinto},
  C., \& {Ponti}, G. 2011, \aap, 534, A40

\bibitem[{{Elvis}(2000)}]{elvis2000}
{Elvis}, M. 2000, \apj, 545, 63

\bibitem[{{Ferland} {et~al.}(1998){Ferland}, {Korista}, {Verner}, {Ferguson},
  {Kingdon}, \& {Verner}}]{ferland1998}
{Ferland}, G.~J., {Korista}, K.~T., {Verner}, D.~A., {Ferguson}, J.~W.,
  {Kingdon}, J.~B., \& {Verner}, E.~M. 1998, \pasp, 110, 761

\bibitem[{{Gallo} {et~al.}(2004){Gallo}, {Boller}, {Brandt}, {Fabian}, \&
  {Vaughan}}]{gallo2004}
{Gallo}, L.~C., {Boller}, T., {Brandt}, W.~N., {Fabian}, A.~C., \& {Vaughan},
  S. 2004, \aap, 417, 29

\bibitem[{{Ganguly} {et~al.}(2001){Ganguly}, {Charlton}, \&
  {Eracleous}}]{ganguly2001}
{Ganguly}, R., {Charlton}, J.~C., \& {Eracleous}, M. 2001, \apjl, 556, L7

\bibitem[{{Gibson} {et~al.}(2005){Gibson}, {Marshall}, {Canizares}, \&
  {Lee}}]{gibson2005}
{Gibson}, R.~R., {Marshall}, H.~L., {Canizares}, C.~R., \& {Lee}, J.~C. 2005,
  \apj, 627, 83

\bibitem[{{Gofford} {et~al.}(2013){Gofford}, {Reeves}, {Tombesi}, {Braito},
  {Turner}, {Miller}, \& {Cappi}}]{gofford2013}
{Gofford}, J., {Reeves}, J.~N., {Tombesi}, F., {Braito}, V., {Turner}, T.~J.,
  {Miller}, L., \& {Cappi}, M. 2013, \mnras, 430, 60

\bibitem[{{Gofford} {et~al.}(2011){Gofford}, {Reeves}, {Turner}, {Tombesi},
  {Braito}, {Porquet}, {Miller}, {Kraemer}, \& {Fukazawa}}]{gofford2011}
{Gofford}, J., {Reeves}, J.~N., {Turner}, T.~J., {Tombesi}, F., {Braito}, V.,
  {Porquet}, D., {Miller}, L., {Kraemer}, S.~B., \& {Fukazawa}, Y. 2011,
  \mnras, 414, 3307

\bibitem[{{Grevesse} \& {Sauval}(1998)}]{grevesse1998}
{Grevesse}, N. \& {Sauval}, A.~J. 1998, ssr, 85, 161

\bibitem[{{Halpern}(1984)}]{halpern1984}
{Halpern}, J.~P. 1984, \apj, 281, 90

\bibitem[{{Holczer} \& {Behar}(2012)}]{holczer&behar2012}
{Holczer}, T. \& {Behar}, E. 2012, \apj, 747, 71

\bibitem[{{Hopkins} \& {Elvis}(2010)}]{hopkins&elvis2010}
{Hopkins}, P.~F. \& {Elvis}, M. 2010, \mnras, 401, 7

\bibitem[{{Iwasawa} \& {Taniguchi}(1993)}]{iwasawa1993}
{Iwasawa}, K. \& {Taniguchi}, Y. 1993, \apjl, 413, L15

\bibitem[{{Kaastra} {et~al.}(2011{\natexlab{a}}){Kaastra}, {de Vries},
  {Steenbrugge}, {Detmers}, {Ebrero}, {Behar}, {Bianchi}, {Costantini},
  {Kriss}, {Mehdipour}, {Paltani}, {Petrucci}, {Pinto}, \&
  {Ponti}}]{kaastra2011b}
{Kaastra}, J.~S., {de Vries}, C.~P., {Steenbrugge}, K.~C., {Detmers}, R.~G.,
  {Ebrero}, J., {Behar}, E., {Bianchi}, S., {Costantini}, E., {Kriss}, G.~A.,
  {Mehdipour}, M., {Paltani}, S., {Petrucci}, P.-O., {Pinto}, C., \& {Ponti},
  G. 2011{\natexlab{a}}, \aap, 534, A37

\bibitem[{{Kaastra} {et~al.}(2012){Kaastra}, {Detmers}, {Mehdipour}, {Arav},
  {Behar}, {Bianchi}, {Branduardi-Raymont}, {Cappi}, {Costantini}, {Ebrero},
  {Kriss}, {Paltani}, {Petrucci}, {Pinto}, {Ponti}, {Steenbrugge}, \& {de
  Vries}}]{Kaastra2012}
{Kaastra}, J.~S., {Detmers}, R.~G., {Mehdipour}, M., {Arav}, N., {Behar}, E.,
  {Bianchi}, S., {Branduardi-Raymont}, G., {Cappi}, M., {Costantini}, E.,
  {Ebrero}, J., {Kriss}, G.~A., {Paltani}, S., {Petrucci}, P.-O., {Pinto}, C.,
  {Ponti}, G., {Steenbrugge}, K.~C., \& {de Vries}, C.~P. 2012, \aap, 539, A117

\bibitem[{{Kaastra} {et~al.}(2000){Kaastra}, {Mewe}, {Liedahl}, {Komossa}, \&
  {Brinkman}}]{kaastra2000}
{Kaastra}, J.~S., {Mewe}, R., {Liedahl}, D.~A., {Komossa}, S., \& {Brinkman},
  A.~C. 2000, \aap, 354, L83

\bibitem[{{Kaastra} {et~al.}(2011{\natexlab{b}}){Kaastra}, {Petrucci}, {Cappi},
  {Arav}, {Behar}, {Bianchi}, {Bloom}, {Blustin}, {Branduardi-Raymont},
  {Costantini}, {Dadina}, {Detmers}, {Ebrero}, {Jonker}, {Klein}, {Kriss},
  {Lubi{\'n}ski}, {Malzac}, {Mehdipour}, {Paltani}, {Pinto}, {Ponti}, {Ratti},
  {Smith}, {Steenbrugge}, \& {de Vries}}]{Kaastra2011a}
{Kaastra}, J.~S., {Petrucci}, P.-O., {Cappi}, M., {Arav}, N., {Behar}, E.,
  {Bianchi}, S., {Bloom}, J., {Blustin}, A.~J., {Branduardi-Raymont}, G.,
  {Costantini}, E., {Dadina}, M., {Detmers}, R.~G., {Ebrero}, J., {Jonker},
  P.~G., {Klein}, C., {Kriss}, G.~A., {Lubi{\'n}ski}, P., {Malzac}, J.,
  {Mehdipour}, M., {Paltani}, S., {Pinto}, C., {Ponti}, G., {Ratti}, E.~M.,
  {Smith}, R.~A.~N., {Steenbrugge}, K.~C., \& {de Vries}, C.~P.
  2011{\natexlab{b}}, \aap, 534, A36

\bibitem[{{Kaastra} {et~al.}(2002){Kaastra}, {Steenbrugge}, {Raassen}, {van der
  Meer}, {Brinkman}, {Liedahl}, {Behar}, \& {de Rosa}}]{kaastra2002}
{Kaastra}, J.~S., {Steenbrugge}, K.~C., {Raassen}, A.~J.~J., {van der Meer},
  R.~L.~J., {Brinkman}, A.~C., {Liedahl}, D.~A., {Behar}, E., \& {de Rosa}, A.
  2002, \aap, 386, 427

\bibitem[{{Kalberla} {et~al.}(2005){Kalberla}, {Burton}, {Hartmann}, {Arnal},
  {Bajaja}, {Morras}, \& {P{\"o}ppel}}]{kalberla2005}
{Kalberla}, P.~M.~W., {Burton}, W.~B., {Hartmann}, D., {Arnal}, E.~M.,
  {Bajaja}, E., {Morras}, R., \& {P{\"o}ppel}, W.~G.~L. 2005, \aap, 440, 775

\bibitem[{{Kallman} {et~al.}(2004){Kallman}, {Palmeri}, {Bautista}, {Mendoza},
  \& {Krolik}}]{kallman2004}
{Kallman}, T.~R., {Palmeri}, P., {Bautista}, M.~A., {Mendoza}, C., \& {Krolik},
  J.~H. 2004, \apjs, 155, 675

\bibitem[{{Kaspi} {et~al.}(2002){Kaspi}, {Brandt}, {George}, {Netzer},
  {Crenshaw}, {Gabel}, {Hamann}, {Kaiser}, {Koratkar}, {Kraemer}, {Kriss},
  {Mathur}, {Mushotzky}, {Nandra}, {Peterson}, {Shields}, {Turner}, \&
  {Zheng}}]{kaspi2002}
{Kaspi}, S., {Brandt}, W.~N., {George}, I.~M., {Netzer}, H., {Crenshaw}, D.~M.,
  {Gabel}, J.~R., {Hamann}, F.~W., {Kaiser}, M.~E., {Koratkar}, A., {Kraemer},
  S.~B., {Kriss}, G.~A., {Mathur}, S., {Mushotzky}, R.~F., {Nandra}, K.,
  {Peterson}, B.~M., {Shields}, J.~C., {Turner}, T.~J., \& {Zheng}, W. 2002,
  \apj, 574, 643

\bibitem[{{Kaspi} {et~al.}(2001){Kaspi}, {Brandt}, {Netzer}, {George},
  {Chartas}, {Behar}, {Sambruna}, {Garmire}, \& {Nousek}}]{kaspi2001}
{Kaspi}, S., {Brandt}, W.~N., {Netzer}, H., {George}, I.~M., {Chartas}, G.,
  {Behar}, E., {Sambruna}, R.~M., {Garmire}, G.~P., \& {Nousek}, J.~A. 2001,
  \apj, 554, 216

\bibitem[{{Kaspi} {et~al.}(2000){Kaspi}, {Brandt}, {Netzer}, {Sambruna},
  {Chartas}, {Garmire}, \& {Nousek}}]{kaspi2000}
{Kaspi}, S., {Brandt}, W.~N., {Netzer}, H., {Sambruna}, R., {Chartas}, G.,
  {Garmire}, G.~P., \& {Nousek}, J.~A. 2000, \apjl, 535, L17

\bibitem[{{Kaspi} {et~al.}(2005){Kaspi}, {Maoz}, {Netzer}, {Peterson},
  {Vestergaard}, \& {Jannuzi}}]{kaspi2005}
{Kaspi}, S., {Maoz}, D., {Netzer}, H., {Peterson}, B.~M., {Vestergaard}, M., \&
  {Jannuzi}, B.~T. 2005, \apj, 629, 61

\bibitem[{{Kaspi} {et~al.}(2004){Kaspi}, {Netzer}, {Chelouche}, {George},
  {Nandra}, \& {Turner}}]{kaspi2004}
{Kaspi}, S., {Netzer}, H., {Chelouche}, D., {George}, I.~M., {Nandra}, K., \&
  {Turner}, T.~J. 2004, \apj, 611, 68

\bibitem[{{King} \& {Pounds}(2003)}]{king2003}
{King}, A.~R. \& {Pounds}, K.~A. 2003, \mnras, 345, 657

\bibitem[{{Kinkhabwala} {et~al.}(2002){Kinkhabwala}, {Sako}, {Behar}, {Kahn},
  {Paerels}, {Brinkman}, {Kaastra}, {Gu}, \& {Liedahl}}]{Kinkhabwala2002}
{Kinkhabwala}, A., {Sako}, M., {Behar}, E., {Kahn}, S.~M., {Paerels}, F.,
  {Brinkman}, A.~C., {Kaastra}, J.~S., {Gu}, M.~F., \& {Liedahl}, D.~A. 2002,
  \apj, 575, 732

\bibitem[{{Kraemer} {et~al.}(2004){Kraemer}, {Ferland}, \&
  {Gabel}}]{kraemer2004}
{Kraemer}, S.~B., {Ferland}, G.~J., \& {Gabel}, J.~R. 2004, \apj, 604, 556

\bibitem[{{Kraemer} {et~al.}(2005){Kraemer}, {George}, {Crenshaw}, {Gabel},
  {Turner}, {Gull}, {Hutchings}, {Kriss}, {Mushotzky}, {Netzer}, {Peterson}, \&
  {Behar}}]{kraemer2005}
{Kraemer}, S.~B., {George}, I.~M., {Crenshaw}, D.~M., {Gabel}, J.~R., {Turner},
  T.~J., {Gull}, T.~R., {Hutchings}, J.~B., {Kriss}, G.~A., {Mushotzky}, R.~F.,
  {Netzer}, H., {Peterson}, B.~M., \& {Behar}, E. 2005, \apj, 633, 693

\bibitem[{{Krolik} \& {Kriss}(2001)}]{Krolik01}
{Krolik}, J.~H. \& {Kriss}, G.~A. 2001, \apj, 561, 684

\bibitem[{{Krongold} {et~al.}(2003){Krongold}, {Nicastro}, {Brickhouse},
  {Elvis}, {Liedahl}, \& {Mathur}}]{krongold2003}
{Krongold}, Y., {Nicastro}, F., {Brickhouse}, N.~S., {Elvis}, M., {Liedahl},
  D.~A., \& {Mathur}, S. 2003, \apj, 597, 832

\bibitem[{{Krongold} {et~al.}(2007){Krongold}, {Nicastro}, {Elvis},
  {Brickhouse}, {Binette}, {Mathur}, \&
  {Jim{\'e}nez-Bail{\'o}n}}]{krongold2007}
{Krongold}, Y., {Nicastro}, F., {Elvis}, M., {Brickhouse}, N., {Binette}, L.,
  {Mathur}, S., \& {Jim{\'e}nez-Bail{\'o}n}, E. 2007, \apj, 659, 1022

\bibitem[{{Krongold} {et~al.}(2005){Krongold}, {Nicastro}, {Elvis},
  {Brickhouse}, {Mathur}, \& {Zezas}}]{Krongold2005}
{Krongold}, Y., {Nicastro}, F., {Elvis}, M., {Brickhouse}, N.~S., {Mathur}, S.,
  \& {Zezas}, A. 2005, \apj, 620, 165

\bibitem[{{Lee} {et~al.}(2001){Lee}, {Ogle}, {Canizares}, {Marshall}, {Schulz},
  {Morales}, {Fabian}, \& {Iwasawa}}]{lee2001}
{Lee}, J.~C., {Ogle}, P.~M., {Canizares}, C.~R., {Marshall}, H.~L., {Schulz},
  N.~S., {Morales}, R., {Fabian}, A.~C., \& {Iwasawa}, K. 2001, \apjl, 554, L13

\bibitem[{{Liedahl} \& {Paerels}(1996)}]{Liedahl1996}
{Liedahl}, D.~A. \& {Paerels}, F. 1996, \apjl, 468, L33

\bibitem[{{Lira} {et~al.}(2011){Lira}, {Ar{\'e}valo}, {Uttley}, {McHardy}, \&
  {Breedt}}]{Lira11}
{Lira}, P., {Ar{\'e}valo}, P., {Uttley}, P., {McHardy}, I., \& {Breedt}, E.
  2011, \mnras, 415, 1290

\bibitem[{{Lobban} {et~al.}(2011){Lobban}, {Reeves}, {Miller}, {Turner},
  {Braito}, {Kraemer}, \& {Crenshaw}}]{lobban2011}
{Lobban}, A.~P., {Reeves}, J.~N., {Miller}, L., {Turner}, T.~J., {Braito}, V.,
  {Kraemer}, S.~B., \& {Crenshaw}, D.~M. 2011, \mnras, 414, 1965

\bibitem[{{Longinotti} {et~al.}(2009){Longinotti}, {Bianchi}, {Ballo}, {de La
  Calle}, \& {Guainazzi}}]{longinotti2009}
{Longinotti}, A.~L., {Bianchi}, S., {Ballo}, L., {de La Calle}, I., \&
  {Guainazzi}, M. 2009, \mnras, 394, L1

\bibitem[{{Longinotti} {et~al.}(2010){Longinotti}, {Costantini}, {Petrucci},
  {Boisson}, {Mouchet}, {Santos-Lleo}, {Matt}, {Ponti}, \& {Gon{\c
  c}alves}}]{longinotti2010}
{Longinotti}, A.~L., {Costantini}, E., {Petrucci}, P.~O., {Boisson}, C.,
  {Mouchet}, M., {Santos-Lleo}, M., {Matt}, G., {Ponti}, G., \& {Gon{\c
  c}alves}, A.~C. 2010, \aap, 510, A92

\bibitem[{{Macchetto} {et~al.}(1990){Macchetto}, {Colina}, {Golombek},
  {Perryman}, \& {di Serego Alighieri}}]{macchetto1990}
{Macchetto}, F., {Colina}, L., {Golombek}, D., {Perryman}, M.~A.~C., \& {di
  Serego Alighieri}, S. 1990, \apj, 356, 389

\bibitem[{{McKernan} {et~al.}(2003){McKernan}, {Yaqoob}, {George}, \&
  {Turner}}]{mckernan2003}
{McKernan}, B., {Yaqoob}, T., {George}, I.~M., \& {Turner}, T.~J. 2003, \apj,
  593, 142

\bibitem[{{McKernan} {et~al.}(2007){McKernan}, {Yaqoob}, \&
  {Reynolds}}]{mckernan2007}
{McKernan}, B., {Yaqoob}, T., \& {Reynolds}, C.~S. 2007, \mnras, 379, 1359

\bibitem[{{Mineo} \& {Stewart}(1993)}]{mineo1993}
{Mineo}, T. \& {Stewart}, G.~C. 1993, \mnras, 262, 817

\bibitem[{{Monier} {et~al.}(2001){Monier}, {Mathur}, {Wilkes}, \&
  {Elvis}}]{monier2001}
{Monier}, E.~M., {Mathur}, S., {Wilkes}, B., \& {Elvis}, M. 2001, \apj, 559,
  675

\bibitem[{{Mor} {et~al.}(2009){Mor}, {Netzer}, \& {Elitzur}}]{Mor09}
{Mor}, R., {Netzer}, H., \& {Elitzur}, M. 2009, \apj, 705, 298

\bibitem[{{Nahar} \& {Pradhan}(2003)}]{Nahar03}
{Nahar}, S.~N. \& {Pradhan}, A.~K. 2003, \apjs, 149, 239

\bibitem[{{Nandra} {et~al.}(1997){Nandra}, {George}, {Mushotzky}, {Turner}, \&
  {Yaqoob}}]{nandra1997}
{Nandra}, K., {George}, I.~M., {Mushotzky}, R.~F., {Turner}, T.~J., \&
  {Yaqoob}, T. 1997, \apjl, 488, L91

\bibitem[{{Netzer}(1996)}]{Netzer1996}
{Netzer}, H. 1996, \apj, 473, 781

\bibitem[{{Netzer}(2004)}]{netzer2004}
---. 2004, \apj, 604, 551

\bibitem[{{Netzer} {et~al.}(2003){Netzer}, {Kaspi}, {Behar}, {Brandt},
  {Chelouche}, {George}, {Crenshaw}, {Gabel}, {Hamann}, {Kraemer}, {Kriss},
  {Nandra}, {Peterson}, {Shields}, \& {Turner}}]{netzer2003}
{Netzer}, H., {Kaspi}, S., {Behar}, E., {Brandt}, W.~N., {Chelouche}, D.,
  {George}, I.~M., {Crenshaw}, D.~M., {Gabel}, J.~R., {Hamann}, F.~W.,
  {Kraemer}, S.~B., {Kriss}, G.~A., {Nandra}, K., {Peterson}, B.~M., {Shields},
  J.~C., \& {Turner}, T.~J. 2003, \apj, 599, 933

\bibitem[{{Netzer} \& {Laor}(1993)}]{netzer&laor1993}
{Netzer}, H. \& {Laor}, A. 1993, \apjl, 404, L51

\bibitem[{{Ogle} {et~al.}(2004){Ogle}, {Mason}, {Page}, {Salvi}, {Cordova},
  {McHardy}, \& {Priedhorsky}}]{ogle2004}
{Ogle}, P.~M., {Mason}, K.~O., {Page}, M.~J., {Salvi}, N.~J., {Cordova}, F.~A.,
  {McHardy}, I.~M., \& {Priedhorsky}, W.~C. 2004, \apj, 606, 151

\bibitem[{{Orr} {et~al.}(2001){Orr}, {Barr}, {Guainazzi}, {Parmar}, \&
  {Young}}]{Orr2001}
{Orr}, A., {Barr}, P., {Guainazzi}, M., {Parmar}, A.~N., \& {Young}, A.~J.
  2001, \aap, 376, 413

\bibitem[{{Page} {et~al.}(2004){Page}, {O'Brien}, {Reeves}, \&
  {Turner}}]{page2004}
{Page}, K.~L., {O'Brien}, P.~T., {Reeves}, J.~N., \& {Turner}, M.~J.~L. 2004,
  \mnras, 347, 316

\bibitem[{{Pan} {et~al.}(1990){Pan}, {Stewart}, \& {Pounds}}]{pan1990}
{Pan}, H.~C., {Stewart}, G.~C., \& {Pounds}, K.~A. 1990, \mnras, 242, 177

\bibitem[{{Papadakis} {et~al.}(2007){Papadakis}, {Brinkmann}, {Page},
  {McHardy}, \& {Uttley}}]{papadakis2007}
{Papadakis}, I.~E., {Brinkmann}, W., {Page}, M.~J., {McHardy}, I., \& {Uttley},
  P. 2007, \aap, 461, 931

\bibitem[{{Patrick} {et~al.}(2012){Patrick}, {Reeves}, {Porquet}, {Markowitz},
  {Braito}, \& {Lobban}}]{patrick2012}
{Patrick}, A.~R., {Reeves}, J.~N., {Porquet}, D., {Markowitz}, A.~G., {Braito},
  V., \& {Lobban}, A.~P. 2012, \mnras, 426, 2522

\bibitem[{{Peterson} {et~al.}(2004){Peterson}, {Ferrarese}, {Gilbert}, {Kaspi},
  {Malkan}, {Maoz}, {Merritt}, {Netzer}, {Onken}, {Pogge}, {Vestergaard}, \&
  {Wandel}}]{peterson2004}
{Peterson}, B.~M., {Ferrarese}, L., {Gilbert}, K.~M., {Kaspi}, S., {Malkan},
  M.~A., {Maoz}, D., {Merritt}, D., {Netzer}, H., {Onken}, C.~A., {Pogge},
  R.~W., {Vestergaard}, M., \& {Wandel}, A. 2004, \apj, 613, 682

\bibitem[{{Phillips}(1980)}]{phillips1980}
{Phillips}, M.~M. 1980, \apjl, 236, L45

\bibitem[{{Piconcelli} {et~al.}(2005){Piconcelli}, {Jimenez-Bail{\'o}n},
  {Guainazzi}, {Schartel}, {Rodr{\'{\i}}guez-Pascual}, \&
  {Santos-Lle{\'o}}}]{piconcelli2005}
{Piconcelli}, E., {Jimenez-Bail{\'o}n}, E., {Guainazzi}, M., {Schartel}, N.,
  {Rodr{\'{\i}}guez-Pascual}, P.~M., \& {Santos-Lle{\'o}}, M. 2005, \aap, 432,
  15

\bibitem[{{Ponti} {et~al.}(2013){Ponti}, {Cappi}, {Costantini}, {Bianchi},
  {Kaastra}, {De Marco}, {Fender}, {Petrucci}, {Kriss}, {Steenbrugge}, {Arav},
  {Behar}, {Branduardi-Raymont}, {Dadina}, {Ebrero}, {Lubi{\'n}ski},
  {Mehdipour}, {Paltani}, {Pinto}, \& {Tombesi}}]{Ponti2013}
{Ponti}, G., {Cappi}, M., {Costantini}, E., {Bianchi}, S., {Kaastra}, J.~S.,
  {De Marco}, B., {Fender}, R.~P., {Petrucci}, P.-O., {Kriss}, G.~A.,
  {Steenbrugge}, K.~C., {Arav}, N., {Behar}, E., {Branduardi-Raymont}, G.,
  {Dadina}, M., {Ebrero}, J., {Lubi{\'n}ski}, P., {Mehdipour}, M., {Paltani},
  S., {Pinto}, C., \& {Tombesi}, F. 2013, \aap, 549, A72

\bibitem[{{Ponti} {et~al.}(2009){Ponti}, {Cappi}, {Vignali}, {Miniutti},
  {Tombesi}, {Dadina}, {Fabian}, {Grandi}, {Kaastra}, {Petrucci}, {Bianchi},
  {Matt}, {Maraschi}, \& {Malaguti}}]{Ponti2009}
{Ponti}, G., {Cappi}, M., {Vignali}, C., {Miniutti}, G., {Tombesi}, F.,
  {Dadina}, M., {Fabian}, A.~C., {Grandi}, P., {Kaastra}, J., {Petrucci},
  P.~O., {Bianchi}, S., {Matt}, G., {Maraschi}, L., \& {Malaguti}, G. 2009,
  \mnras, 394, 1487

\bibitem[{{Porquet} \& {Dubau}(2000)}]{Porquet2000}
{Porquet}, D. \& {Dubau}, J. 2000, \aaps, 143, 495

\bibitem[{{Porquet} {et~al.}(2010){Porquet}, {Dubau}, \&
  {Grosso}}]{Porquet2010}
{Porquet}, D., {Dubau}, J., \& {Grosso}, N. 2010, \ssr, 157, 103

\bibitem[{{Porquet} {et~al.}(2004){Porquet}, {Reeves}, {O'Brien}, \&
  {Brinkmann}}]{porquet2004}
{Porquet}, D., {Reeves}, J.~N., {O'Brien}, P., \& {Brinkmann}, W. 2004, \aap,
  422, 85

\bibitem[{{Pounds} {et~al.}(2001){Pounds}, {Reeves}, {O'Brien}, {Page},
  {Turner}, \& {Nayakshin}}]{pounds2001}
{Pounds}, K., {Reeves}, J., {O'Brien}, P., {Page}, K., {Turner}, M., \&
  {Nayakshin}, S. 2001, \apj, 559, 181

\bibitem[{{Pounds} \& {Reeves}(2009)}]{PR2009}
{Pounds}, K.~A. \& {Reeves}, J.~N. 2009, \mnras, 397, 249

\bibitem[{{Pounds} {et~al.}(2004{\natexlab{a}}){Pounds}, {Reeves}, {King}, \&
  {Page}}]{pounds2004a}
{Pounds}, K.~A., {Reeves}, J.~N., {King}, A.~R., \& {Page}, K.~L.
  2004{\natexlab{a}}, \mnras, 350, 10

\bibitem[{{Pounds} {et~al.}(2003){Pounds}, {Reeves}, {King}, {Page}, {O'Brien},
  \& {Turner}}]{pounds2003b}
{Pounds}, K.~A., {Reeves}, J.~N., {King}, A.~R., {Page}, K.~L., {O'Brien},
  P.~T., \& {Turner}, M.~J.~L. 2003, \mnras, 345, 705

\bibitem[{{Pounds} {et~al.}(2004{\natexlab{b}}){Pounds}, {Reeves}, {Page}, \&
  {O'Brien}}]{pounds2004b}
{Pounds}, K.~A., {Reeves}, J.~N., {Page}, K.~L., \& {O'Brien}, P.~T.
  2004{\natexlab{b}}, \apj, 616, 696

\bibitem[{{Proga} \& {Kallman}(2004)}]{proga&kallman2004}
{Proga}, D. \& {Kallman}, T.~R. 2004, \apj, 616, 688

\bibitem[{{Raimundo} {et~al.}(2012){Raimundo}, {Fabian}, {Vasudevan}, {Gandhi},
  \& {Wu}}]{Raimundo2012}
{Raimundo}, S.~I., {Fabian}, A.~C., {Vasudevan}, R.~V., {Gandhi}, P., \& {Wu},
  J. 2012, \mnras, 419, 2529

\bibitem[{{Ram{\'{\i}}rez} {et~al.}(2008){Ram{\'{\i}}rez}, {Komossa},
  {Burwitz}, \& {Mathur}}]{ramirez2008}
{Ram{\'{\i}}rez}, J.~M., {Komossa}, S., {Burwitz}, V., \& {Mathur}, S. 2008,
  \apj, 681, 965

\bibitem[{{Reeves} {et~al.}(2010){Reeves}, {Gofford}, {Braito}, \&
  {Sambruna}}]{Reeves2010}
{Reeves}, J.~N., {Gofford}, J., {Braito}, V., \& {Sambruna}, R. 2010, \apj,
  725, 803

\bibitem[{{Reeves} {et~al.}(2004){Reeves}, {Nandra}, {George}, {Pounds},
  {Turner}, \& {Yaqoob}}]{reeves2004a}
{Reeves}, J.~N., {Nandra}, K., {George}, I.~M., {Pounds}, K.~A., {Turner},
  T.~J., \& {Yaqoob}, T. 2004, \apj, 602, 648

\bibitem[{{Reeves} {et~al.}(2009){Reeves}, {O'Brien}, {Braito}, {Behar},
  {Miller}, {Turner}, {Fabian}, {Kaspi}, {Mushotzky}, \& {Ward}}]{reeves2009}
{Reeves}, J.~N., {O'Brien}, P.~T., {Braito}, V., {Behar}, E., {Miller}, L.,
  {Turner}, T.~J., {Fabian}, A.~C., {Kaspi}, S., {Mushotzky}, R., \& {Ward}, M.
  2009, \apj, 701, 493

\bibitem[{{Reeves} \& {Turner}(2000)}]{reeves&turner2000}
{Reeves}, J.~N. \& {Turner}, M.~J.~L. 2000, \mnras, 316, 234

\bibitem[{{Reynolds}(1997)}]{reynolds1997}
{Reynolds}, C.~S. 1997, \mnras, 286, 513

\bibitem[{{Ricker} {et~al.}(1978){Ricker}, {Clark}, {Doxsey}, {Dower},
  {Jernigan}, {Delvaille}, {MacAlpine}, \& {Hjellming}}]{ricker1978}
{Ricker}, G.~R., {Clark}, G.~W., {Doxsey}, R.~E., {Dower}, R.~G., {Jernigan},
  J.~G., {Delvaille}, J.~P., {MacAlpine}, G.~M., \& {Hjellming}, R.~M. 1978,
  \nat, 271, 35

\bibitem[{{Risaliti} {et~al.}(2005){Risaliti}, {Bianchi}, {Matt}, {Baldi},
  {Elvis}, {Fabbiano}, \& {Zezas}}]{risaliti2005}
{Risaliti}, G., {Bianchi}, S., {Matt}, G., {Baldi}, A., {Elvis}, M.,
  {Fabbiano}, G., \& {Zezas}, A. 2005, \apjl, 630, L129

\bibitem[{{Sako} {et~al.}(2001){Sako}, {Kahn}, {Behar}, {Kaastra}, {Brinkman},
  {Boller}, {Puchnarewicz}, {Starling}, {Liedahl}, {Clavel}, \&
  {Santos-Lleo}}]{sako2001}
{Sako}, M., {Kahn}, S.~M., {Behar}, E., {Kaastra}, J.~S., {Brinkman}, A.~C.,
  {Boller}, T., {Puchnarewicz}, E.~M., {Starling}, R., {Liedahl}, D.~A.,
  {Clavel}, J., \& {Santos-Lleo}, M. 2001, \aap, 365, L168

\bibitem[{{Schmidt} {et~al.}(2006){Schmidt}, {Schippers}, {M{\"u}ller},
  {Lestinsky}, {Sprenger}, {Grieser}, {Repnow}, {Wolf}, {Brandau}, {Luki{\'c}},
  {Schnell}, \& {Savin}}]{Schmidt2006}
{Schmidt}, E.~W., {Schippers}, S., {M{\"u}ller}, A., {Lestinsky}, M.,
  {Sprenger}, F., {Grieser}, M., {Repnow}, R., {Wolf}, A., {Brandau}, C.,
  {Luki{\'c}}, D., {Schnell}, M., \& {Savin}, D.~W. 2006, \apjl, 641, L157

\bibitem[{{Scott} {et~al.}(2011){Scott}, {Stewart}, {Mateos}, {Alexander},
  {Hutton}, \& {Ward}}]{scott2011}
{Scott}, A.~E., {Stewart}, G.~C., {Mateos}, S., {Alexander}, D.~M., {Hutton},
  S., \& {Ward}, M.~J. 2011, \mnras, 417, 992

\bibitem[{{Scott} {et~al.}(2004){Scott}, {Kriss}, {Lee}, {Arav}, {Ogle},
  {Roraback}, {Weaver}, {Alexander}, {Brotherton}, {Green}, {Hutchings},
  {Kaiser}, {Marshall}, {Oegerle}, \& {Zheng}}]{Scott2004}
{Scott}, J.~E., {Kriss}, G.~A., {Lee}, J.~C., {Arav}, N., {Ogle}, P.,
  {Roraback}, K., {Weaver}, K., {Alexander}, T., {Brotherton}, M., {Green},
  R.~F., {Hutchings}, J., {Kaiser}, M.~E., {Marshall}, H., {Oegerle}, W., \&
  {Zheng}, W. 2004, \apjs, 152, 1

\bibitem[{{Shu} {et~al.}(2010){Shu}, {Yaqoob}, \& {Wang}}]{shu2010}
{Shu}, X.~W., {Yaqoob}, T., \& {Wang}, J.~X. 2010, \apjs, 187, 581

\bibitem[{{Smith} {et~al.}(2007){Smith}, {Page}, \&
  {Branduardi-Raymont}}]{smith2007}
{Smith}, R.~A.~N., {Page}, M.~J., \& {Branduardi-Raymont}, G. 2007, \aap, 461,
  135

\bibitem[{{Steenbrugge} {et~al.}(2009){Steenbrugge}, {Fenov{\v c}{\'{\i}}k},
  {Kaastra}, {Costantini}, \& {Verbunt}}]{Steenbrugge2009}
{Steenbrugge}, K.~C., {Fenov{\v c}{\'{\i}}k}, M., {Kaastra}, J.~S.,
  {Costantini}, E., \& {Verbunt}, F. 2009, \aap, 496, 107

\bibitem[{{Steenbrugge} {et~al.}(2005{\natexlab{a}}){Steenbrugge}, {Kaastra},
  {Crenshaw}, {Kraemer}, {Arav}, {George}, {Liedahl}, {van der Meer},
  {Paerels}, {Turner}, \& {Yaqoob}}]{steenbrugge2005}
{Steenbrugge}, K.~C., {Kaastra}, J.~S., {Crenshaw}, D.~M., {Kraemer}, S.~B.,
  {Arav}, N., {George}, I.~M., {Liedahl}, D.~A., {van der Meer}, R.~L.~J.,
  {Paerels}, F.~B.~S., {Turner}, T.~J., \& {Yaqoob}, T. 2005{\natexlab{a}},
  \aap, 434, 569

\bibitem[{{Steenbrugge} {et~al.}(2005{\natexlab{b}}){Steenbrugge}, {Kaastra},
  {Sako}, {Branduardi-Raymont}, {Behar}, {Paerels}, {Blustin}, \&
  {Kahn}}]{steenbrugge2005b}
{Steenbrugge}, K.~C., {Kaastra}, J.~S., {Sako}, M., {Branduardi-Raymont}, G.,
  {Behar}, E., {Paerels}, F.~B.~S., {Blustin}, A.~J., \& {Kahn}, S.~M.
  2005{\natexlab{b}}, \aap, 432, 453

\bibitem[{{Tarter} {et~al.}(1969){Tarter}, {Tucker}, \&
  {Salpeter}}]{tarter1969}
{Tarter}, C.~B., {Tucker}, W.~H., \& {Salpeter}, E.~E. 1969, \apj, 156, 943

\bibitem[{{Tatum} {et~al.}(2013){Tatum}, {Turner}, {Miller}, \&
  {Reeves}}]{tatum2013}
{Tatum}, M.~M., {Turner}, T.~J., {Miller}, L., \& {Reeves}, J.~N. 2013, \apj,
  762, 80

\bibitem[{{Terashima} {et~al.}(2009){Terashima}, {Gallo}, {Inoue}, {Markowitz},
  {Reeves}, {Anabuki}, {Fabian}, {Griffiths}, {Hayashida}, {Itoh}, {Kokubun},
  {Kubota}, {Miniutti}, {Takahashi}, {Yamauchi}, \& {Yonetoku}}]{Terashima2009}
{Terashima}, Y., {Gallo}, L.~C., {Inoue}, H., {Markowitz}, A.~G., {Reeves},
  J.~N., {Anabuki}, N., {Fabian}, A.~C., {Griffiths}, R.~E., {Hayashida}, K.,
  {Itoh}, T., {Kokubun}, N., {Kubota}, A., {Miniutti}, G., {Takahashi}, T.,
  {Yamauchi}, M., \& {Yonetoku}, D. 2009, \pasj, 61, 299

\bibitem[{{Tombesi} {et~al.}(2012){Tombesi}, {Cappi}, {Reeves}, \&
  {Braito}}]{tombesi2012a}
{Tombesi}, F., {Cappi}, M., {Reeves}, J.~N., \& {Braito}, V. 2012, \mnras, 422,
  L1

\bibitem[{{Tombesi} {et~al.}(2013){Tombesi}, {Cappi}, {Reeves}, {Nemmen},
  {Braito}, {Gaspari}, \& {Reynolds}}]{Tombesi2013}
{Tombesi}, F., {Cappi}, M., {Reeves}, J.~N., {Nemmen}, R.~S., {Braito}, V.,
  {Gaspari}, M., \& {Reynolds}, C.~S. 2013, \mnras, 430, 1102

\bibitem[{{Tombesi} {et~al.}(2011){Tombesi}, {Cappi}, {Reeves}, {Palumbo},
  {Braito}, \& {Dadina}}]{tombesi2011a}
{Tombesi}, F., {Cappi}, M., {Reeves}, J.~N., {Palumbo}, G.~G.~C., {Braito}, V.,
  \& {Dadina}, M. 2011, \apj, 742, 44

\bibitem[{{Tombesi} {et~al.}(2010{\natexlab{a}}){Tombesi}, {Cappi}, {Reeves},
  {Palumbo}, {Yaqoob}, {Braito}, \& {Dadina}}]{tombesi2010a}
{Tombesi}, F., {Cappi}, M., {Reeves}, J.~N., {Palumbo}, G.~G.~C., {Yaqoob}, T.,
  {Braito}, V., \& {Dadina}, M. 2010{\natexlab{a}}, \aap, 521, A57

\bibitem[{{Tombesi} {et~al.}(2010{\natexlab{b}}){Tombesi}, {Sambruna},
  {Reeves}, {Braito}, {Ballo}, {Gofford}, {Cappi}, \&
  {Mushotzky}}]{tombesi2010b}
{Tombesi}, F., {Sambruna}, R.~M., {Reeves}, J.~N., {Braito}, V., {Ballo}, L.,
  {Gofford}, J., {Cappi}, M., \& {Mushotzky}, R.~F. 2010{\natexlab{b}}, \apj,
  719, 700

\bibitem[{{Turner} {et~al.}(2004){Turner}, {Fabian}, {Lee}, \&
  {Vaughan}}]{turner2004}
{Turner}, A.~K., {Fabian}, A.~C., {Lee}, J.~C., \& {Vaughan}, S. 2004, \mnras,
  353, 319

\bibitem[{{Turner} {et~al.}(2005){Turner}, {Kraemer}, {George}, {Reeves}, \&
  {Bottorff}}]{turner2005}
{Turner}, T.~J., {Kraemer}, S.~B., {George}, I.~M., {Reeves}, J.~N., \&
  {Bottorff}, M.~C. 2005, \apj, 618, 155

\bibitem[{{Turner} {et~al.}(2008){Turner}, {Reeves}, {Kraemer}, \&
  {Miller}}]{turner2008}
{Turner}, T.~J., {Reeves}, J.~N., {Kraemer}, S.~B., \& {Miller}, L. 2008, \aap,
  483, 161

\bibitem[{{Urry} \& {Padovani}(1995)}]{urry&padovani1995}
{Urry}, C.~M. \& {Padovani}, P. 1995, \pasp, 107, 803

\bibitem[{{Vasudevan} \& {Fabian}(2009)}]{VF2009}
{Vasudevan}, R.~V. \& {Fabian}, A.~C. 2009, \mnras, 392, 1124

\bibitem[{{Weisskopf} {et~al.}(2000){Weisskopf}, {Hester}, {Tennant}, {Elsner},
  {Schulz}, {Marshall}, {Karovska}, {Nichols}, {Swartz}, {Kolodziejczak}, \&
  {O'Dell}}]{weisskopf2000}
{Weisskopf}, M.~C., {Hester}, J.~J., {Tennant}, A.~F., {Elsner}, R.~F.,
  {Schulz}, N.~S., {Marshall}, H.~L., {Karovska}, M., {Nichols}, J.~S.,
  {Swartz}, D.~A., {Kolodziejczak}, J.~J., \& {O'Dell}, S.~L. 2000, \apjl, 536,
  L81

\bibitem[{{Wilms} {et~al.}(2000){Wilms}, {Allen}, \& {McCray}}]{wilms2000}
{Wilms}, J., {Allen}, A., \& {McCray}, R. 2000, \apj, 542, 914

\bibitem[{{Yaqoob} {et~al.}(2003){Yaqoob}, {McKernan}, {Kraemer}, {Crenshaw},
  {Gabel}, {George}, \& {Turner}}]{yaqoob2003}
{Yaqoob}, T., {McKernan}, B., {Kraemer}, S.~B., {Crenshaw}, D.~M., {Gabel},
  J.~R., {George}, I.~M., \& {Turner}, T.~J. 2003, \apj, 582, 105

\bibitem[{{Zhang} {et~al.}(2011){Zhang}, {Ji}, {Marshall}, {Longinotti},
  {Evans}, \& {Gu}}]{Zhang2011}
{Zhang}, S.~N., {Ji}, L., {Marshall}, H.~L., {Longinotti}, A.~L., {Evans}, D.,
  \& {Gu}, Q.~S. 2011, \mnras, 410, 2274

\end{thebibliography}

\clearpage

\clearpage

\begin{deluxetable}{lccccc}
\tabletypesize{\small}
\tablecaption{Summary of MR\,2251-178 Observations}
\tablewidth{0pt}
\tablehead{
\colhead{Mission} & \colhead{Obsid} & \colhead{Start Date/Time$^{a}$} & \colhead{Inst} 
& \colhead{Exposure (ks)} & \colhead{Net Rate s$^{-1}$}}

\startdata
  XMM-Newton & 0670120201 & 2011-11-11 07:58:17 & RGS\,1 & 133.1 & $0.518\pm0.002$ \\
& -- & -- & RGS\,2 & -- & $0.562\pm0.002$ \\
& 0670120301 & 2011-11-13 18:50:38 & RGS\,1 & 127.8 & $0.506\pm0.002$ \\
& -- & -- & RGS\,2 & -- & $0.544\pm0.002$ \\
& 0670120401 & 2011-11-15 18:42:16 & RGS\,1 & 128.0 & $0.464\pm0.002$ \\
& -- & -- & RGS\,2 & -- & $0.501\pm0.002$\\
& Total & -- & RGS\,1 & 389.1$^{b}$ & $0.496\pm0.001$ \\
&  & -- & RGS\,2 & 389.1$^{b}$ & $0.535\pm0.001$ \\
\hline
 Chandra & 2977 & 2002-09-11 00:52:46 & MEG & 146.3 & $0.317\pm0.001$ \\
         & -- & -- & HEG & -- & $0.164\pm0.001$ \\
\hline           
Chandra & 12828 & 2011-09-26 20:34:38 & MEG & 163.1 \\
& -- & -- & HEG & -- \\
& 12829 & 2011-09-29 07:03:04 & MEG & 187.6 \\
& -- & -- & HEG & -- \\
& 12830 & 2011-10-01 22:53:18 & MEG & 48.6 \\
& -- & -- & HEG & -- \\
& Total & -- & MEG & 392.9 & $0.485\pm0.001$ \\
& -- & -- & HEG & -- & $0.245\pm0.001$ \\
\enddata

\tablenotetext{a}{Observation Start/End times are in UT.} 
\tablenotetext{b}{Net exposure time, after screening and deadtime correction, in ks.}
\label{observations}
\end{deluxetable}

\clearpage

\begin{deluxetable}{lcrcc}
\tabletypesize{\small}
\tablecaption{Soft X-ray absorption lines in 2011 Chandra HETG Spectrum}
\tablewidth{0pt}
\tablehead{
\colhead{ID$^{a}$} &\colhead{E$_{\rm atomic}^{b}$}  &\colhead{E$_{\rm quasar}^{c}$} & \colhead{EW$^{c}$} &  

\colhead{$\Delta C$$^{d}$}}
\startdata
O\,\textsc{viii} &  653.5 [18.972] &$654.4^{+0.1}_{-0.3}$  [18.946]& $-0.8\pm0.3$ & 9.8 \\
O\,\textsc{viii} &  -- &$733.3^{+0.5}_{-0.9}$  [16.908]& $-1.6\pm0.5$ & 20.4 \\
Fe\,\textsc{ix-x} $2p-3d$ & -- &$750.4\pm1.2$  [16.522]& $-1.5\pm0.6$ & 18.0 \\
Ne\,\textsc{v} &871.4 [14.228], 873.7  [14.191]& $873.3\pm0.5$  [14.197]& $-1.2\pm0.4$ & 15.6 \\
Ne\,\textsc{vi} &  885.0 [14.010], 883.3  [14.036]&$884.5\pm0.5$  [14.017]& $-1.7\pm0.4$ & 27.3 \\
Ne\,\textsc{vii} & 898.2 [13.804] &$897.7\pm0.4$ [13.811]& $-1.6\pm0.5$ & 29.9 \\
Ne\,\textsc{viii} & 909.2 [13.637]&$908.6\pm0.3$ [13.646]& $-1.0\pm0.3$ & 12.1 \\
Fe\,\textsc{xix} $2p-3d$ &  918.0 [13.506]&$918.6^{+0.2}_{-0.4}$ [13.497]& $-1.0\pm0.4$ & 15.2 \\
Ne\,\textsc{ix} & 922.0 [13.447]&$922.6\pm0.4$ [13.439]& $-1.5\pm0.5$ & 25.7 \\
Fe\,\textsc{xx} $2p-3d$ &  967.3 [12.818]&$966.4\pm0.7$ [12.829]& $-1.3\pm0.5$ & 17.2 \\
Fe\,\textsc{xx} $2p-3d$ &  987.8 [12.552]&$986.4\pm0.8$ [12.569] & $-1.2\pm0.4$ & 14.7 \\
Fe\,\textsc{xxi} $2p-3d$ &  1000.9 [12.387]& $1010.1\pm0.8$ [12.274] & $-0.9\pm0.4$ & 9.8 \\
Ne\,\textsc{x} & 1021.5 [12.137], 1022.0 [12.132]& $1022.8\pm0.4$ [12.122] & $-1.8\pm0.3$ & 56.6\\
Fe\,\textsc{xxii} $2p-3d$ &  1053.6 [11.768]& $1052.9\pm1.0$ [11.766] & $-1.0\pm0.4$ & 13.6 \\ 
Ne\,\textsc{ix} $1s-3p$ &  1073.8 [11.546]&$1074.1\pm0.7$ [11.543] & $-1.0\pm0.3$ & 18.5 \\
Ne\,\textsc{ix} $1s-4p$ & 1127.1 [11.000]&$1128.6\pm0.6$ [10.986] & $-1.2\pm0.3$ & 36.3 \\
Ne\,\textsc{ix} $1s-6p$ &  1165.0 [10.642]&$1165.5\pm0.7$ [10.638] & $-0.7\pm0.3$ & 11.2 \\
Ne\,\textsc{x} $1s-3p$ &  1210.9 [10.239]& $1211.7\pm0.5$ [10.232]& $-0.6\pm0.3$ &11.6 \\
Mg\,\textsc{vi} &  1276.8 [9.711]&$1276.9\pm0.9$ [9.710] & $-0.7\pm0.3$ & 12.6 \\
Mg\,\textsc{vii} &  1291.6 [9.599]&$1294.5\pm1.3$ [9.578] & $-0.7\pm0.3$ & 10.8 \\
Mg\,\textsc{viii} &  1306.4 [9.491], (1304.2 [9.507]) &$1306.4\pm0.7$ [9.491] & $-1.4\pm0.3$ & 41.8 \\
Mg\,\textsc{ix} &  1323.1 [9.371] &$1322.6\pm1.1$ [9.374] & $-0.6\pm0.3$ & 9.7 \\
Mg\,\textsc{xi} & 1353.1 [9.163] & $1352.7\pm0.8$ [9.166] & $-0.8\pm0.3$ & 18.2 \\
Mg\,\textsc{xii} &  1472.6 [8.419], 1471.7 [8.425] &$1473.0\pm0.5$ [8.417] & $-1.2\pm0.2$ & 47.3 \\
Mg\,\textsc{xi} $1s-3p$ &  1579.3 [7.851] &$1581.3\pm0.9$ [7.841] & $-0.6\pm0.3$ & 12.6 \\
Si\,\textsc{viii} &  1772.8 [6.994] &$1772.6\pm0.6$ [6.994] & $-1.3\pm0.3$ & 38.8 \\
Si\,\textsc{ix} &  1792.2 [6.918], (1788.2 [6.933]) &$1791.9^{+0.9}_{-1.2}$ [6.919] & $-0.9\pm0.3$ & 19.2 \\
Si\,\textsc{x} &  1810.3 [6.849], 1807.3 [6.860] & $1809.6^{+1.4}_{-2.2}$ [6.851] & $-0.8\pm0.3$ & 15.5 \\
Si\,\textsc{xi} &  1830.6 [6.773] &$1830.2\pm1.1$ [6.774] & $-0.7\pm0.3$ & 14.1 \\
Si\,\textsc{xiii} &  1866.4  [6.643]&$1866.0\pm0.9$ [6.644] & $-1.0\pm0.3$ & 23.5 \\
Si\,\textsc{xiv} &  2006.1 [6.180], (2004.8 [6.184]) &$2007.0\pm1.2$ [6.178] & $-0.8\pm0.4$ & 12.3 \\
\enddata
\tablenotetext{a}{\,Line identification. Lines correspond to the $1s-2p$ 
transition unless stated.}
\tablenotetext{b}{Known atomic line energy in eV. Values are from www.nist.gov, 
\cite{behar2001} for Fe M-shell UTA and \cite{behar2002}  
for inner shell Ne, Mg, Si. The corresponding wavelength in
 \AA~ is given
within brackets.}
\tablenotetext{c}{Measured line energy and equivalent width in quasar
  rest frame, units eV. The corresponding mean wavelength value in
  \AA~ is given
within brackets.}
\tablenotetext{d}{Improvement in C-statistic upon adding line to model.} 
\label{hetg-lines}
\end{deluxetable}

\clearpage

\begin{deluxetable}{lrrr}
\tabletypesize{\small}
\tablecaption{Soft X-ray Lines identified in 2011 XMM-Newton RGS.}
\tablewidth{0pt}
\tablehead{\colhead{Line ID$^{a}$} &  \colhead{E$_{\rm lab}^{b}$}& \colhead{E$_{\rm quasar}^{c}$}& \colhead{E$_{\rm obs}^{d}$}} 
\startdata
1. N\,\textsc{vi} $1s\to2p$ & 430.7 [28.787]& 430.3 [28.813] & 405.1 [30.606] \\
2. C\,\textsc{vi} $1s\to3p$ & 435.5 [28.469]& 435.6 [28.463] & 409.8 [30.255]  \\ 
3. C\,\textsc{vi} $1s\to4p$ & 459.4 [26.988]& 459.0 [27.012] & 432.3 [28.680]  \\
4. C\,\textsc{vi} $1s\to5p$ & 470.4 [26.357]& 470.7 [26.340] & 443.2 [27.975]  \\
5. C\,\textsc{vi} $1s\to6p$ & 476.4 [26.025]& 476.0 [26.047]  & 448.1 [27.669]  \\
6. C\,\textsc{vi} K-edge & 489.9 [25.308]& 489.7 [25.318] & 460.0 [26.953]  \\
7. N\,\textsc{vi} $1s\to3p$ & 496.7 [24.962] & 497.3 [24.931] & 467.4 [26.526]  \\
8. N\,\textsc{vii}$^{e}$ $2p\to1s$ & 500.4 [24.777]& 500.3 [24.782]& 470.2  [26.368]  \\ 
9. N\,\textsc{vii} $1s\to2p$ & 500.4 [24.777]& 501.1 [24.742] & 470.9 [26.329]  \\
10. N\,\textsc{vi} $1s\to4p$ & 521.6 [23.770]& 521.8 [23.761] & 491.1 [25.246]  \\ 
11. O\,\textsc{v} $1s\to2p$ & 554.5 [22.360]& 554.2 [22.372] & 521.8 [23.761]  \\
12. O\,\textsc{vii}$^{e}$ & & 564 [21.983] & 530 [23.393]  \\
13. O\,\textsc{vi} $1s\to2p$ & 562.6 [22.038]& 564.0 [21.983] & 530.1 [23.389]  \\
14. O\,\textsc{vi} $1s\to2p$ & 568.2 [21.821]& 568.6 [21.805]& 535.2 [23.166]  \\ 
15. N\,\textsc{vii} $1s\to3p$ & 592.9 [20.911]& 592.9 [20.911] & 558.1 [22.215]  \\ 
16. N\,\textsc{vii} $1s\to4p$ & 625.4 [19.825]& 625.0 [19.837] & 588.3 [21.075]  \\ 
17. N\,\textsc{vii} $1s\to5p$ & 640.4 [19.360]& 640.2 [19.366] & 602.6 [20.575]  \\ 
18. O\,\textsc{viii} $1s\to2p$ & 653.5 [18.972] & 653.3 [18.978]& 614.9 [20.163]   \\ 
19. O\,\textsc{viii}$^{e}$ $2p\to1s$ & 653.5 [18.972] & 654.5 [18.943] & 615.1 [20.157]  \\ 
20. O\,\textsc{vii} $1s\to3p$ & 665.6 [18.627] & 665.9 [18.619] & 626.8 [19.781]  \\ 
21. O\,\textsc{vii} $1s\to4p$ & 697.1 [17.786]& 697.6 [17.773] & 656.6 [18.883]  \\ 
22. O\,\textsc{vii} $1s\to5p$ & 712.7 [17.396]& 712.8 [17.394] & 670.9 [18.480]  \\
23. O\,\textsc{vii} $1s\to6p$ & 720.7 [17.203]  & 721.2 [17.191] & 678.8 [18.265]\\
24. Fe\,M UTA \\
25. O\,\textsc{viii} $1s\to3p$ & 774.6 [16.006]& 774.4 [16.010] & 728.9 [17.010]  \\
26. Fe\,\textsc{xvii} & & 812.4 [15.261] & 764.7 [16.213] \\ 
27. O\,\textsc{viii} $1s\to4p$ & 816.9 [15.177] & 816.4 [15.187] & 768.4 [16.135] \\
28. Fe\,\textsc{xvii-xix} & & 825.9 [15.012] & 777.4 [15.949] \\ 
29. O\,\textsc{viii} $1s\to5p$ & 836.6 [14.820] & 836.2 [14.827] & 787.1 [15.752] \\ 
30. O\,\textsc{viii} $1s\to6p$ & 847.2 [14.635] & 846.8 [14.642] & 797.0 [15.556]   \\
31. Ne\,\textsc{v}$^{i}$ & 873.7 [14.191] & 874.2 [14.183] & 821.7 [15.085]  \\
32. Ne\,\textsc{vii}$^{i}$ & 898.8 [13.794]& 898.1 [13.805] & 844.1 [14.688]  \\
33. Ne\,\textsc{ix}$^{e}$ & & 922.1 [13.446] & 866.6 [14.307] \\
34. Ne\,\textsc{viii}$^{i}$ & 909.2 [13.637]& 909.4 [13.634] & 854.7 [14.506]  \\
35. Fe\,\textsc{xix} & & 916.8 [13.524] &  862.9 [14.368] \\
36. Ne\,\textsc{ix} $1s\to2p$ & 922.0 [13.447] & 921.9 [13.449] & 867.7 [14.289]  \\
37. Fe\,\textsc{xx} & & 965.5 [12.841] & 908.9 [13.641] \\
38. Ne\,\textsc{x} $1s\to2p$ & 1021.5 [12.137] & 1021.5 [12.137] & 961.5 [12.895]  \\
39. Mg\,\textsc{viii}$^{i}$ & 1306.4 [9.491] & 1306.4 [9.491] & 1227.9 [10.097]  \\
40. Mg\,\textsc{ix}$^{i}$ & 1323.1 [9.371]  & 1322.5 [9.375] & 1243.0 [9.975] \\
41. Mg\,\textsc{xi} $1s\to2p$ & 1353.3 [9.162] & 1352.4 [9.169] & 1272.9 [9.687]  \\
42. Mg\,\textsc{xii} $1s\to2p$ & 1472.3 [8.421] & 1471.9 [8.423]& 1385.4 [8.949]   \\
\enddata
\tablenotetext{a}{\,Line identification. Line number corresponds to
  those marked Figure~\ref{rgs-panels1} and  
Figure~\ref{rgs-panels2}.}
\tablenotetext{b}{Known atomic/lab frame energy of line in units eV.
  The corresponding wavelength in  \AA~ is given
within brackets. Values are from www.nist.gov}
\tablenotetext{c}{Measured line energy in the quasar rest frame in
  eV.  The corresponding wavelength in
 \AA~ is given
within brackets.}
\tablenotetext{d}{Measured line energy in the observed frame in
  eV. Typical uncertainty is within $\pm 1$\,eV.  The corresponding wavelength in
 \AA~ is given
within brackets.}
\tablenotetext{e}{Possible emission line.}
\tablenotetext{i}{Possible inner-shell absorption line. Known atomic energy taken from 
\cite{behar2002}.} 
\label{rgs-lines}
\end{deluxetable}

\clearpage

\begin{deluxetable}{llcccc}
\tabletypesize{\small}
\tablecaption{Gaussian Fits to Velocity Profiles of H and He-like Absorption Lines}
\tablewidth{0pt}
\tablehead{
\colhead{Line} & \colhead{Instrument} & 
\colhead{$\sigma_{\rm obs}$$^{a}$} & \colhead{$\sigma_{\rm int}$$^{b}$} 
& \colhead{$v_{\rm out}$$^{c}$} & \colhead{$\Delta \chi^{2}$$^d$}} 
\startdata
H-like:-\\
C\,\textsc{vi} Ly-$\beta$ (a) & RGS & $320\pm80$ & $<270$ & $-444\pm73$ & 43.0 \\
C\,\textsc{vi} Ly-$\beta$ (b)$^{e}$ & RGS & -- & -- & $-1840\pm100^{e}$ & 16.0 \\
N\,\textsc{vii} Ly-$\alpha$ (a) & RGS & $480\pm95$ & $360^{+120}_{-140}$ & $-450\pm94$ & 57.5 \\
N\,\textsc{vii} Ly-$\alpha$ (b)$^{e}$ & RGS & --  & -- & $-2020\pm120^{e}$ & 8.2 \\ 
O\,\textsc{viii} Ly-$\alpha$ & RGS & $297\pm65$ & $<200$ & $-353\pm62$ & 47.0 \\
Ne\,\textsc{x} Ly-$\alpha$ & HEG & $415\pm135$ & $395^{+170}_{-145}$ & $-227\pm123$ & 33.1 \\
Mg\,\textsc{xii} Ly-$\alpha$ & HEG & $<120$ & -- & $<40$ & 31.2 \\
Si\,\textsc{xiv} Ly-$\alpha$ & HEG & $<125$ & -- & $<340$ & 8.7 \\
\hline
He-like:-\\
N\,\textsc{vi} He-$\alpha$ (a) & RGS & $320\pm90$ & $<280$ & $-428\pm88$ & 26.2 \\
N\,\textsc{vi} He-$\alpha$ (b)$^{e}$ & RGS & -- & -- & $-1990\pm120^{e}$ & 12.1 \\
O\,\textsc{vii} He-$\beta$ & RGS & $600\pm180$ & $460^{+220}_{-280}$ & $-900\pm180$ & 22.3 \\
Ne\,\textsc{ix} He-$\beta$ & HEG & $440\pm190$ & $420^{+180}_{-200}$ & $<234$ & 14.1 \\
Mg\,\textsc{xi} He-$\alpha$ & HEG & $310\pm175$ & $<455$ & $<170$ & 7.0 \\
Si\,\textsc{xiii} He-$\alpha$ & HEG & $<120$ & -- & $<163$ & 8.8 \\
\enddata
\tablenotetext{a}{Observed $1\sigma$ width of absorption line in km\,s$^{-1}$.}
\tablenotetext{a}{Intrinsic $1\sigma$ width of absorption line in km\,s$^{-1}$ after correcting 
for instrument spectral resolution.}
\tablenotetext{c}{Velocity shift of absorption line in km\,s$^{-1}$. Negative 
values donate blue-shift.}
\tablenotetext{d}{Improvement in fit statistic after modeling Gaussian absorption profile.}
\tablenotetext{e}{Possible higher velocity component of absorption line.}
\label{line-widths}
\end{deluxetable}

\clearpage

\begin{deluxetable}{lrrrr}
\tabletypesize{\small}
\tablecaption{Warm Absorber Parameters from RGS and HETG Spectra}
\tablewidth{0pt}
\tablehead{
\colhead{Component} & \colhead{Parameter} & \colhead{RGS 2011} & \colhead{HETG 2011} 
& \colhead{HETG 2002}} 
\startdata
Power-Law & $\Gamma$ & $2.32\pm0.08$ & $2.13^{+0.11}_{-0.10}$ & $=2011^{t}$ \\
(uncovered) & $f_{\rm uncov}$$^{a}$ & $0.39^{+0.03}_{-0.02}$ & $0.23\pm0.03$ & $0.18\pm0.02$ \\
\hline
Warm Absorber & $N_{\rm H}$$^{b}$ & $2.12\pm0.07$ & $2.10^{+0.19}_{-0.23}$ & $1.43^{+0.34}_{-0.37}$\\
(Component 1) & $\logxi$$^{c}$ & $1.27\pm0.02$ & $1.15\pm0.05$ & $0.91\pm0.16$ \\
& $v_{\rm out}$$^{d}$ & $-480\pm40$ & $-315\pm40$ & $-290\pm150$ \\
$\Delta C$ or $\Delta\chi^{2}$$^{e}$ & -- & 2065 & 176.2 & --\\
\hline
Warm Absorber & $N_{\rm H}$$^{b}$ & $1.50\pm0.20$ & $1.5^{+0.3}_{-0.5}$ & $1.2^{+0.9}_{-0.7}$\\
(Component 2) & $\logxi$$^{c}$ & $2.04^{+0.04}_{-0.07}$ & $2.14^{+0.10}_{-0.11}$ & $2.03^{+0.23}_{-0.13}$ \\
& $v_{\rm out}$$^{d}$ & $-470\pm60$ & $-260^{+30}_{-60}$ & $-150^{+130}_{-140}$ \\
$\Delta C$ or $\Delta \chi^{2}$$^{e}$ & -- & 244.6 & 148.1 & --\\
\hline
Warm Absorber & $N_{\rm H}$$^{b}$ & $3.6\pm1.3$ & $1.7^{+0.7}_{-0.6}$ & $2.3^{+2.5}_{-1.4}$\\
(Component 3) & $\logxi$$^{c}$ & $2.80^{+0.05}_{-0.07}$ & $2.88^{+0.11}_{-0.14}$ & $2.9^{+0.4}_{-0.3}$ \\
& $v_{\rm out}$$^{d}$ & $<130$ & $<70$ & $-380^{+200}_{-220}$ \\
$\Delta C$ or $\Delta \chi^{2}$$^{e}$ & -- & 18.5 & 33.7 & --\\
\hline
Partial Coverer & $N_{\rm H}$$^{b}$ & $60.0^{f}$ & $55^{+2}_{-3}$ & $=2011^{t}$\\
(pc 1) & $\logxi$$^{c}$ & $1.0^{f}$ & $1.04^{+0.08}_{-0.11}$ & $=2011^{t}$ \\
& $f_{\rm cov1}$$^{g}$ & $0.61\pm0.05$ & $0.40\pm0.10$ & $0.39\pm0.07$ \\
$\Delta C$ or $\Delta \chi^{2}$$^{e}$ & -- & 213.1 & 193.1 & --\\
\hline
Partial Coverer & $N_{\rm H}$$^{b}$ & -- & $690^{+90}_{-100}$ & $=2011^{t}$\\
(pc 2) & $\logxi$$^{c}$ & -- & $1.04^{f}$ & $=2011^{t}$ \\
& $f_{\rm cov2}$$^{g}$ & -- & $0.37\pm0.10$ & $0.43\pm0.11$ \\
$\Delta C$ or $\Delta \chi^{2}$$^{e}$ & -- & -- & 31.6 & --\\
\hline
Total Flux$^{h}$ & $F_{0.5-2.0}$ & $1.80\pm0.01$ & $1.33\pm0.01$ & $0.75\pm0.01$ \\
& $F_{2.0-10.0}$ & -- & $3.8\pm0.1$ & $2.5\pm0.1$ \\
\enddata
\tablenotetext{a}{Uncovered fraction of power-law component.}
\tablenotetext{b}{Hydrogen column density, units $\times10^{21}$\,cm$^{-2}$.}
\tablenotetext{c}{\,Log ionization parameter.} 
\tablenotetext{d}{Outflow velocity in units km\,s$^{-1}$. Negative values indicate outflow.} 
\tablenotetext{e}{Improvement in either C-statistic (HETG) or $\chi^{2}$ (RGS) upon the addition of the 
component to the model.} 
\tablenotetext{f}{Indicates parameter is fixed.} 
\tablenotetext{g}{Covering fraction of partial covering component}
\tablenotetext{h}{0.5--2.0\,keV or $2-10$\,keV band flux, units $\times10^{-11}$\,erg\,cm$^{-2}$\,s$^{-1}$.}
\tablenotetext{t}{Parameter is tied to the 2011 HETG value.}
\label{absorbers}
\end{deluxetable}

\clearpage

\begin{deluxetable}{lrccccc}
\tabletypesize{\small}
\tablecaption{Soft X-ray Emission Lines in 2011 RGS and Chandra HETG Spectra}
\tablewidth{0pt}
\tablehead{
\colhead{Line ID} & \colhead{E$_{\rm quasar}^{a}$} & \colhead{Flux$^{b}$} & \colhead{EW$^{c}$} 
& \colhead{$\sigma_{\rm v}$$^{d}$} & \colhead{FWHM$^{e}$} 
& \colhead{$\Delta \chi^{2}$ or $\Delta C$$^{g}$}}
\startdata
RGS:- \\
C\,\textsc{vi} Ly-$\alpha$ & $363^{+2.5}_{-3.5}$ [34.155] & $34^{+14}_{-11}$ & $2.5^{+1.0}_{-0.8}$ & 
$4400^{+500}_{-600}$$^{t}$ & $10200^{+1200}_{-1400}$$^{t}$ & 43.9 \\
N\,\textsc{vi} & $419.3^{+1.5}_{-2.0}$ [29.569] & $9.2^{+2.6}_{-3.2}$ & $1.1\pm0.4$ & 
$1800^{+900}_{-600}$ & $4200^{+2100}_{-1400}$ & 16.6 \\
N\,\textsc{vii} Ly-$\alpha$  & $498.7\pm0.2$ [24.861] & $11.3^{+4.5}_{-1.8}$ & $0.9^{+0.4}_{-0.2}$ & 
$340\pm130$ & $780\pm300$ & 75.1 \\
O\,\textsc{vii} (broad) & $564.5\pm0.9$ [21.964] & $38.3^{+4.2}_{-4.9}$ & $8.3^{+0.9}_{-1.1}$ & 
$4400^{+500}_{-600}$ & $10200^{+1200}_{-1400}$ & 345.1 \\
O\,\textsc{vii} (narrow) & $561\pm1$ [22.100] & $4.8\pm2.0$ & $0.9\pm0.4$ & 
$<530$ & $<1250$ & 16.0 \\
O\,\textsc{viii} Ly-$\alpha$ & $654.5\pm1.0$ [18.943] & $6.4\pm1.5$ & $1.7\pm0.4$ & 
$1650^{+1400}_{-700}$ & $3900^{+3300}_{-1600}$ & 28.8 \\
Ne\,\textsc{ix} & $905.1\pm1.3$ [13.698] & $1.2\pm0.5$ & $0.9\pm0.4$ & 
$<1260$ & $<2700$ & 9.7\\
\hline
O\,\textsc{vii} line$^{h}$:-\\
O\,\textsc{vii} (f) & $561.0^{f}$ [22.100] & $26\pm4$ & $5.1\pm0.8$ & $3160^{+400}_{-600}$ 
& $7300^{+1000}_{-1500}$ & --\\
O\,\textsc{vii} (i) & $568.6^{f}$ [21.805] & $9\pm6$ & $1.7\pm1.1$ & -- & -- & --\\
O\,\textsc{vii} (r) & $573.9^{f}$ [21.604] & $<9$ & $<1.7$ & -- & -- & --\\
\hline
HETG:- \\
O\,\textsc{vii} & $567\pm5$ [21.867] & $20^{+11}_{-9}$ & $7.4^{+4.1}_{-3.3}$ & 
$4300^{+2000}_{1600}$ & $9900^{+4600}_{-3700}$ & 17.0 \\
Ne\,\textsc{ix} (narrow) & $905.3\pm0.9$ [13.695] & $0.9^{+0.6}_{-0.5}$ & $1.0\pm0.6$ & 
$<600$ & $<1400$ & 10.6\\
Ne\,\textsc{ix} (broad) & $940^{+6}_{-30}$ [13.190] & $2.6^{+3.2}_{-1.6}$ & $3.0^{+3.7}_{-1.8}$ & 
$2100^{+1900}_{-1300}$ & $4800^{+4400}_{-3000}$ & 9.8\\
\enddata
\tablenotetext{a}{Measured line energy in quasar rest frame, units
  eV.  The corresponding mean wavelength value in
  \AA~ is given
within brackets.}
\tablenotetext{b}{\,Line photon flux, units $\times10^{-5}$\,photons\,cm$^{-2}$\,s$^{-1}$}
\tablenotetext{c}{Equivalent width in quasar rest frame, units eV.}
\tablenotetext{d}{$1 \sigma$ velocity width, units km\,s$^{-1}$.}
\tablenotetext{e}{FWHM velocity width, units km\,s$^{-1}$.}
\tablenotetext{f}{Indicates parameter is fixed.}
\tablenotetext{g}{Improvement in C-statistic or $\Delta \chi^{2}$ upon adding line to model.}
\tablenotetext{h}{RGS deconvolution of broad O\,\textsc{vii} line into forbidden, intercombination 
and resonance line components. } 
\tablenotetext{t}{\,Line velocity width of C\,\textsc{vi} line tied to broad O\,\textsc{vii} line.} 
\label{emission-lines}
\end{deluxetable}

\begin{deluxetable}{lcccc}
\tabletypesize{\small}
\tablecaption{Properties of the fully covered warm absorber
  components (``components 1, 2, and 3'') and the highly ionized component (``component
  high'') discussed
  in $\S$\ref{sec:zonehigh}. The values of $N_{\rm H}$, $\log \xi$ and $v_{\rm
    out}$ are those inferred from the 2011 RGS observations (see
  Table~\ref{absorbers}) for components 1, 2, and 3, and from the HETG
  observations for the highly ionized absorber. 
See text for full definitions of the parameters. }
\tablewidth{0pt}
\tablehead{
\colhead{Parameters} & \colhead{component 1} & \colhead{component 2} & \colhead{component 3}& \colhead{component high}}
\startdata
$N_{\rm H}$ ($\times$10$^{21}$\,cm$^{-2}$) & 2.12$\pm$0.07 &
1.50$\pm$0.20 & 3.6$\pm$1.3 & $>$150\\
$\logxi$ & $1.27\pm0.02$ &
$2.04^{+0.04}_{-0.07}$ & $2.80^{+0.05}_{-0.07}$ & $4.8^{+1.0}_{-0.8}$\\
$v_{\rm out}$$^{d}$ (km\,s$^{-1}$)  & $-480\pm40$ &$-470\pm60$
&$<130$ & $-15600\pm2400$\\
$r_{\mathrm{min}}$ (cm)/(pc)  &  2.8$\times$10$^{19}$/9.0 &
2.9$\times$10$^{19}$/9.4& 3.8$\times$10$^{20}$/ 122 & 2.6$\times$10$^{16}$/0.008\\
$r_{\mathrm{max}}$$^{a}$ (cm)/(pc) & 5.3$\times$10$^{19}$/17.2 &
1.2$\times$10$^{22}$/3940&  8.8$\times$10$^{20}$/285 &2.1$\times$10$^{17}$/0.068  \\
$\dot{M}_{\mathrm{out}}$ ($\times$10$^{25}$\,g/s) &  [1.9-3.6] &
[1.4-560] & [11.8-27.3] & $>4.0$\\
$\dot{M}_{\mathrm{out}}$ ($\dot{M}$/yr) & [0.3-0.6] & [0.2-89] &
[1.9-4.3] & $>0.6$\\
$\dot{E}_\mathrm{K}$/$L_{\rm bol}$$^{b}$ ($\%$)  & [$5.1\times10^{-4}$-$9.8\times10^{-4}$]
 &
[$3.5\times10^{-4}$-0.14] & [$2.4\times10^{-4}$-$5.4\times10^{-4}$] & $>1.1$ \\
$\dot{P}_\mathrm{out}$/$\dot{P}_\mathrm{rad}$ ($\%$)& [0.63-1.2]& [0.45-184] & [1.1-2.5]& $>46$ \\
\hline
\enddata
\tablenotetext{a}{$r_{\mathrm{max}}$ is inferred from
  eq.~\ref{eq:rmax}, except for component 1 for which $r_{\mathrm{max}}$
  corresponds to $r_{\mathrm{var}}$ inferred from recombination
  time-scale, see $\S$\ref{sec:recomb}.}
\tablenotetext{b}{$L_{\rm bol}=4.3\times 10^{45}$\,erg\,s$^{-1}$ \citep{dunn2008},
  and $L_{\rm Edd}=3.0\times 10^{46}$\,erg\,s$^{-1}$. }
\label{energetics}
\end{deluxetable}

\end{document}